**И. З. ШКУРЧЕНКО**

# ДВИЖЕНИЕ ТВЁРДЫХ ТЕЛ В ЖИДКОСТЯХ И ГАЗАХ С ТОЧКИ ЗРЕНИЯ МЕХАНИКИ БЕЗЫНЕРТНОЙ МАССЫ

**г. Нижнекамск**

**начато: 14 сентября 1974 г.**
**закончено: 9 ноября 1974 г.**

**СОДЕРЖАНИЕ:**

стр.

## ОТ АВТОРА

Название работы выглядит несколько суховато и кажется в определённой степени устаревшим. Ведь самолёты летают, корабли плавают, насосы и турбины работают. Всё это так и есть лишь до тех пор, пока кому-нибудь не захочётся сделать самостоятельно что-либо плавающее, летающее и т.д. В этом случае такой человек в первую очередь обращается к теории, к расчётам всего этого плавающего, летающего, чтобы выбрать и рассчитать для себя необходимый вариант. И в конечном итоге он убеждается, что всё сводится к продувкам и проливкам, к пересчёту, а не к расчёту. После чего ему остаётся лишь сожалеть о потраченном времени. Ибо даже сам смысл того, как создавать самое быстроходное, самое компактное, в общем, «самое, самое» не улавливается в существующей теории механики жидкости и газа. Потому что там его нет.

Поэтому человек, пытающийся действовать и создавать по существующей теории, остается в недоумении. Если же он обратится за разъяснением к инженеру, уже работающему в области движения твёрдых тел в жидкостях и газах, или к доктору наук, или даже к академику, то он не получит ни от кого толковых разъяснений, которые необходимы ему для его деятельности. Оказавшись в таком положении, человек поступает просто: он отказывается от изучения теории и берёт за прототип какую-либо конструкцию, которую пытается усовершенствовать на свой манер, пользуясь отчасти практическими советами других. Иногда у него что-то получается, а в большинстве случаев у него ничего не получается. Но в сумме получается всегда то, что и этот человек, и академик после проделанной работы будут считать себя крупными специалистами.

Поэтому, когда случается разговор между подобными людьми, то в своей аргументации они надеются на силу своих голосовых связок, а не на знание необходимых законов природы. Вы, конечно, догадываетесь о конце разговора. Доктор наук или академик будут стараться задавить не в меру любопытного человека своим авторитетом, чтобы сохранить его, но сущность дела от этого не изменится и не прояснится. Современные академики поступают даже проще в подобных ситуациях: они просто не разговаривают с такого рода людьми и тем самым поддерживают свой авторитет.

Всё дело здесь заключается в том, что современная механика жидкости и газа базируется не на своих законах, а на законах механики твёрдого тела. По этой причине современная механика жидкости и газа просто неверна, что ставит вышеназванного человека и академика на один уровень. Ибо один делал что-то практически, а другой пропагандировал неверные знания. В конечном итоге, как говорится, оба слышали звон, да не знают, откуда он.

В данной работе мы дадим положения о движении твёрдых тел в жидкостях и газах в соответствии с естественными законами природы (см.[1]). Поэтому каждый прочитавший предлагаемую работу и осмыслённо усвоивший её положения сможет самостоятельно создавать любые технические устройства, как плавающие, так и летающие. Например, если такой пожелает создать летательный аппарат с машущими крыльями, то он найдет здесь качественное понимание машущего полёта и все необходимые количественные расчёты, то есть всё, что необходимо человеку для целенаправлённой деятельности. Также понятие об обтекаемости твёрдого тела жидкостями и газами в данной работе отличается качественно и количественно от понятий, которые в настоящее время считаются общепринятыми.

Основные законы теории механики жидкости и газа, или безынертной массы, с декабря 1969 года находятся в виде заявок на предполагаемое открытие в Комитете по делам изобретений и открытий при Совете министров СССР. «Механика жидкости и газа, или механика безынертной массы» была написана мной в 1971 году, и с 5 января 1972 года находится в Институте проблем механики АН СССР (Москва, Ленинградский проспект, д. № 7). К большому сожалению, до настоящего времени все вышеперечислённые работы находятся без движения.<...>

После того, как меня уволили с работы по сокращению штатов, несмотря на то, что я проработал девять лет по специальности «двигатели летательных аппаратов», мне пришлось искать работу. Всё это длилось в течение полугода. После чего мне удалось устроиться на работу в воронежский филиал союзной организации «Оргнефтезаводы». Эта организация является пусконаладочной организацией, поэтому работа в ней связана с длительными командировками. Так я оказался в городе Нижнекамске, поселённым в общежитие. У меня не осталось условий не только для работы, но и для элементарного человеческого существования.

Всё вышеизложенное дано для того, чтобы объяснить всем тем, кто когда-либо прочитает мою работу и не найдет в её выкладках учёта вязкости и сопротивления жидкостей и газов, которое принято считать сопротивлением трения, почему эти выкладки даны, как принято считать в таких случаях, в идеальной форме.

До настоящего времени самолётостроение, кораблестроение и прочее подобное «строение» основано на эксперименте, а не на теории. По вязкости и по сопротивлению трения накоплен богатый экспериментальный материал. Но мне в моих условиях просто невозможно им воспользоваться. В будущем, если у меня появится хот бы небольшая возможность, я устраню этот пробел. Для этого не потребуется большого труда, а лишь необходимо иметь соответствующие данные и некоторое количество времени. В то же время нашей основной задачей является необходимость сделать теорию руководством к действию, а эксперимент – лишь проверкой расчёта, так как сейчас всё делается наоборот.

Далее идут трудности такого характера, что мне приходится каждый раз переписывать основные законы и положения механики безынертной массы, когда я приступаю к очередной работе.

За неимением времени мне придется дать в данной работе основные законы и положения механики безынертной массы в предельно сжатой форме, т.е. без дополнительных пояснений. Кого эта механика заинтересует в полной форме, тот может обратиться в Институт проблем механики АН СССР. <...>

## ВВЕДЕНИЕ
## ОСНОВНЫЕ ЗАКОНЫ И ПОЛОЖЕНИЯ МЕХАНИКИ БЕЗЫНЕРТНОЙ МАССЫ

В самом общем случае в механике безынертной массы мы принимаем жидкости и газы как невязкие, несжимаемые жидкости, то есть как идеальную жидкость. Это значит, что единственным физическим различием для них служит плотность. Объёмно мы воспринимаем идеальные жидкости как пространство, полностью заполненное такой жидкостью. Это объёмное восприятие жидкостей и газов мы называем средой.

Далее жидкости и газы образующие среду мы определяем как сплошную однородную массу, не делящуюся даже на молекулы или атомы по всему объёму среды. Тогда движение идеальной жидкости мы можем представить себе в общем виде в следующей форме:

1. что при движении перемещается вся масса жидкости, заключенная в исследуемом пространстве;
2. что границы этого пространства определяют нам форму потока движущейся массы;
3. что количественной характеристикой любого движущегося потока является расход массы в единицу времени.

Это значит, что пространственно любое движение жидкости мы можем представить себе по границам, в которых размещается движущийся объём жидкости, и который мы называем потоком.

Непосредственно само движение жидкостей мы должны воспринимать как вытеснение текучей массы из объёма, в отличие от движения твёрдого тела, которое мы воспринимаем как перемещение в пространстве. Ведь понятие среды включает в себя пространство, а при движении жидкости границы среды совпадают с границами потока.

Теперь мы можем перейти к основным законам механики безынертной массы. Первый её закон называется “Закон сохранения состояния”. Он формулируется таким образом:

***Жидкости и газы сохраняют энергию покоя и установившегося движения только в силовом поле и изменяют её лишь при изменении этого поля.***

Этот закон определяет общность обращения с жидкостями и газами. Всем хорошо известно, что местным или сосредоточенным приложением силы нельзя удержать жидкости и газы в состоянии покоя или установившегося движения. Для этих целей к ним надо приложить распределённые силы, которые можно получить с помощью силового поля. В природе существует множество силовых полей. Одни из них взаимодействуют с жидкостями и газами, другие – нет. Нас будут интересовать те из них, которые взаимодействуют с жидкостями и газами. Далее разделим интересующие нас силовые поля на векторные и скалярные. Примером векторного силового поля служит гравитационное поле Земли, примером скалярного силового поля могут служить границы сосуда, где сжатый воздух находится под давлением.

Второй закон механики безынертной массы называется “Уравнение расходного вида движения”. Запишем его формулировку:

***Динамические силы давления ($P_{дин}$) равны произведению расхода массы в единицу времени (M) на линейную скорость (W) и делённому на площадь сечения потока (F), или***

$$P_{дин} = \frac{M}{F}\,W. \qquad (1)$$

Этот закон определяет общность сил, которые во всех случаях действуют в жидкостях и газах.

Вы сразу же заметили, что в уравнении сил (1) отсутствует ускорение, а есть скорость и расход массы[1]. Отсутствие ускорения указывает на то, что масса не может запасать механическую энергию. В этом можно убедиться на многочислённых примерах. Например, жидкость движется в трубах лишь в том случае, когда её гонит насос. Как только мы останавливаем насос, то движение жидкости сразу прекращается, т.е. движение по инерции отсутствует. Далее вы знаете о существовании центростремительных турбин и принцип их работы. Рабочее тело, имеющее инерцию, не может осуществлять рабочий цикл по такой схеме. Безынертность массы жидкостей и газов, т.е. подобное её состояние возможно лишь в пространстве среды.

Третий закон механики безынертной массы называется “Формальный принцип связи вида движения с формой уравнения движения”. Запишем этот закон:

***Отсутствие или наличие в уравнении движения параметров пространства и времени определяет его связь с тем или иным видом движения***

Смысл этого закона заключается в том, что он дает возможность по количеству сочетаний параметров пространства и времени определить полное число видов движения жидкостей и газов. Таких сочетаний можно составить четыре. Это значит, что жидкости и газы имеют четыре вида движения, а не бесчисленное множество, как принято сейчас считать. Перечислим все виды движения жидкостей и газов:

1. **установившийся вид движения**
   (отсутствуют параметры пространства и времени);
2. **плоский установившийся вид движения**
   (отсутствует параметр времени и присутствует параметр пространства);
3. **расходный вид движения**
   (отсутствует параметр пространства и присутствует параметр времени);
4. **акустический вид движения**
   (присутствуют параметры пространства и времени).

Ниже мы разберём каждый из этих видов движения, и вы поймете, что они собой представляют.

Отметим ещё один важный принцип движения жидкостей и газов. Он гласит:

**жидкости и газы могут двигаться одновременно либо только в прямом направлении, либо во взаимно перпендикулярных направлениях.[2]**

Для любого механического движения должны быть записаны необходимые зависимости. Для движущихся жидкостей таких уравнений должно быть два. Одно из них будет описывать непосредственно само движение жидкости и газа[3], другое – силы, действующие на жидкости и газы при конкретном их движении.

---

[1] В оригинале рукописи расход массы обозначен буквой $M$ с точкой сверху. По техническим причинам редактор обозначил расход массы в единицу времени буквой $M$ без точки. Параметр массы будет обозначен буквой m.

[2] см. работу [1] Реальное движение среды, в т.ч. по окружности, может представлять сложную комбинацию четырёх видов движения, которые рассматриваются ниже, и к которым относится данный принцип (прим.-ред.)

[3] см. работу [1] - движением массы среды (текучестью) является расход массы в единицу времени, скорость является лишь характеристикой расхода. При уменьшении площади сечения потока его скорость увеличивается, но не потому, что вещество среды ускоряется, как инертные тела, они же - инерционные. Увеличение скоростей не является ускорением в принципе. О чём и речь. Силы давления вызывают расход массы на площади сечения, т.е. в реальности на площади действия сил, который должен быть неизменным при неизменной величине силового воздействия на весь поток. Поэтому при изменении величины площади сечения соответственно изменяется характеристика расхода – скорость, поскольку изменяется соотношение потенциальной и кинетической энергии силового поля, т.е. изменяются величины статических и динамических сил давления, что проявляется в изменении скорости. Каждой площади сечения соответствует своя скорость, хоть эти сечения, как место действия сил, как бы мнимые, но факт, что силы действуют во всём пространстве среды, между плоскостями их действия нет промежутков. Все уравнения движения являются уравнениями расхода массы в единицу времени. Вещество как бы выдавливается из плоскости в плоскость. Но не всё так просто, коль и сами силы могут как бы передвигаться в пространстве среды (напр. акустический поток, гидроудар) (прим. ред.).

Прежде чем приступить к записи этих уравнений, мы сначала должны оговорить способы их условной записи. В комплекс приёмов записи и изображения входят:

система единиц измерения, система координат и приёмы изображения движения и равновесия механических состояний жидкостей и газов. Также мы должны будем оговорить условия движения и состояния покоя жидкостей и газов.

Для механики безынертной массы принимаем следующие условия движения и состояния покоя:

1. пространство считаем неподвижным и полностью заполненным жидкостью или газом, что выше мы определили как среду;
2. время считаем непрерывным и постоянным в своем движении;
3. различные дополнительные факторы берутся из конкретных условий движения и равновесия.

Комплекс приёмов записи и изображения движения и равновесия будет заключаться в следующем:

1. за систему единиц принимаются общепризнанные системы единиц для записи пространства, времени, силы, массы, скорости и т.д.;
2. для изображения условий движения и равновесия принимается плоскость (поверхность), для которой конкретно записываются условия движения и равновесия. В соответствии со вторым законом механики безынертной массы, эта плоскость (поверхность) исследования располагается обязательно перпендикулярно направлению движения;
3. условия движения и равновесия рассматриваются в одних плоскостях относительно других, то есть в каждом конкретном случае для исследования движения потока мы размещаем плоскости (поверхности) в необходимых для исследования местах. Затем делаем записи пространственного расположения этих плоскостей относительно избранной системы координат.

Теперь перейдем непосредственно к записи необходимых зависимостей для четырёх видов движения жидкостей и газов.

## 1. УСТАНОВИВШИЙСЯ ВИД ДВИЖЕНИЯ

Наглядной формой такого движения может служить движение жидкости на прямолинейных участках труб. При этом эти участки могут иметь плавно сужающиеся или расширяющиеся границы. Такой участок установившегося потока покажем на *рис.1*. На этом рисунке показано, что сечение установившегося потока может быть как круглой, так и прямоугольной формы, то есть сечение потока обязательно должно быть, как минимум, симметричным относительно двух взаимно перпендикулярных плоскостей.

Исследуемый поток может иметь материальные границы, но в этом случае мы полагаем, что они не обладают сопротивлением, которое принято называть сопротивлением трения. Здесь происходят процессы взаимодействия между материальной границей потока и движущейся жидкостью, которые связаны с взаимодействием их силовых полей. Поэтому мы чисто условно называем это сопротивление сопротивлением трения.[4]

Так как мы в своей работе идеализируем условия движения жидкости, то мы это сопротивление не учитываем. Поэтому в этом случае материальные границы потока будут совпадать с границами потока, т.е границами среды. Данное положение относится ко всем четырём видам движения жидкости.

На рис. 1 показан установившийся поток жидкости общего характера. Также показаны плоскости $S$, $S_1$, $S_2$, $S_3$ и $S_4$, которые обозначают места наших исследований потока при выводе уравнений движения и сил этого потока. Все эти плоскости являются чисто условными, или мнимыми. Они означают, что мы в этих местах потока составили условия равновесия или движения. Буквой $W$ обозначены линейные скорости потока по его сечениям. Буквой $F$ обозначены площади сечения потока, которые могут быть либо квадратными, либо круглыми. Эти обозначения мы будем применять и для остальных видов движения.

---

[4] В связи с краткостью изложения основных положений механики безынертной массы автор называет сопротивление из-за взаимодействия границ потока с потоком сопротивлением трения условно, т.к. это сопротивление имеет другую природу, см. [1]. К такому же приёму условности автор прибегает при описании акустического вида движения. См. сноску 6.

Теперь без всяких выводов, которые даны в работе [1], запишем уравнения движения и сил для установившегося потока жидкости.

*уравнение движения:*

$$M = F\rho W = \text{const}. \qquad (2)$$

Оно означает, что массовый расход жидкости в единицу времени $M$ в любом сечении потока постоянен и не зависит от времени и пространственных характеристик. Движение по сечениям потока различается по линейной скорости $W$ и по площади сечения $F$. Уравнение (2) хорошо известно. Оно называется уравнением неразрывности в древней механике жидкости и газа.

*уравнения сил установившегося потока:*

$$FP_{\text{пр.ст}} + FP_{\text{пр.дин}} = FP_{\text{ст}} + FP_{\text{дин}}, \qquad (3)$$

$$P_{\text{пр.ст}} + P_{\text{пр.дин}} = P_{\text{ст}} + P_{\text{дин}}, \qquad (4)$$

$$P_{\text{пр.дин}} = \rho W^2 . \qquad (5)$$

Уравнения (3), (4) и (5) являются уравнениями сил установившегося потока жидкости. Эти уравнения получены методом мысленного рассечения потока плоскостью исследования и составлением условий равновесия.

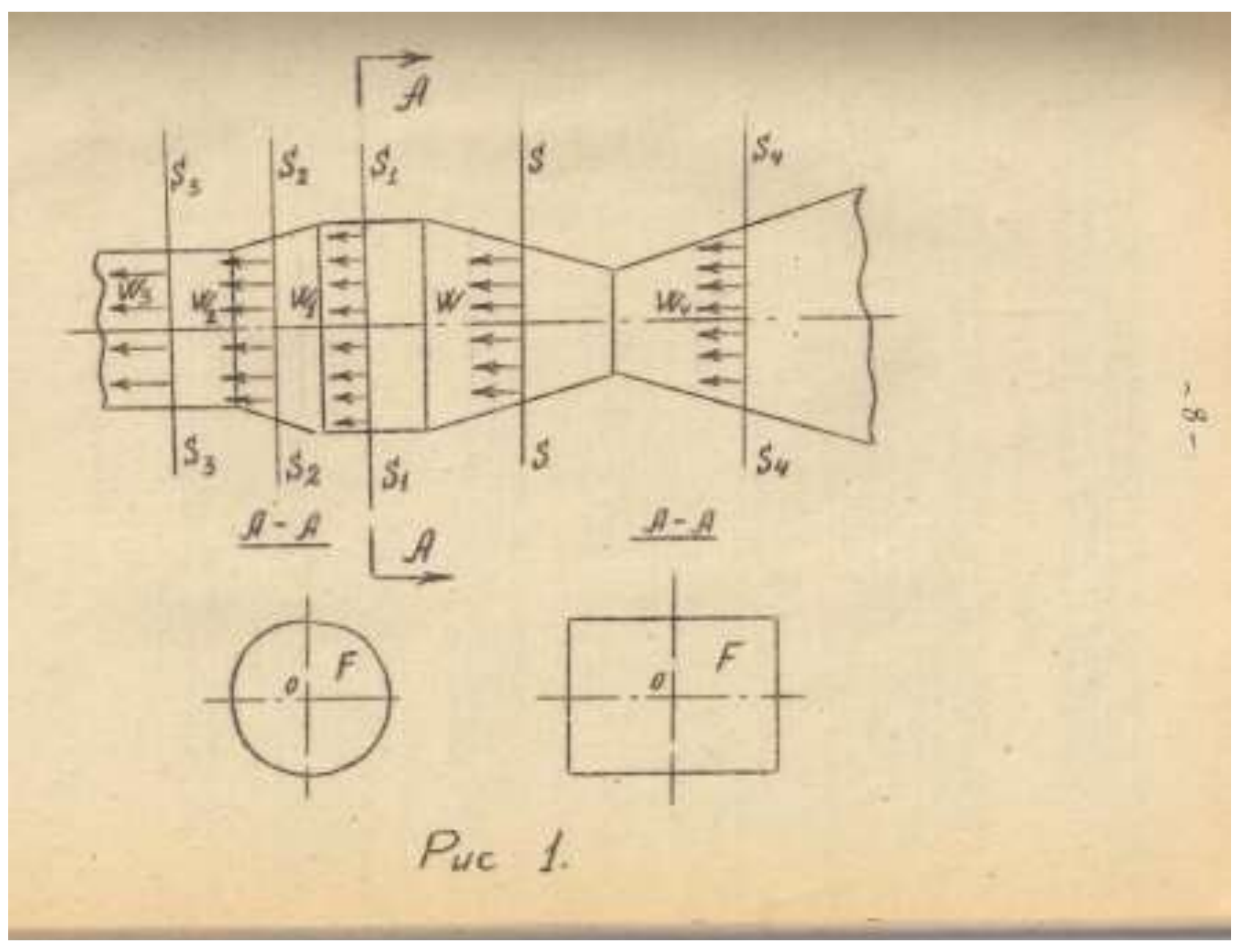

Рис 1.

В потоке жидкости существуют динамические и статические силы давления. Поэтому мы должны были компенсировать действия сил мысленно отброшенной части потока *принятыми* динамическими и статическими силами давления, чтобы сохранить условия равновесия в потоке.

Уравнение (3) является полным уравнением сил установившегося потока. В левой части с индексом $_{пр}$ стоят принятые статические и динамические давления, умноженные на площадь сечения потока $F$. В правой части уравнения (3) стоят действительные силы давления потока.

Уравнение (4) тоже является полным уравнением сил, но в этом случае оно записано относительно единицы площади сечения потока.

Уравнение (5) является частным уравнением сил установившегося потока для единицы площади его сечения. Непосредственно динамические силы давления выражены здесь через характеристики потока, которые преображаются в соответствующий вид с помощью второго закона механики безынертной массы и уравнения движения (2) этого потока.

Величина статических сил давления определяется по уравнениям энергии установившегося вида движения жидкости, на которых мы остановимся ниже. После этого цепь необходимых зависимостей установившегося вида движения замкнётся, то есть с их помощью мы сможем получать любые интересующие нас характеристики установившегося потока.

## 2. ПЛОСКИЙ УСТАНОВИВШИЙСЯ ВИД ДВИЖЕНИЯ

Плоский установившийся вид движения вам тоже хорошо знаком. Его можно наблюдать в виде водоворотов на реке, в виде смерчей и циклонов в атмосфере и т.д. На рис. 2 изобразим плоский установившийся поток и запишем для него уравнения движения и сил.

Плоский установившийся поток имеет цилиндрический объём, ограниченный внутренней и наружной цилиндрическими поверхностями. Особенно наглядно эти границы видны в турбинах и центробежных насосах. Жидкость может подходить к плоскому установившемуся потоку через наружную его поверхность, а выходить через внутреннюю граничную поверхность в виде установившегося потока. Или наоборот – может входить в объём потока через внутреннюю поверхность, а выходить – через наружную. В первом случае мы будем иметь плоский установившийся поток турбинного типа, во втором – насосного типа. Непосредственно само движение жидкости мы рассматриваем только в цилиндрическом объёме плоского установившегося потока.

Следующей особенностью этого потока является то обстоятельство, что он образуется на границе двух сред с различными энергетическими уровнями.

На рис. 2, согласно характеру внешнего притока жидкости к плоскому установившемуся потоку и характеру оттока жидкости от него, показан поток турбинного типа.

Расход массы в единицу времени в плоском установившемся потоке является постоянным и не зависит от времени.

Жидкость в объёме этого потока движется одновременно в двух взаимно перпендикулярных направлениях: в радиальном и тангенциальном, то есть по направлению радиуса и по окружности. Соответственно для всех характеристик потока: скоростей, площадей сечения и т.д, применяются обозначения $r$ и $tg$.

Тангенциальные площади сечения $F_{tg}$ имеют форму плоскости, а радиальные площади сечения $F_r$ – форму цилиндра. Тангенциальные площади сечения $F_{tg}$ в любом сечении одинаковы и равны между собой, а величины радиальных площадей сечения $F_r$ уменьшаются к внутренней поверхности и увеличиваются к наружной граничной поверхности. Соответственно ведут себя и линейные скорости движения жидкости в объёме потока. Тангенциальные скорости движения $W_{tg}$ жидкости одинаковы по всему сечению потока и равны определённой величине. Радиальные скорости движения жидкости $W_r$ изменяются от сечения к сечению потока, поскольку величины площади в радиальном направлении различны. Соответственно будут записаны уравнения движения и сил для плоского установившегося потока.

<u>*уравнения движения:*</u>

*для радиальной площади сечения:* $M_r = F_r \rho W_r = 2\pi r_n h \cdot \rho W_r;$ (6)

*для тангенциальной площади сечения:* $M_{tg} = F_{tg}\ \rho W_{tg}$ (7)

при этом**:** $M_r = M_{tg} = \text{const}.$

<u>*уравнения сил:*</u>

*для радиальной площади сечения:* $F_r P_{\text{пр}\ r} = F_r P_{\text{ст}\ r} + F_r \rho W_r^2;$ (8)

*для тангенциальной площади сечения:* $F_{tg} P_{\text{пр}\ tg} = F_{tg} P_{\text{ст}\ tg} + F_{tg} \rho W_{tg}^2.$ (9)

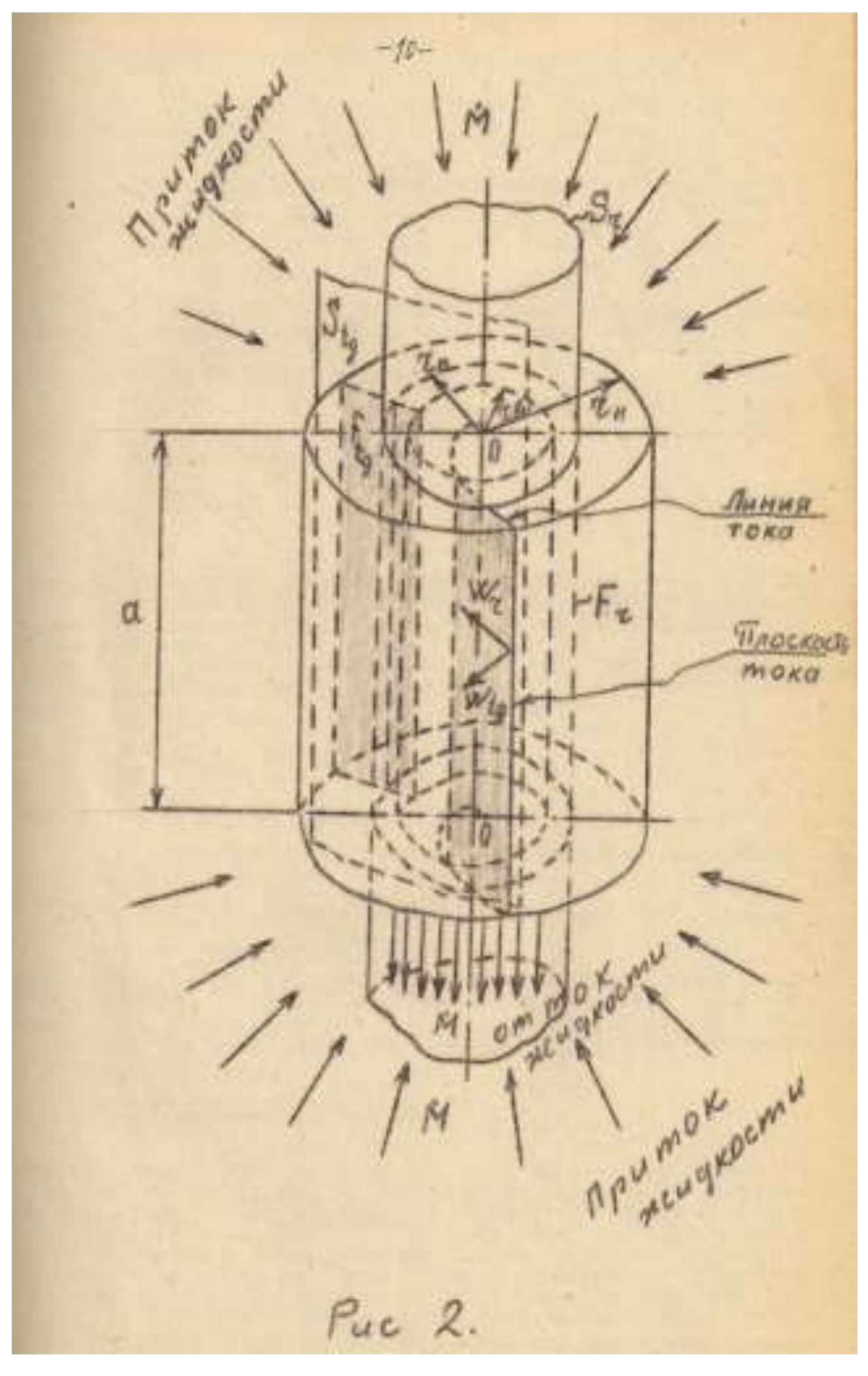

Рис 2.

Запись уравнений сил только для динамических сил давления для единицы площади будет иметь вид:

*для радиальной площади сечения:* $P_{\text{пр дин}\ r} = \rho W_r^2;$ (10)

*для тангенциальной площади сечения:* $P_{\text{пр дин}\ tg} = \rho W_{tg}^2.$ (11)

В этих уравнениях сил динамические силы давления сразу записаны через характеристики потока. Вы видите, что особой разницы между уравнениями плоского установившегося вида движения и просто установившегося вида движения нет.

Нам осталось записать зависимость для линий тока плоского установившегося потока. Они будут иметь зависимость логарифмической спирали:

$$r = ae^{\kappa\varphi},$$

где $\varphi$ – угол полярной системы координат; $a$ – некоторая величина, относящаяся к конкретной спирали; $\kappa = \text{ctg}\ \alpha$. В нашем случае угол $\alpha = 45°$, поэтому ctg $\alpha = 1$.

Линии тока плоского установившегося потока видны на его торцевых поверхностях в виде логарифмической спирали. В самом объёме потока они существуют в виде плоскостей тока, образующими для которых служат логарифмические спирали.

Мы получили необходимые зависимости для плоского установившегося потока.

## 3. РАСХОДНЫЙ ВИД ДВИЖЕНИЯ

Наглядной формой расходного потока может служить футбольная или какая-нибудь другая камера, накачиваемая насосом. Расходный поток существует в любом объёме, где нагнетается или откачивается жидкость или газ под давлением. В более общей форме расходный поток представляет собой объём с подвижными граничными стенками, куда через множество штуцеров одновременно подводится и отводится жидкость с различной энергией. Покажем расходный поток на рис. 3.

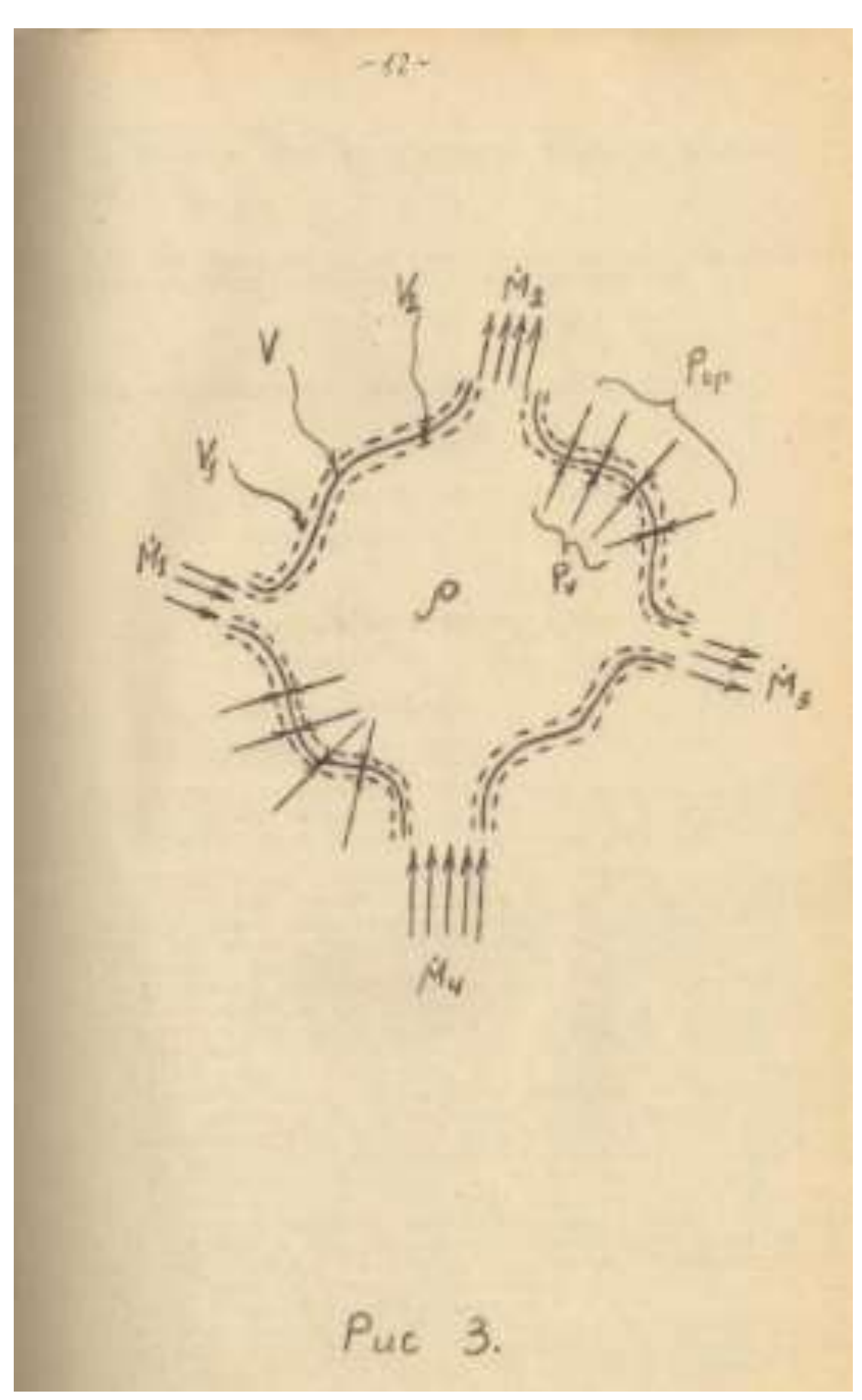

Рис 3.

На этом рисунке мы приняли объём расходного потока с подвижными границами.

***Уравнение движения*** расходного вида движения жидкостей будет иметь следующий вид:

$$m = \rho V = t\sum_{1}^{n} M_i - t\sum_{1}^{n} M_j \,, \qquad (12)$$

где $m$ – масса, заключенная в объёме потока; $V$ – объём потока; $M$ – расход массы в единицу времени; $\rho$ – плотность.

Для газов это уравнение движения будет иметь несколько иной вид, поскольку оно должно учитывать сжимаемость газов. Запишем его:

$$\rho \frac{dV_t}{dt} + V \frac{d\rho_t}{dt} = \sum_{1}^{n} M_i - \sum_{1}^{n} M_j \,. \qquad (13)$$

*уравнение сил расходного потока:*

$$F_t P_{t\,\text{дин}} = F_t\,(P_{t\,v} - P_{t\,\text{ср}}) = \sum_{1}^{n} M_{ti} W_{tj} - \sum_{1}^{n} M_t W_t \,. \qquad (14)$$ [5]

## 4. АКУСТИЧЕСКИЙ ВИД ДВИЖЕНИЯ

Акустический вид движения называют ещё волновым. С помощью этого вида движения звуки и шумы передаются на расстояние, образуются ударные волны, различные волнения на поверхности морей и океанов. Акустические волны разделяются на волны сжатия и волны разряжения.[6]

Для получения необходимых зависимостей мы воспользуемся пластиной с бесконечно большой площадью, совершающей возвратно-поступательное движение. Непосредственно для вывода уравнений мы взяли лишь часть площади движущейся пластины с радиусом $r$, как показано на рис. 4.

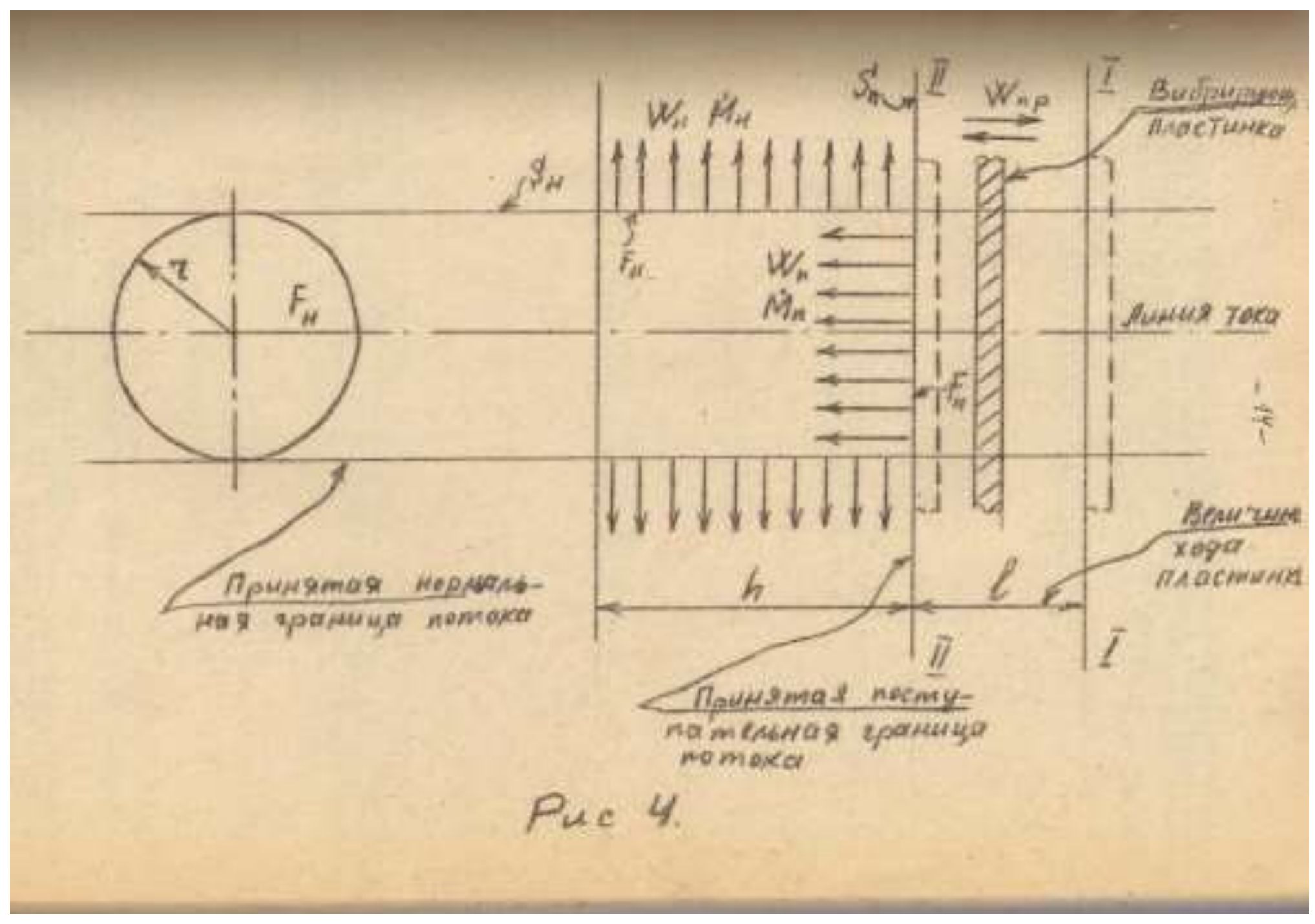

Рис 4.

[5] В уравнениях (13), (14) индексы не разборчивы для редактора. Их нужно уточнять по работе [1]

[6] Для краткого описания акустического вида движения автор воспользовался положениями существующей гидромеханики. В данном случае термины «волна сжатия/разрежения» лишь выполняют назначение аналогий для тех, кто не знаком с работой [1] (прим.ред)

Далее принимаем, что вибрирующая пластина совершает возвратно-поступательное движение со скоростью $W_{пл}$ в пределах границ I и II на расстояние $l$, которое является величиной хода пластины.

На рис. 4 показано образование акустической волны только с одной стороны пластины, которая расположена в направлении II. При движение пластины из положения I в положение II будет образовываться акустическая волна сжатия, а при движение пластины из положения II в положение I будет образовываться волна разрежения. Нормальной плоскостью $S_н$ и поступательной плоскостью $S_п$ выделены поверхности исследования. Чтобы получить уравнения движения и сил, мы рассмотрим движение пластины поэтапно: сначала из положения I в положение II, затем из положения II в положение I. Для этих двух положений пластины составляем уравнения движения и сил, как для волны сжатия, так и для волны разрежения. Перед началом движения вибрирующей пластинки среда, в которой она размещается, находится в состоянии покоя

С началом движения пластинки происходит возмущение среды одновременно в двух взаимно перпендикулярных направлениях: поступательном и нормальном. В поступательном направлении площадь потока $F_п$ не изменяется за всё время движения пластинки. В нормальном – она непрерывно увеличивается. Это происходит по той причине, что возмущение в среде происходит со скоростью звука ($C$). Вот эта скорость определяет нам площадь нормального потока за время движения пластины. На рис. 4 увеличивающаяся площадь нормального потока $F_н$ обозначена буквой $h$.[7]

Тогда уравнения движения и сил запишутся в таком виде:

*уравнения движения для волны сжатия*

*в поступательном направлении:* $$M_n^{I}(t) = F_n \rho W_n(t)\ ; \qquad (15)$$

*в нормальном направлении:* $$M_н^{I}(t) = -\rho W_н(t) \cdot 2\pi r (C \cdot l / W^{I}_{пл}(t) - l). \qquad (16)$$

*уравнения движения для волны разрежения:*

*в поступательном направлении:* $$M_n^{II}(t) = -F_n \rho W_n(t); \qquad (17)$$

*в нормальном направлении:* $$M_н^{II} = \rho W_н(t) \cdot 2\pi r (C \cdot l / W^{II}_{пл}(t) + l), \qquad (18)$$

*уравнения сил для волн сжатия и разрежения:*

*в поступательном направлении:* $$F_n P_{n\,пр}(t) = F_n P_{n.ст}(t) + M_n W_n(t); \qquad (19)$$

*в нормальном направлении:* $$F_н(t) P_{н.пл}(t) = F_н(t) P_{н.ст}(t) + M_н(t) W_н(t). \qquad (20)$$

где $C$ – скорость звука,

$l$ – величина хода пластины,

$W_{пл}$ – скорость движения пластины,

$r$ – радиус поступательной принятой площади сечения $F_п$ потока.

В этих уравнениях некоторые характеристики имеют знак ($t$). Он означает, что эти величины могут быть переменными во времени. Эта переменность зависит от того, с постоянной скоростью ($W_{пл}$) или нет движется вибрирующая пластинка. Нормальная площадь сечения потока и скорость всегда являются переменными во времени.

---

[7] Конечно, в смысле, что площадь нормального потока ($F_н = 2\pi r h$), увеличивается за счёт увеличения $h$.(прим.ред.)

Будем считать, что мы познакомились с четырьмя видами движения жидкости и газа: получили для них уравнения движения и сил. Но этих зависимостей ещё недостаточно, чтобы решать практические задачи. Для этого необходимо знать зависимости работы и энергии.

## 5. РАБОТА И ЭНЕРГИЯ

Под работой мы всегда понимаем работу, которая уже выполнена кем-то или чем-то. Под энергией мы тоже понимаем работу, но располагаемую, то есть такую, которую мы можем получить от имеющегося источника энергии. Работа совершаемая в настоящий момент называется мощностью. Поэтому мощность количественно определяется тоже работой, совершенной в единицу времени. Все эти характеристики при одинаковой размерности имеют различное назначение в науке и технике. Поэтому определяются различными зависимостями.

Переходим непосредственно к работе. Работа означает новое количественное понятие, которое не зависит от переменности движения жидкостей и газов и определяет в общей сумме прошедшее движение. Поэтому выше мы определили работу как относящуюся к прошлому движению жидкостей и газов. Это новое понятие не даётся при помощи каких-либо выводов, а берётся непосредственно из практических наблюдений. Запишем определение работы:

**Работой (*L*) сил давления (*P*) на неподвижной поверхности или плоскости потока называется произведение объёма жидкости (*V*) на силы давления (*P*), или:**

$$\boldsymbol{L = VP}. \tag{21}$$

Отсюда следует, что работа ($L$) не зависит ни от плотности, ни от температуры, ни от каких-либо других характеристик жидкостей и газов. По этой причине работа является очень удобной величиной.

Работу, совершенную в настоящий момент, мы называем мощностью. Мощность определяется количеством работы в единицу времени. Мощность ($N$) можно выразить через работу ($L$) следующей зависимостью:

$$N = \frac{L}{t} = \frac{VP}{t}. \tag{22}$$

Что касается энергии, или располагаемой работы, то зависимость для неё была давно получена Даниилом Бернулли. Нам остаётся только переписать её:

$$VP_{\text{пол}} = VP_{\text{ст}} + \frac{1}{2}VP_{\text{дин.}} \tag{23}$$

Смысл этой записи вам хорошо известен: полная энергия объёма $V$ равна сумме потенциальной ($VP_{\text{ст}}$) и кинетической ($1/2\,VP_{\text{дин}}$) энергий потока.

Для единицы объёма уравнение Бернулли имеет такой вид:

$$P_{\text{пол}}(1/\text{м}^3) = P_{\text{ст}}(1/\text{м}^3) + \frac{1}{2}\rho W^2(1/\text{м}^3). \tag{24}$$

Для газового потока, где необходимо учитывать энергию сжатия газа, к этим уравнениям добавляется ещё один член. В термодинамике энергия сжатия газа учитывается такой зависимостью:

$$L = \int_{V_1}^{V_2} PdV\,. \tag{25}$$

Если мы подставим уравнение (25) в уравнение (23), то получим уравнение энергии для газа в таком виде:

$$VP_{\text{пол}} = VP_{\text{ст}} + \frac{1}{2}VP_{\text{дин}} + \int_{V_1}^{V_2} PdV\,. \tag{26}$$

Мы получили все необходимые зависимости для работы и энергии. Теперь определим их отношение к каждому из четырёх видов движения.

1. Установившийся вид движения с механической точки зрения служит для переноса и передачи энергии на расстояние. По этой причине уравнение энергии записывается как уравнение (23) или (26).

2. Плоский установившийся вид движения характеризует переход жидкостей и газов из среды с большей энергией в среду с меньшей энергией, который осуществляется с выделением механической работы. Или, наоборот, этот вид движения характеризует переход жидкостей и газов из среды с меньшей энергией в среду с большей энергией, который происходит при непосредственном подводе механической работы к жидкости в общем объёме.

3. При акустическом движении происходит переход работы механического движения пластины в энергию волны сжатия или разрежения.

4. Объём потока расходного вида движения может быть либо накопителем энергии, либо объёмом совершенной работы в зависимости от того, какое назначение он имеет.

Отметим, что энергия покоя жидкостей и газов земной атмосферы и гидросферы определяется следующей зависимостью:

$$E_{\text{в}} = Vh\rho w^2 \text{ [2].} \tag{27}$$

Уравнение (27) означает, что энергия состояния покоя ($E_{\text{в}}$) для площади сечения застывшего потока расположенной на глубине ($h$) равна произведению объёма потока ($V$), который расположен над площадью сечения, глубины потока ($h$), плотности жидкости или газа ($\rho$) и квадрата скорости ($w$) постоянной для жидкостей и газов в поле земного тяготения - $w$ = 3,132 м/сек [1]. Начало отсчёта глубины $h$ находится на высоте порядка 25 км для земной атмосферы [см. 2].

Вот всё, что мы можем кратко сообщить о механике безынертной массы. Далее нам придётся использовать сравнительно небольшое количество преобразующих положений, вполне доступных для понимания любому человеку. Поскольку приходится писать быстро, без каких-либо дополнительных редакций и уточнений, то в работе могут возникнуть неточности, которые будет нетрудно устранить самостоятельно, если вы разобрались с законами и положениями механики безынертной массы [1].

Переходим к непосредственному изучению движения твёрдых тел в жидкостях и газах.

## *ЧАСТЬ 1*

## ДВИЖЕНИЕ ТВЁРДЫХ ТЕЛ В СРЕДЕ, ПРИ КОТОРОМ ВНОСИМАЯ ИХ ДВИЖЕНИЕМ УДЕЛЬНАЯ ЭНЕРГИЯ ВОЗМУЩЕНИЯ НЕ ПРЕВЫШАЕТ ПО ВЕЛИЧИНЕ УДЕЛЬНУЮ ЭНЕРГИЮ СРЕДЫ

Для начала отметим некоторые особенности движения твёрдых тел в жидкостях и газах.

Рассматривая выше виды движения жидкостей и газов, мы определили их движение как поток, границы которого могут быть выполнены даже из твёрдого материала. При движении твёрдых тел в жидкостях и газах мы будем иметь как бы движение материальной границы относительно неподвижных в среде жидкостей или газов, то есть в данном случае мы будем иметь дело как бы с обратной картиной движения. Это значит, что при перемещении твёрдого тела в среде оно должно создавать эффект такого потока, часть которого оно выражает в своём движении. Этим мы лишь хотели подчеркнуть, что при движении жидкости относительно границ потока или при движении границ потока относительно жидкости изменяется лишь форма принципа относительности, а сущность видов движения в обоих случаях остается одинаковой. Следовательно, при движении твёрдых тел в жидкостях и газах положения и выкладки по видам движения жидкостей и газов будут применимы и к ним.

На примере акустического вида движения мы увидели, что вибрирующая пластина вносит энергию в среду, где жидкость находится в состоянии покоя, уже обладая определённой величиной энергии. Так и любое твёрдое тело при своем движении тоже вносит в среду определённое количество энергии. Энергию, отнесенную к единице объёма жидкости, мы назвали удельной энергией. Величина удельной энергии позволяет без дополнительных геометрических характеристик среды и твёрдого тела делать различие в их энергетических уровнях. В данной части работы мы будем рассматривать движение твёрдых тел в среде, когда вносимая ими удельная энергия будет меньше по величине, чем удельная энергия среды. Если величина вносимой удельной энергии превышает величину удельной энергии среды, то возникают дополнительные физические эффекты, которые мы рассмотрим в другом разделе данной работы.

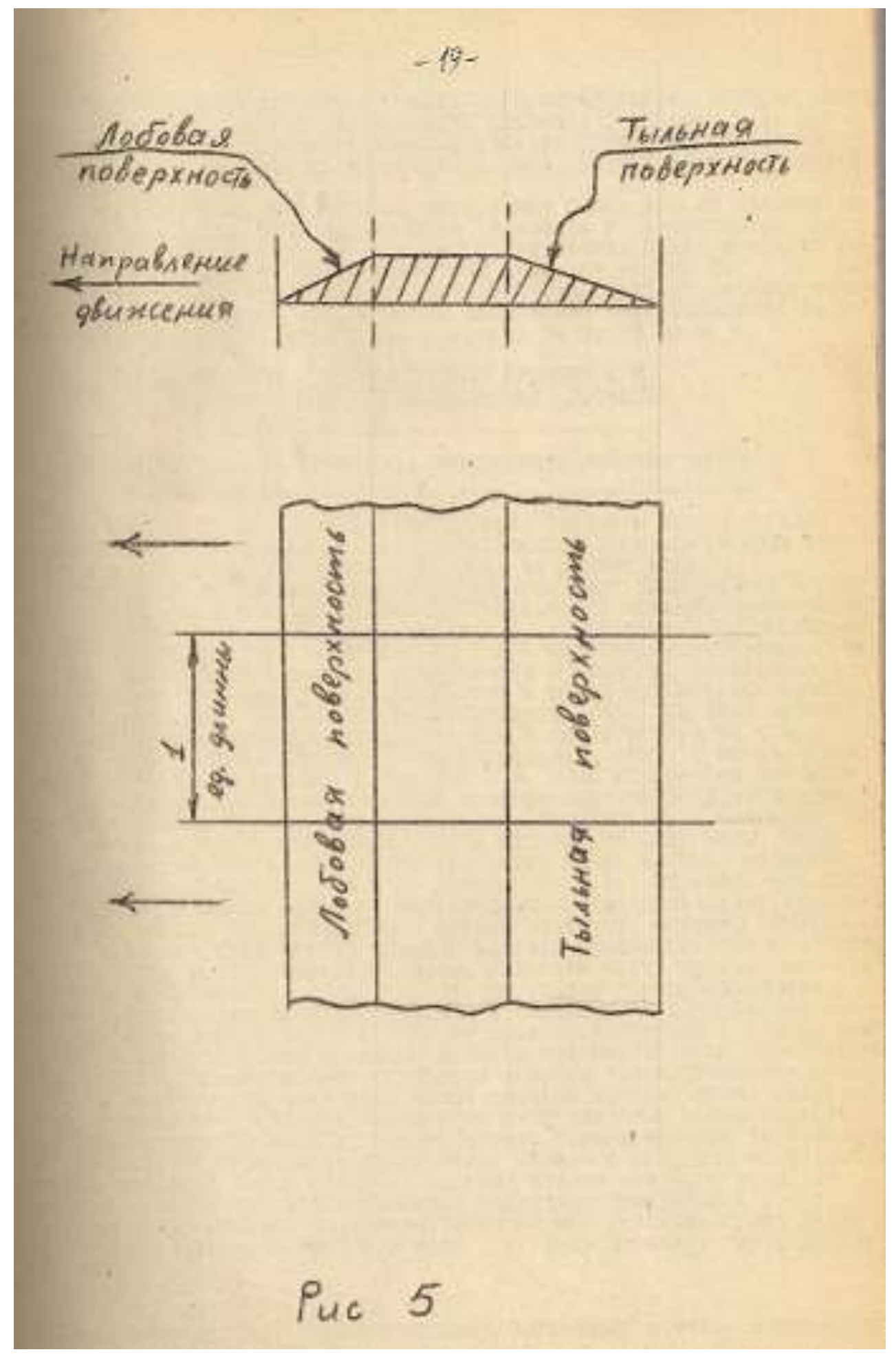

Рис 5

Для характеристики движущихся в среде твёрдых тел приняты, прежде всего, понятия их обтекаемости и необтекаемости. В настоящее время мы различаем эти понятия количественно, как различие в силе при перемещении обтекаемых и необтекаемых тел в среде, без осмысления их физической сущности. В данном разделе мы разберём движение твёрдых тел в среде относительно понятий их обтекаемости и необтекаемости, чтобы придать этим понятиям присущий им физический смысл. А затем уже рассмотрим все остальные положения, которые определяют движение твёрдых тел.

Движение твёрдых тел мы понимаем как их перемещение в пространстве. В нашем случае пространством для их перемещения является среда. За основное движение твёрдых тел в среде примем равномерное прямолинейное движение. В связи с тем, что для жидкостей среды не существует ускорения, а только скорость, то нет смысла рассматривать переменность движения твёрдых тел во времени. Возможное их криволинейное движение мы будем оговаривать в необходимых случаях. Зная направление движения и имея геометрические размеры твёрдого тела, мы можем располагать его относительно направления движения и изучать его взаимодействие со средой, как это делается, например, в аэромеханике или аэродинамике. В своей работе мы обойдемся без всяких “аэро-”.

В первую очередь мы рассмотрим движение плоского тела, которое большими своими плоскостями будет располагаться параллельно направлению движения (рис. 5). При движении твёрдого тела в среде мы будем различать лобовую и тыльную поверхности твёрдого тела, которые располагаются перпендикулярно направлению движения. При определении взаимодействия твёрдого тела со средой, это взаимодействие мы будем определять раздельно для лобовой и тыльной поверхностей.

Далее принимаем, что твёрдое тело имеет сколь угодно большой размер в перпендикулярном движению направлении. По этой причине для своих исследований мы будем вынуждены выделить погонную единицу длины, как это показано на рисунке 5.

Теперь рассмотрим выделенный элемент относительно понятий обтекаемости и необтекаемости. Сначала рассмотрим взаимодействие лобовой части профиля со средой при движении твёрдого тела.

## *Глава I*

## ЛОБОВОЕ СОПРОТИВЛЕНИЕ ПРОФИЛЯ ПРИ РАВНОМЕРНОМ И ПРЯМОЛИНЕЙНОМ ДВИЖЕНИИ ТВЁРДОГО ТЕЛА

### I.1. ПОНЯТИЕ О НЕОБТЕКАЕМОЙ ПОВЕРХНОСТИ ЛОБОВОЙ ЧАСТИ ТВЁРДОГО ТЕЛА

Под необтекаемой лобовой поверхностью твёрдого тела мы будем понимать его лобовую плоскость, расположенную перпендикулярно направлению движения. Так это или не так, мы узнаем ниже.

Поскольку мы приняли движение твёрдого тела равномерным и прямолинейным, то для характеристики обтекаемости лобовой поверхности нам придётся воспользоваться положениями и зависимостями установившегося вида движения жидкости на основании принципа относительности движения. На рис. 6 покажем лобовую поверхность с потоком, обтекающим её.

При движении твёрдого тела в жидкостях и газах организуется поток, который оказывает сопротивление движущемуся телу. По этой причине для обеспечения движения твёрдого тела с соответствующей скоростью к нему должна прикладываться соответствующая сила. В нашем сравнительно коротком примере (рис. 6) дано конкретное движение твёрдого тела. При таком движении непосредственное сопротивление его движению по его лобовой плоскости оказывает среда, находящаяся в состоянии покоя.

Чтобы преодолеть силу сопротивления среды, мы должны будем приложить силу, которая на рис. 6 обозначена как $R_л$. Индекс ($_л$) означает, что эта сила является силой лобового сопротивления. Она будет уравновешиваться силами установившегося потока жидкости, площадь сечения которого $f_п$ будет равна площади лобовой поверхности твёрдого тела. Здесь мы под установившимся потоком понимаем часть возмущенной среды, которая оказывает сопротивление движению

Для исследования этого потока мы также расположили в необходимых местах потока поступательную плоскость $S_п$ и нормальные плоскости $S_н$. Ибо при таком их расположении мы можем выявить все необходимые характеристики потока.

Непосредственно площадь сечения установившегося потока $f_п$ в нашем примере будет равна площади $s_л$ лобовой части твёрдого тела, потому что жидкости безынертны и их движение можно вызвать только непосредственным воздействием. Взаимодействие же происходит по лобовой поверхности твёрдого тела. В связи с тем, что мы рассматриваем движение одной погонной единицы длины твёрдого тела, площадь сечения потока мы обозначили прописной буквой $f_п$. Тогда площадь сечения поступательного потока будет равна произведению толщины твёрдого тела $\delta$ на погонную единицу длины, то есть

$$f_п = \delta \cdot 1. \qquad (28)$$

Мы получили одну характеристику поступательного потока. Характеристики потока нам нужны для того, чтобы записать уравнения движения и сил для нашего потока. Только в этом случае мы сможем количественно судить о потоке[8] жидкости, как о потоке.

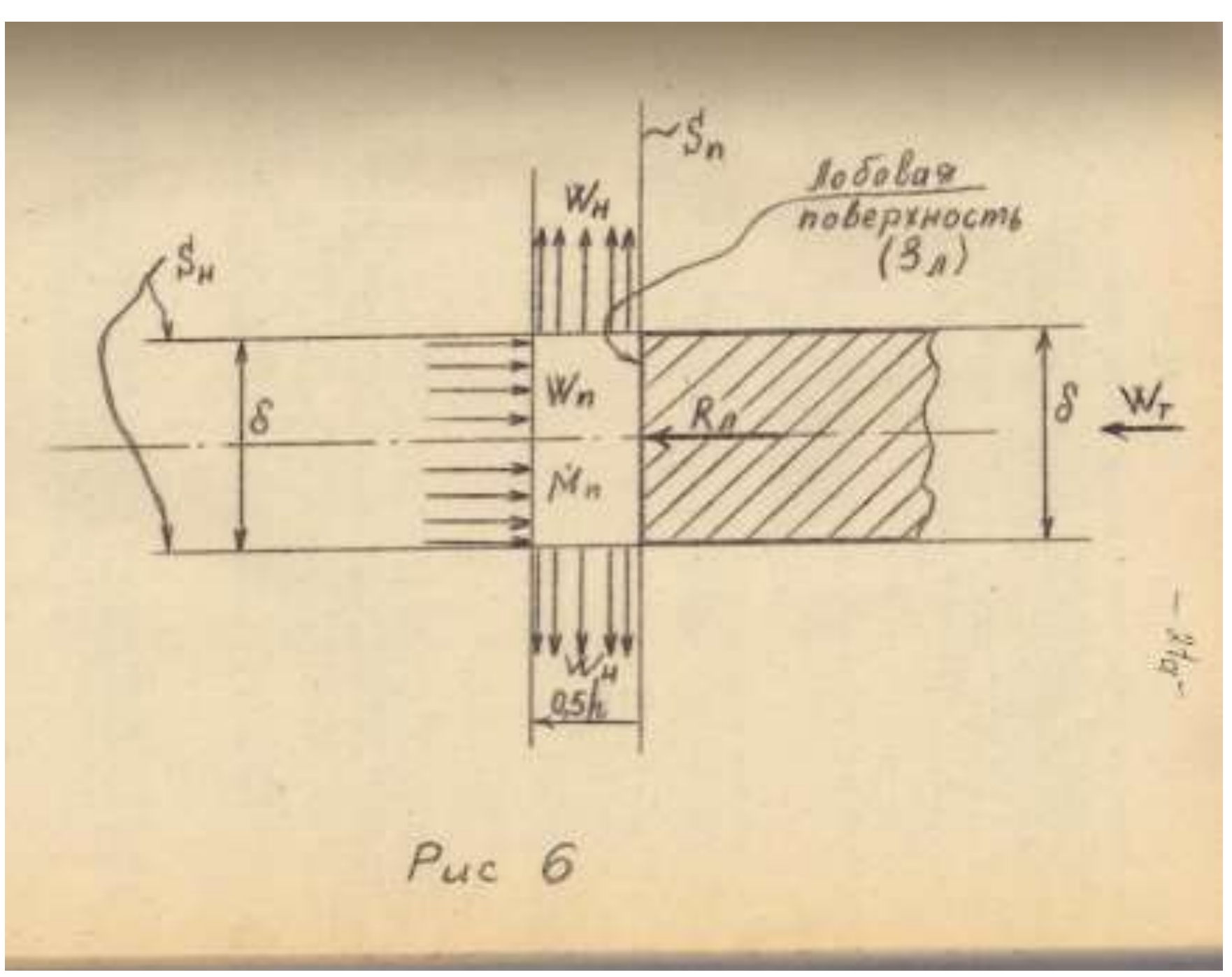

Рис 6

Поступательный поток мы отнесли к установившемуся виду движения, для которого нам известны уравнения движения и сил. Чтобы записать эти уравнения для нашего потока, нам необходимо знать величину скорости движения $W_п$ жидкости в потоке. В нашем случае она будет равна скорости тела, то есть $W_п = W_т$. Тогда уравнение движения для поступательного потока запишется в таком виде:

$$M_п = f_п \rho W_п, \qquad (29)$$

[8] в смысле, судить об относительном потоке как потоке (прим.ред.)

где $M_п$ – расход массы в единицу времени для поступательного потока; $\rho$ – плотность жидкости среды.

Уравнение сил определим тоже из условия равновесия потока на плоскости $S_п$. Здесь мы видим, что сила $R_л$, движущая твёрдое тело, уравновешивается принятыми силами ($f_п P_{пр.дин}$ ) как динамическими силами потока:

$$f_п P_{пр.дин} = M_п W_п = R_л. \tag{30}$$

В уравнении (30) мы можем заменить расход массы в единицу времени $M_п$ через характеристики потока по уравнению движения (29). Тогда получим:

$$R_л = f_п \rho W_п^2. \tag{31}$$

В уравнении сил мы записали только динамические силы давления лишь потому, что величина статических сил давления здесь определяется величиной статических сил давления неподвижной среды, которая обладает соответствующим энергетическим уровнем. Действие статических сил давления не является направленным, то есть оно относится к разряду скалярных величин. По этой причине данные силы не препятствуют силе $R_л$, движущей твёрдое тело.

При своем перемещении в среде твёрдое тело совершает работу $L_т$, которая равна произведению силы $R_л$ на путь $Д$, то есть:

$$L_т = R_л Д. \tag{32}$$

Эта работа твёрдого тела соответственно преобразуется в работу поступательного потока жидкости. Для поступательного потока жидкости работа $L_п$ равна произведению сил давления $P$ на объём $V$ вытесненной жидкости, то есть:

$$L_п = PV. \tag{33}$$

В данном случае работа твёрдого тела $L_т$ должна быть равна работе поступательного потока, то есть:

$$L_т = L_п. \tag{34}$$

Для уравнения работы поступательного потока (33) силы давления определяются уравнением (30), а объём жидкости $V$ может быть сколь угодно большим. Он зависит от количества времени, в течение которого твёрдое тело совершает работу. Чтобы работа потока приобрела конкретный смысл, отнесем её к единице объёма, то есть запишем как удельную работу. Тогда получим:

$$l_п(1/м^3) = P \cdot 1. \tag{35}$$

Удельная работа, как и удельная энергия, равны по величине силам давления.

Далее мы видим, что поступательный поток не совершает работу, так как он существует в замкнутом объёме, а отток жидкости, или работа, производится через нормальный поток (рис. 6). Это значит, что в поступательном потоке работа твёрдого тела преобразуется в энергию поступательного потока $U_п$. Затем энергия поступательного потока преобразуется в работу нормального потока $L_н$. Тогда мы должны будем записать, что удельная работа нормального потока $l_н$ является удельной энергией $U_п$ поступательного потока, то есть:

$$l_н(1/м^3) = U_п. \tag{36}$$

Это значит, что в поступательном потоке происходит прирост энергии, то есть удельная энергия среды $U_{\text{с}}$ увеличивается на величину удельной энергии поступательного потока $U_{\text{п}}$. Отсюда следует, что полная энергия в поступательном потоке $U_{\text{п.пол}}$ равна сумме энергии среды $U_{\text{ср}}$ и энергии поступательного потока $U_{\text{п}}$:

$$U_{\text{п.пол}} = U_{\text{ср}} + U_{\text{п}} \tag{37}$$

Мы получили полную удельную энергию поступательного потока. Также мы можем получить количественные составляющие этой энергии, поскольку величина составляющих удельных энергий равна силам давления. Для среды величину её сил мы всегда можем замерить, а для поступательного потока прирост удельной энергии $U_{\text{п}}$ мы определим по уравнению (30).

Будем считать, что мы получили количественные зависимости для поступательного потока. Нам остаётся разобраться с нормальным потоком.

Нормальный поток жидкости образуется в результате того, что между объёмом поступательного потока и средой образуется разность энергий. Ведь нормальный поток движется в среде, которая имеет одинаковую с величиной энергии поступательного потока величину энергии. Это значит, что разность энергий образуется за счёт прироста энергии в поступательном потоке $U_{\text{п}}$. Вот это количество энергии перемещает жидкость в нормальном потоке. По отношению к нормальному потоку эта энергия будет выступать в роли полной энергии. Поэтому для этого потока мы должны записать её в форме уравнения энергии Бернулли как состоящую из потенциальной и кинетической энергии, то есть

$$U_{\text{пол.н}}(1/\text{м}^3) = U_{\text{ст.н}} + U_{\text{кин.н}}, \tag{38}$$

где $U_{\text{пол.н}}$ – полная удельная энергия нормального потока, величина которой определяется уравнением (30) как[9] <…>

$U_{\text{ст.н}}$ – удельная статическая, или потенциальная энергия потока, она равна $P_{\text{ст.н}}$, то есть статическим силам давления потока

$U_{\text{кин.н}}$ – удельная кинетическая энергия нормального потока. Она равна <…>

Тогда уравнение (38) мы можем записать в таком виде:

$$U_{\text{пол.н}}\,(1/\text{м}^3) = \rho W_{\text{п}}^2 = P_{\text{ст.н}}\,(1/\text{м}^3) + \frac{1}{2}\rho W_{\text{н}}^2. \tag{39}$$

В уравнении (42) мы знаем или можем определить только величину полной энергии нормального потока, а её составляющие – потенциальная и кинетическая энергии – нам неизвестны. Но нам известно одно замечательное свойство уравнения энергии Бернулли, что кинетическая часть энергии может быть либо меньше, либо равной потенциальной части энергии, то есть:

$$P_{\text{ст.н}}(1/\text{м}^3) \geq \frac{1}{2}\rho W_{\text{н}}^2. \tag{40}$$

Вот этим замечательным свойством мы и воспользуемся. Во-первых, раньше мы приняли, что жидкость среды идеальна и не обладает вязкостью и сжимаемостью. Во-вторых, наш нормальный поток не имеет материальных границ и поэтому может быть ограничен только максимально допустимыми скоростями движения для этого потока, а они ограничиваются только уравнением энергии (40). Исходя из этих условий, мы можем сказать, что для нормального потока потенциальная и кинетическая составляющие энергии будут равны между собой, то есть:

$$P_{ст.н}\,(1/м^3) = \frac{1}{2}\rho W_н^2. \qquad (41)$$

[9] Оригинал – это машинописный текст, в который уравнения автор вписывал от руки. Пояснение к уравнению (38) не вписано.

Получив равенство (41), мы можем получить необходимую для нас величину скорости нормального потока $W_н$ по уравнению (39):

$$\frac{1}{2}\rho W_п^2 = \frac{1}{2}\rho W_н^2. \qquad (42)$$

Отсюда следует, что скорость движения поступательного потока $W_п$ равна скорости движения нормального потока $W_н$, то есть $W_п = W_н$.

Для определения зависимостей поступательного потока мы воспользовались известными для нас площадью сечения поступательного потока и скоростью движения твёрдого тела. Теперь, определив величину скорости нормального потока из уравнения энергий, мы можем найти интересующие нас величины для этого потока.

Прежде всего, по уравнению движения нормального потока мы найдем величину $h$ (рис. 6), которая определит нам длину объёма поступательного потока и площадь сечения нормального потока:

$$M_п = M_н = f_н \rho W_н, \qquad (43)$$

где площадь сечения нормального потока $f_н$ равна ширине $h$, умноженной на единицу погонной длины твёрдого тела, то есть:

$$f_н = h \cdot 1.$$

Отсюда следует, что величины площадей сечения поступательного и нормального потоков будут одинаковы, то есть $f_п = f_н$. Уравнения сил, записанные относительно динамических сил давления, тоже будут одинаковыми, то есть:

$$R_н = f_п \rho W_п^2 = f_н \rho W_н^2. \qquad (44)$$

На этом, будем считать, что мы разобрались с количественными величинами нормального и поступательного потоков, которые через необтекаемую поверхность твёрдого тела противодействуют движению этого тела. Нам осталось качественно уяснить характер этого взаимодействия.

Он выражается в том, что при необтекаемой лобовой поверхности твёрдого тела происходит его взаимодействие со средой с одновременным движением жидкости в двух взаимно перпендикулярных направлениях. В поступательном потоке динамические силы движения твёрдого тела преобразуются в статические силы давления потока, то есть кинетическая энергия тела преобразуется в полную энергию поступательного потока, которая выражается приростом энергии в среде. Затем прирост энергии поступательного потока затрачивается на выброс жидкости через нормальный поток. Во всех этих случаях происходит прямое без потерь, преобразование кинетической энергии движения твёрдого тела.

В то же время мы утверждаем, что подобная лобовая поверхность твёрдого тела необтекаема. Это действительно так и есть, ведь на движение твёрдого тела с такой лобовой поверхностью приходится затрачивать силу большую, чем на движение твёрдого тела с любым иным профилем лобовой поверхности. Почему так происходит, мы рассмотрим ниже при разборе обтекаемого профиля лобовой поверхности твёрдого тела. Сейчас же мы должны убедиться лишь в том, что при необтекаемом профиле лобовой поверхности твёрдого тела не происходит потерь энергии, то есть при таком движении не затрачивается каких-либо дополнительных сил, которые покрывали бы затраты энергии, происходящие не по прямому её назначению.

В том, что в потоке жидкости не происходит потерь энергии, мы можем убедиться следующим образом. Утверждая принцип одновременного взаимно перпендикулярного движения жидкости в потоке, мы исходили не из какого-либо подобия, а из закона сохранения энергии. Это значит, что при таком движении жидкости не происходит потерь энергии. Тем более, что жидкости являются лишь носителями и переносчиками чужой энергии – энергии силового поля, как утверждает первый закон механики безынертной массы, или закон сохранения состояния. Иначе говоря, жидкости являются рабочим телом. В этом можно убедиться, не прибегая к помощи аэродинамических труб, на очень простом эксперименте.

Хорошо известно, что установившийся поток жидкости является переносчиком определённого и постоянного количества энергии. Это значит, что в этом потоке не может происходить какого-либо прироста механической энергии, а могут происходить лишь её потери. Это изменение энергии нетрудно замерить. Поэтому для эксперимента надо взять участок установившегося потока с материальными границами, как показано на рис. 7.*а*. Для сравнения можно взять такой же участок установившегося потока с прямоугольным сечением, но с традиционно выполненным коленом – по радиусу (рис. 7.*б*). Сделав замеры на входе и на выходе потока (рис. 7.*а*), вы убедитесь, что потерь энергии не происходит. При замере энергии на входе и выходе участка установившегося потока (рис. 7.*б*) мы заметим потери энергии.

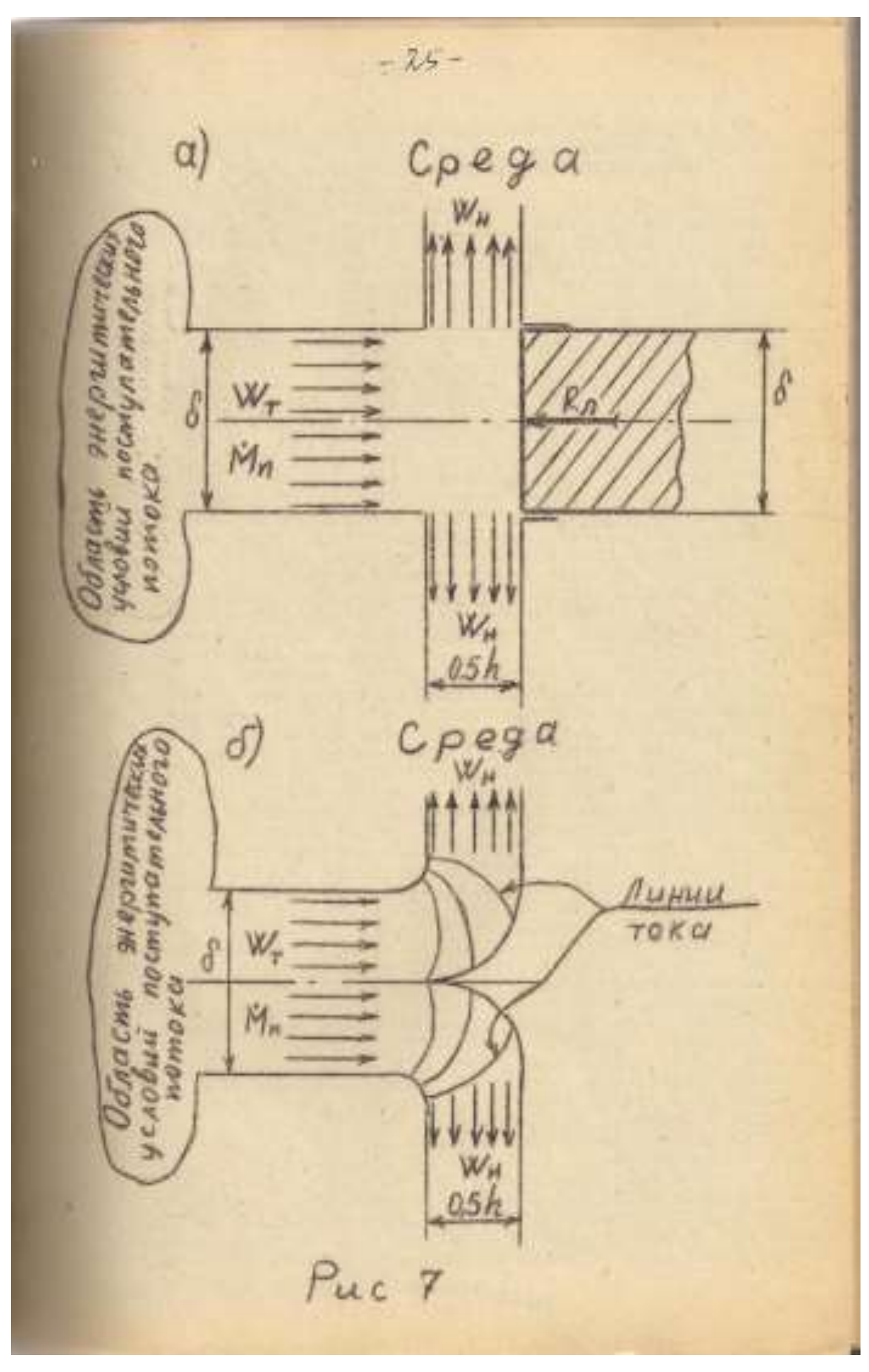

Рис 7

Потери здесь происходят за счёт того, что в коленах, которые выполнены с определённым радиусом, организуется плоский установившийся поток. Линиями тока такого потока являются логарифмические спирали. В то же время этот организующийся поток будет относиться к плоскому установившемуся потоку турбинного типа, в котором жидкость перемещается из области с повышенной энергией в область с пониженной энергией. По этой причине динамические силы давления плоского установившегося потока колена, которые действуют по логарифмической спирали, взаимодействуют с динамическими силами давления в прямых участках трубы. В результате чего происходит частичная потеря энергии потока. Эти явления можно наблюдать сделав материальные границы потока прозрачными. В природе подобные явления наблюдаются по течению рек. На излучинах и поворотах река размывает один берег, а материал размыва выносит на противоположный берег. Статические силы потока непосредственно не участвуют в изменении энергии потока. Потери происходят только за счёт динамических сил давления.

На этих примерах мы убедились, что, действительно, при движении твёрдого тела с необтекаемой лобовой поверхностью никаких потерь энергии не происходит. Отметим, что примеры можно было бы не приводить. Просто данные примеры дают нам возможность ощутить движение жидкости поэтапно, а это положение является очень важным моментом в осмысленном понимании сущности любого явления природы. Именно для этой цели мы приводим подобные примеры и в дальнейшем будем приводить их. Хотя на это приходится затрачивать ни мало времени.

Действительно, мною специальных экспериментов не проводилось, поскольку у меня до сих пор не было такой возможности. Но все положения моих работ получены на основании прямых наблюдений за движением жидкостей и газов, которые проводились для определённых технических нужд при создании определённых гидравлических и газовых машин. С другой стороны, современное развитие гидравлической и газовой техники полностью основано на эксперименте. Поэтому экспериментального материала накоплено достаточно. Следовательно, в настоящее время и в наших примерах не обязательно ставить конкретные количественные величины, так как мы даем пример руководства основными положениями теории механики безынертной массы, которые точно так же применимы ко всему накопленному конкретному экспериментальному опыту.

## I. 2. ПОНЯТИЕ ОБ ОБТЕКАЕМОЙ ПОВЕРХНОСТИ ЛОБОВОЙ ЧАСТИ ТВЁРДОГО ТЕЛА

В настоящее время понятие обтекаемости носит несколько растяжимый характер: как более или менее обтекаемое, поскольку множество различных по форме поверхностей лобовой части твёрдого тела уменьшают величину силы, требуемой на его перемещение или движение. По величине этой силы мы и судим об обтекаемости.

На рис. 8 покажем лобовые поверхности твёрдого тела, которые относят к разряду обтекаемых.

Как видно, все обтекаемые лобовые поверхности мы разделили на две основные группы. К первой группе отнесли лобовые поверхности, которые образованы плоскостью с различным углом наклона $\alpha$ по отношению к поступательной плоскости исследования $S_п$. Лобовые плоскости могут располагаться односторонне (рис. 8.*а*) или симметрично относительно плоскости твёрдого тела (рис. 8.*б*).

Ко второй группе мы отнесли лобовые поверхности, которые образованы плавно искривлёнными поверхностями. Они могут быть односторонними (рис. 8.*в*) или расположенными симметрично относительно плоскости твёрдого тела. Конечно, обтекаемые лобовые поверхности могут иметь и другие, отличные от этих двух групп, формы поверхностей. Мы приняли эти две группы лишь потому, что они дают возможность изучить и понять саму сущность обтекаемости твёрдых тел. А затем уже, используя положения и выкладки для этих поверхностей, можно будет определять обтекаемость для любой поверхности твёрдого тела, так как элементы поверхностей, входящих в эти две группы, составляют любую кривую поверхность.

<u>Рассмотрим обтекаемость лобовых поверхностей первой группы.</u>

Начнём опять с рисунка. На рис. 9 мы показали два твёрдых тела, лобовая плоскость которых наклонена относительно поступательной плоскости $S_п$ под различным углом $\alpha$. На рис. 9.*а* угол наклона лобовой поверхности $\alpha$ либо меньше 45°, либо равен 45°. На рис. 9.*б* угол наклона лобовой поверхности дан более 45°.

Далее для определения степени обтекаемости мы приняли изменение величины силы, которая необходима для движения твёрдого тела в среде. По этой причине мы должны будем рассматривать его обтекаемость относительно необтекаемого твёрдого тела. Поэтому принимаем толщину обтекаемого твёрдого тела $\delta$ одинаковой с толщиной необтекаемого твёрдого тела и одинаковую величину их скоростей движения $W_т$. При этих условиях мы можем выявить различие в действующих силах.

Что касается рисунков, то изображения этих тел отличаются лишь тем, что на рис. 9 мы наклонили лобовую плоскость и тем самым изменили площадь взаимодействия твёрдого тела со средой. Тела с подобным расположением лобовых поверхностей тоже организуют установившийся поток с одновременным взаимно перпендикулярным движением жидкости в нём. Из этого мы и будем исходить.

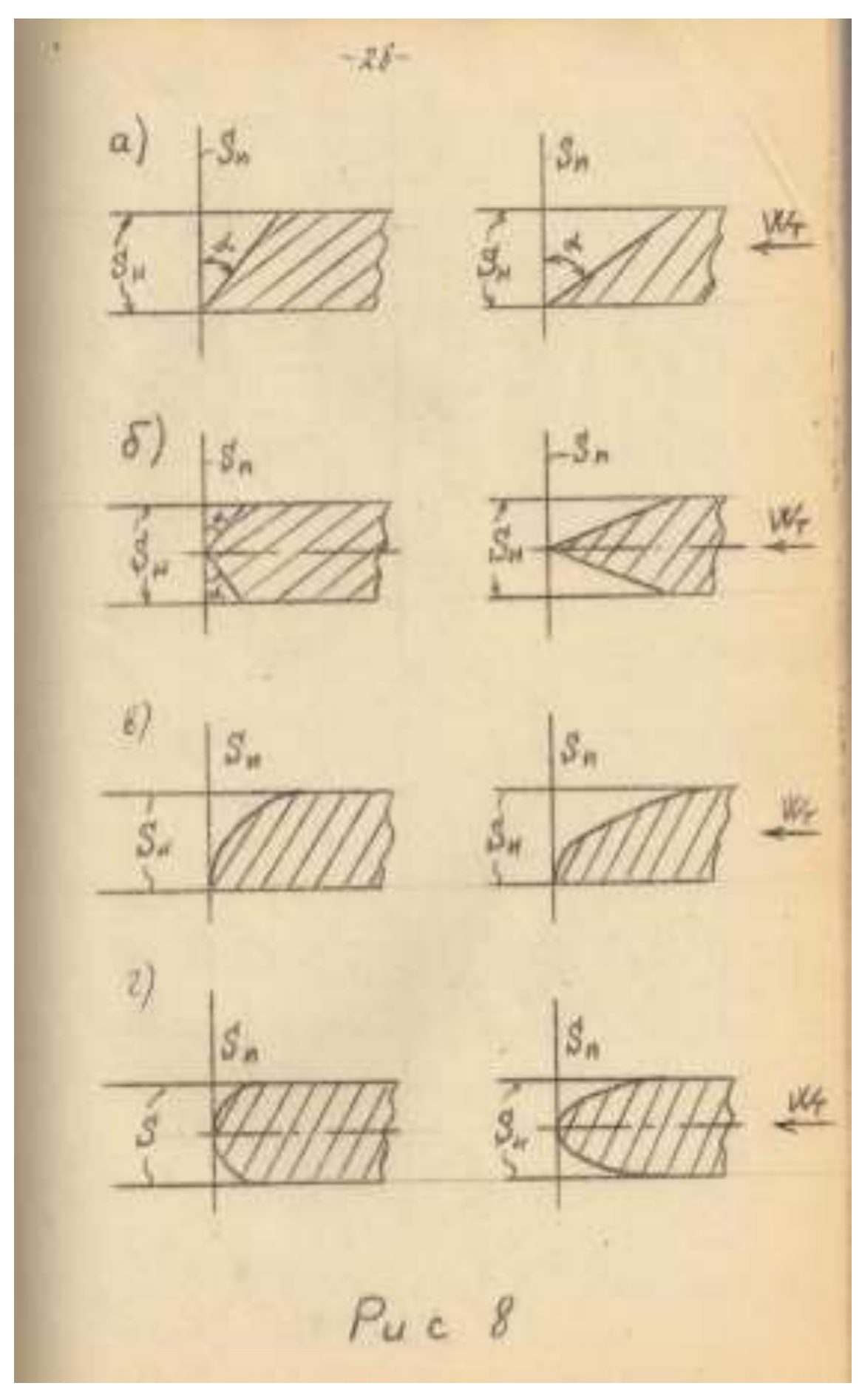

Рис 8

В данном случае площадь сечения поступательного потока $f_п$ остается неизменной в поступательной плоскости исследования $S_п$, а одинаковость скоростей твёрдых тел $W_т$ указывает на то, что удельная энергия $U_п$ поступательного потока в его объёме $V$ остается одинаковой с энергией поступательного потока необтекаемого твёрдого тела.

При подобном движении твёрдого тела определяющим является поступательный поток жидкости. Поэтому расход массы в поступательном и нормальном направлении должен быть одинаковым. определяется расходом массы в поступательном направлении и величины этих расходов должны быть одинаковыми.

Это значит, что уравнения движения, сил и энергии для поступательного потока в исследуемой плоскости $S_{п}$ должны быть одинаковыми с подобными уравнениями для необтекаемого твёрдого тела, которыми мы можем пользоваться в полной количественной мере при подобном взаимодействии твёрдого тела со средой.

Мы выявили одинаковость взаимодействий, теперь остаётся определить их различие.

Начало поступательный поток взаимодействия берет непосредственно от лобовой поверхности твёрдого тела. Ведь этот поток преобразует энергию поступательного движения твёрдого тела в приращение энергии поступательного потока, или в энергию поступательного потока. Отсюда следует, что мы, расположив лобовую поверхность твёрдого тела под углом, тем самым увеличили её площадь $s_{л}$, или увеличили площадь взаимодействия твёрдого тела со средой. Так как мы имеем дело с установившимся потоком, то в этом случае для определения взаимодействия мы должны воспользоваться уравнениями движения и сил этого потока.

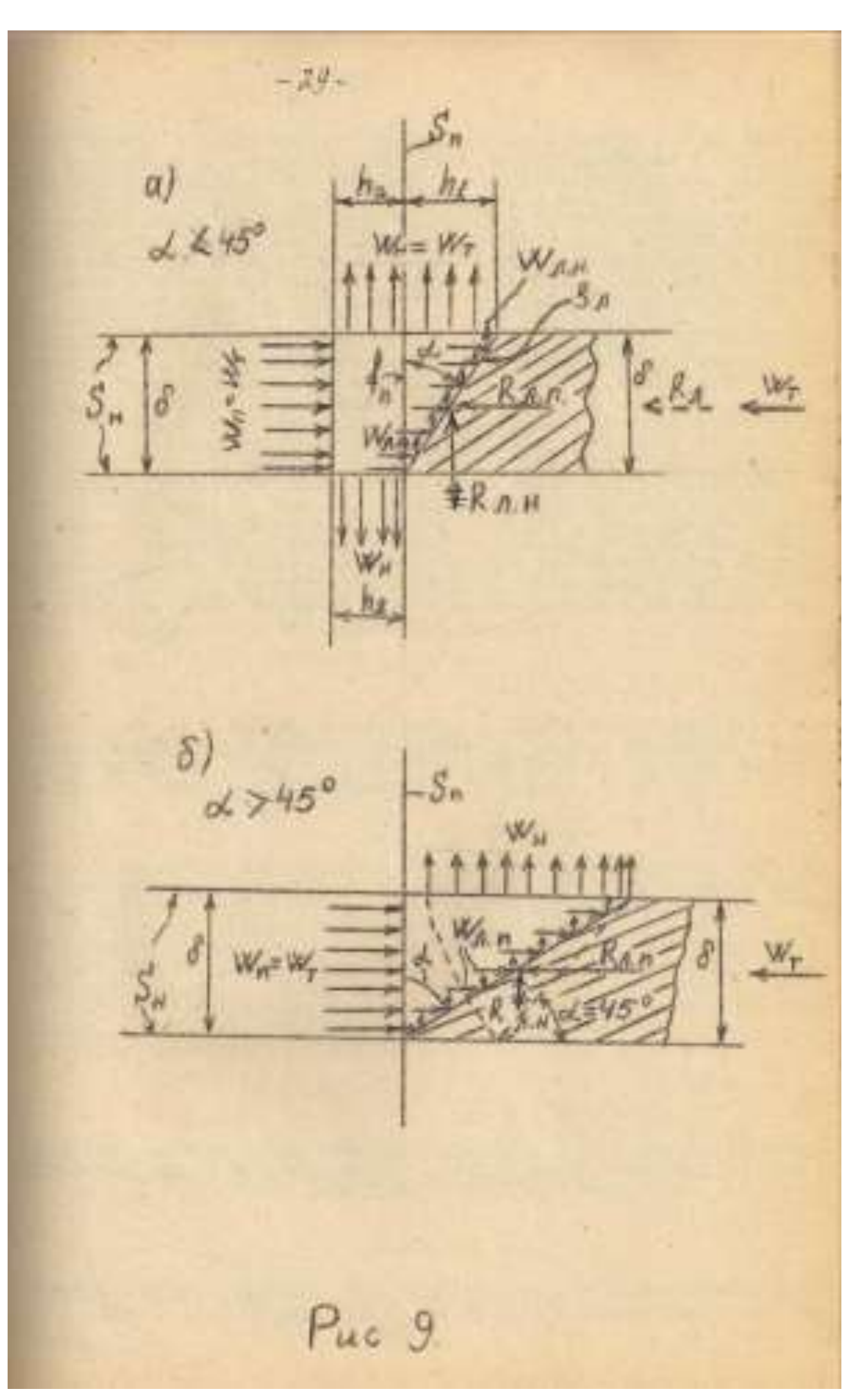

Рис 9

Тогда движение поступательного потока на лобовой поверхности запишется таким уравнением движения:

$$M_{л} = s_{л}\rho W_{л.п}. \qquad (45)$$

Уравнение (45) является уравнением движения взаимодействия поступательного потока с лобовой поверхностью твёрдого тела. В этом уравнении нам известен расход массы в единицу времени $M_{л}$, который мы количественно определяем по уравнению (29),

то есть:

$$M_{п} = f_{п}\rho W_{п}.$$

В этом уравнении нам известны все характеристики потока. Поступательная скорость потока $W_{п}$ в этом случае равна скорости движения твёрдого тела $W_{т}$. Площадь сечения поступательного потока $f_{п}$ равна толщине твёрдого тела $\delta$, умноженной на погонную единицу длины. Плотность среды $\rho$ нам известна из условий задачи. Таким способом мы определили расход массы в поступательном потоке $M_{п}$, а $M_{п} = M_{л}$.

Площадь лобовой поверхности $s_{л}$ в уравнении (45) мы определим как толщину твёрдого тела $\delta$, делённую на $\sin\alpha$ и умноженную на погонную единицу длины, то есть:

$$s_{л} = \frac{\delta}{\sin\alpha} \cdot 1_{п.ед.} \tag{46}$$

Зная эти величины, мы по уравнению движения (45) сможем количественно определить скорость взаимодействия поступательного потока $W_{л.п}$ с лобовой поверхностью твёрдого тела как

$$W_{л.п} = \frac{M_{л}}{\rho s_{л}} \cdot \tag{47}$$

Теперь запишем уравнение сил взаимодействия поступательного потока с лобовой плоскостью $s_{л}$ относительно динамических сил давления. Оно будет иметь вид:

$$R_{л.п} = s_{л} P_{дин.п} = s_{л} \rho W_{л.п}^{2}. \tag{48}$$

В уравнении сил лобового взаимодействия (48) все характеристики потока нам известны. Скорость потока $W_{л.п}$ мы определим по уравнению (47), а величину площади лобовой плоскости – по уравнению (46). Это значит, что по уравнению (48) мы определим величину сил лобового сопротивления поступательного потока, которые при соответствующем движении твёрдого тела должны уравновешиваться лобовой силой $R_{л.п}$, приложенной к твёрдому телу для обеспечения его движения. Это равновесие и записано уравнением (48), то есть мы получили уравнением (48) величину лобового сопротивления.

Теперь эту величину мы можем сравнить с величиной лобового сопротивления (30) для необтекаемой лобовой поверхности твёрдого тела и для тел с лобовой поверхностью, расположенной под углом $\alpha$ к поступательной поверхности $S_{п}$. Видно, что величина сил сопротивления при любой величине угла $\alpha$ расположения лобовой поверхности будет меньше величины сил лобового сопротивления для тел с нормальным расположением лобовой плоскости. Величина сил сопротивления находится в квадратичной зависимости от скорости взаимодействия поступательного потока с лобовой поверхностью.

При нормальном расположении лобовой плоскости эти скорости максимальны, так как они равны скорости движения твёрдого тела, а при наклонном расположении лобовой плоскости они уменьшаются в зависимости от величины угла наклона $\alpha$. Угол $\alpha$ определяет нам величину площади взаимодействия поступательного потока с твёрдым телом. Согласно уравнения (45), это приводит к уменьшению лобовой поступательной скорости потока $W_{л.п}$. Из этого уравнения мы видим, что скорость находится в прямой зависимости от величины площади. Что приводит к квадратичной зависимости падения динамических сил давления на единицу площади взаимодействия. Отсюда следует, что динамические силы давления для единицы площади взаимодействия уменьшаются в квадратичной зависимости, но при этом возрастает площадь взаимодействия. По этой причине уменьшение полной величины сил лобового сопротивления $R_{л}$ будет уменьшаться при увеличении площади лобового сопротивления, но не в квадратичной зависимости, а приблизительно в прямой.

Мы выяснили, что с увеличением угла $\alpha$ непрерывно увеличивается площадь лобового взаимодействия твёрдого тела, а сила лобового сопротивления $R_{л}$ уменьшается. Чем больше угол $\alpha$, тем меньше будет сила сопротивления. Это значит, что твёрдое тело с нормально расположенной лобовой поверхностью требует максимальной величины силы для своего перемещения. В таком случае мы называем тела с нормальной лобовой поверхностью необтекаемыми, а тела с наклонной лобовой поверхностью мы называем обтекаемыми в различной степени в зависимости от величины сил сопротивления.

Будем считать, что мы познакомились в первом приближении с понятиями обтекаемости и необтекаемости твёрдых тел.

За дальнейшими разъяснениями обратимся к уравнениям энергии. Как мы выше установили, при нормальном расположении лобовой плоскости, или при необтекаемой её форме, кинетическая энергия движущегося тела полностью переходит в потенциальную, или статическую, энергию поступательного потока. Далее мы установили, что эта величина энергии является полной энергией поступательного потока и при наклонной лобовой поверхности твёрдого тела. В общем, это очевидно, так как площадь сечения поступательного потока и скорость движения такого твёрдого тела такие же, как у необтекаемого твёрдого тела. В то же время мы знаем, что на перемещение в среде твёрдого тела с наклонной лобовой поверхностью требуется меньшая сила. Это значит, что и меньшая величина энергии, но, мы видим, такого не может быть.

Поэтому мы должны будем расписать энергию поступательного потока для тела с наклонной лобовой поверхностью по уравнению энергии Бернулли:

$$VP_{пол} = VP_{пот} + VP_{дин}. \qquad (49)$$

Уравнение (49) является количественным выражением полной энергии поступательного потока. В этом уравнении полная энергия потока $VP_{пол}$ нам известна, поскольку мы можем её определить как энергию поступательного потока необтекаемого тела, то есть:

$$VP_{пол} = V\rho W_{т}^{2} = V\rho W_{п}^{2}.$$

Одно слагаемое мы определили в уравнении (49). У нас остаются ещё два неизвестных слагаемых. Следовательно, нам необходимо найти величину ещё одного слагаемого, чтобы определить величину третьего слагаемого. Мы можем найти величину потенциальной, или статической, энергии поступательного потока, то есть $VP_{пот}$, поскольку на наклонной лобовой плоскости кинетическая энергия движения твёрдого тела преобразуется в потенциальную, или статическую, энергию поступательного потока. Величину сил давления мы можем определить по уравнению сил (48). Тогда получим зависимость потенциальной энергии поступательного потока в таком виде:

$$VP_{пот} = V\rho W_{л.п}^{2}. \qquad (50)$$

По уравнению (50) мы сможем получить величину потенциальной энергии поступательного потока. При этом нам придётся использовать ещё уравнения (45) и (48). Динамическую, или кинетическую, составляющую энергии поступательного потока мы выразим через характеристики потока. Тогда получим:

$$VP_{\text{дин}} = \frac{1}{2} V\rho W^2_{\text{п.п}}, \qquad (51)$$

где $V$ – объём поступательного потока; $W_{\text{п.п}}$ – скорость поступательного потока.

Теперь запишем уравнение (49) в виде удельной энергии поступательного потока, получим:

$$P_{\text{пол.п}}(1/\text{м}^3) = P_{\text{ст.п}}(1/\text{м}^3) + \frac{1}{2}\rho W^2_{\text{п.п}}. \qquad (52)$$

В этом уравнении все составляющие его величины нам известны, так как мы можем определить величины полной и потенциальной энергии потока по вышеуказанным зависимостям, а величину кинетической энергии мы уже нашли по уравнению (52).

Получив уравнение энергии (52), мы теперь можем рассмотреть его ограничительные действия в отношении поступательного потока, поскольку мы знаем, что в уравнении энергии величина кинетической энергии не должна превышать величину потенциальной энергии, то есть:

$$P_{\text{ст.п}}(1/\text{м}^3) \geq \frac{1}{2}\rho W^2_{\text{п.п}}.$$

В поступательном потоке реализуются предельные ограничивающие свойства уравнения энергии, поэтому мы можем определить предельный, или максимальный, угол наклона $\alpha_{\text{пр.}}$ лобовой поверхности твёрдого тела для невязкой жидкости потока. Для этого нам необходимо будет приравнять потенциальную и кинетическую энергии, то есть:

$$P_{\text{ст.п}}(1/\text{м}^3) = \frac{1}{2}\rho W^2_{\text{п.п}}.$$

Выразим потенциальную энергию через характеристики потока, получим:

$$\rho W^2_{\text{л.п}} = \frac{1}{2}\rho W^2_{\text{п.п}}. \qquad (53)$$

Из уравнения (53) мы можем определить предельную величину лобовой поступательной скорости $W_{\text{л.п}}$, а с её помощью мы уже определим предельный, или максимальный, угол наклона лобовой плоскости.

Эту максимальную величину мы вычислили в данной работе ниже. Она оказалась равной 45°, то есть предельный угол наклона лобовой плоскости $\alpha_{\text{пр}} = 45°$. Это значит, что при величине угла наклона лобовой поверхности твёрдого тела до 45° реализуется взаимно перпендикулярное движение жидкости, то есть образуется нормальный поток, а при величине угла наклона более 45° будет реализовываться только поступательный поток, который уже не зависит от величины угла наклона лобовой поверхности.

Уравнение энергии поступательного потока (52) дает нам возможность, с одной стороны, определить пределы существования нормального потока, с другой стороны, оно дает возможность определить величину нормальной скорости $W_{\text{н}}$, а по её величине мы уже сможем определить геометрические характеристики нормального потока.

Приступим к определению скорости нормального потока $W_{\text{н}}$.

По отношению к нормальному потоку полной его энергией ($VP_{\text{пол.н}}$) в данном случае будет выступать потенциальная, или статическая, энергия поступательного потока, которая определяется уравнением (50). Далее мы будем обязаны расписать эту полную энергию нормального потока по уравнению Бернулли как:

$$VP_{\text{пол.н}} = VP_{\text{ст н}} + \frac{1}{2}V\rho W_{\text{н}}^2. \qquad (54)$$

Запишем уравнение (54) как удельную энергию нормального потока, то есть отнесем её к единице объёма, получим:

$$P_{\text{пол.н}}(1/\text{м}^3) = P_{\text{ст.н}}(1/\text{м}^3) + \frac{1}{2}\rho W_{\text{н}}^2. \qquad (55)$$

Затем, чтобы найти величину скорости нормального потока, мы воспользуемся предельными ограничительными свойствами уравнения энергий. Для невязкой жидкости оно примет вид:

$$\frac{1}{2}P_{\text{пол.н}}(1/\text{м}^3) = P_{\text{ст.н}}(1/\text{м}^3) = \frac{1}{2}\rho W_{\text{н}}^2. \qquad (56)$$

Подставим в уравнение (56) величину полных сил давления по уравнению (50), получим величину скорости нормального потока как:

$$\frac{1}{2}\rho W_{\text{л.п}}^2 = \frac{1}{2}\rho W_{\text{н}}^2. \qquad (57)$$

Из уравнения (57) мы видим, что величина скорости нормального потока $W_{\text{н}}$ равна величине лобовой поступательной скорости $W_{\text{л.п}}$, которая реализуется на площади лобовой поверхности твёрдого тела.

Получив величину скорости нормального потока, мы теперь можем расписать для него все необходимые характеристики и зависимости, используя при этом известную нам величину расхода массы в единицу времени, которая для поступательного и нормального потоков будет одинакова ($M_{\text{п}} = M_{\text{н}}$).

Прежде всего, для нормального потока мы определим величину $h$, которая определяет нам общую поверхность нормального потока. Она будет равна длине лобовой плоскости твёрдого тела, то есть как толщина твёрдого тела $\delta$, делённая на $\sin\alpha$:

$$h = \frac{\delta}{\sin\alpha}. \qquad (58)$$

В нормальном потоке эта величина $h$ по условию движения жидкости (рис. 9.*а*) делится на несколько величин: $h_1$ и $h_2$, так как величина $h_2$ будет располагаться за пределами поступательной плоскости исследования $S_{\text{п}}$, а за пределами этой плоскости нормальный поток реализуется в виде двух симметричных потоков. Но общая суммарная протяженность всех этих "$h$" должна быть равна величине $h$, полученной по зависимости (58), то есть:

$$h = \frac{\delta}{\sin\alpha} = h_1 + h_2 + h_2. \qquad (59)$$

Величины $h$ определили нам геометрию нормального потока и общий объём поступательного потока. Далее для нормального потока мы можем записать уравнения движения и сил. Таким способом мы определим все интересующие нас характеристики. Но делать мы этого не будем, поскольку всё это хорошо понятно.

Остается выяснить, с какой силой нормальный поток действует на наклонную лобовую поверхность твёрдого тела.

Нормальный поток располагается по площади лобовой поверхности твёрдого тела на длине $h_1$ (рис. 9.*а*). Используя эту величину, мы можем определить с помощью уравнения движения, какая часть нормального потока взаимодействует с наклонной лобовой поверхностью в нормальном направлении:

$$M_{л.н} = h_1 \cdot 1 \cdot \rho W_н. \quad (60)$$

По уравнению (60) мы найдем величину расхода массы $M_{л.н}$, которая действует на лобовую поверхность, то есть эта величина расхода массы будет взаимодействовать с площадью лобовой поверхности твёрдого тела. Поэтому мы должны записать уравнение движения для всей лобовой поверхности $s_л$, получим:

$$M_{л.н} = s_л \cdot \rho W_{л.н}. \quad (61)$$

По уравнению (61) мы получим нормальные скорости лобового сопротивления $W_{л.н}$, которые по величине будут меньше нормальных скоростей $W_н$, то есть $W_н > W_{л.н}$. Это понятно, поскольку величина лобовой поверхности больше нормальной площади, которая определяется величиной $h_1$.

Зная скорости, по уравнению сил мы можем определить величину нормальных сил давления, которые действуют на площадь лобовой поверхности. Они будут равны:

$$R_{л.н} = s_л \rho W_{л.н}^2. \quad (62)$$

Распределённые по лобовой поверхности силы мы заменили сосредоточенной силой, обозначенной как $R_{л.н}$ (рис. 9.*а*), и в уравнении (62) записали её как уравновешенную распределёнными нормальными лобовыми силами давления. В аэромеханике нормальную лобовую силу принято называть подъёмной силой. Теперь мы определили все характеристики для нормального потока.

Для первой группы обтекаемых тел с симметрично расположенными лобовыми плоскостями (рис. 8.*б*) все полученные нами выкладки и положения будут верны с поправкой на симметричность. В подобных профилях в силу их симметричности нормальные лобовые силы будут взаимно уравновешены.

<u>Переходим ко второй группе обтекаемых тел.</u>

Например, мы можем построить лобовую поверхность второй группы с помощью отдельных участков плоскостей с различным углом наклона $\alpha$. Покажем построение такого профиля на рис. 10.*а*.

Для каждого выделённого участка с определённым углом $\alpha_1$, $\alpha_2$, $\alpha_3$ и т.д. мы можем определить все необходимые характеристики, а затем их просуммировать.

Мы получили общие характеристики для всей лобовой поверхности движущегося тела. Трудность в вычислении характеристик взаимодействия подобного лобового профиля (рис. 10.*а*) будет выражаться в механических трудностях, ибо вся поверхность разбита на ряд участков, для каждого из которых в отдельности необходимо проделать соответствующие вычисления, чтобы суммированием получить общие характеристики взаимодействия для всей лобовой поверхности.

Как мы видим, распределение скоростей в нормальном потоке $W_н$ имеет ступенчатый характер для лобовых поверхностей такого типа. Этот основной показатель надо уяснить себе. Ибо он определяет назначение подобных лобовых поверхностей в их применении на практике.

Можно подобный профиль лобовой поверхности образовать с помощью математической кривой. Тогда скорости нормального потока будут иметь не ступенчатый, а плавный переход к убыванию (рис. 10.*б*). Для такого профиля лобовой поверхности общие характеристики можно получить при помощи дифференциального исчисления. Тогда дифференциальное решение таких профилей будет точным решением, а решение профилей со ступенчатым наклоном участков под различными углами $\alpha$ будет приближенным с определённой степенью точности. Мы здесь не будем давать решений в дифференциальной форме, поскольку даём решение в общей форме, чтобы выявить назначение различных лобовых поверхностей, а точность решения определяется конкретностью задачи. Пока мы должны уяснить себе, что с помощью полученных зависимостей мы можем найти характеристики взаимодействия для любой формы лобовой поверхности либо точным, либо приближенным решением. Для симметричных профилей лобовой поверхности (рис. 8.*г*) вышеизложенные приемы вычислений характеристик применимы, но с учётом их симметричности.

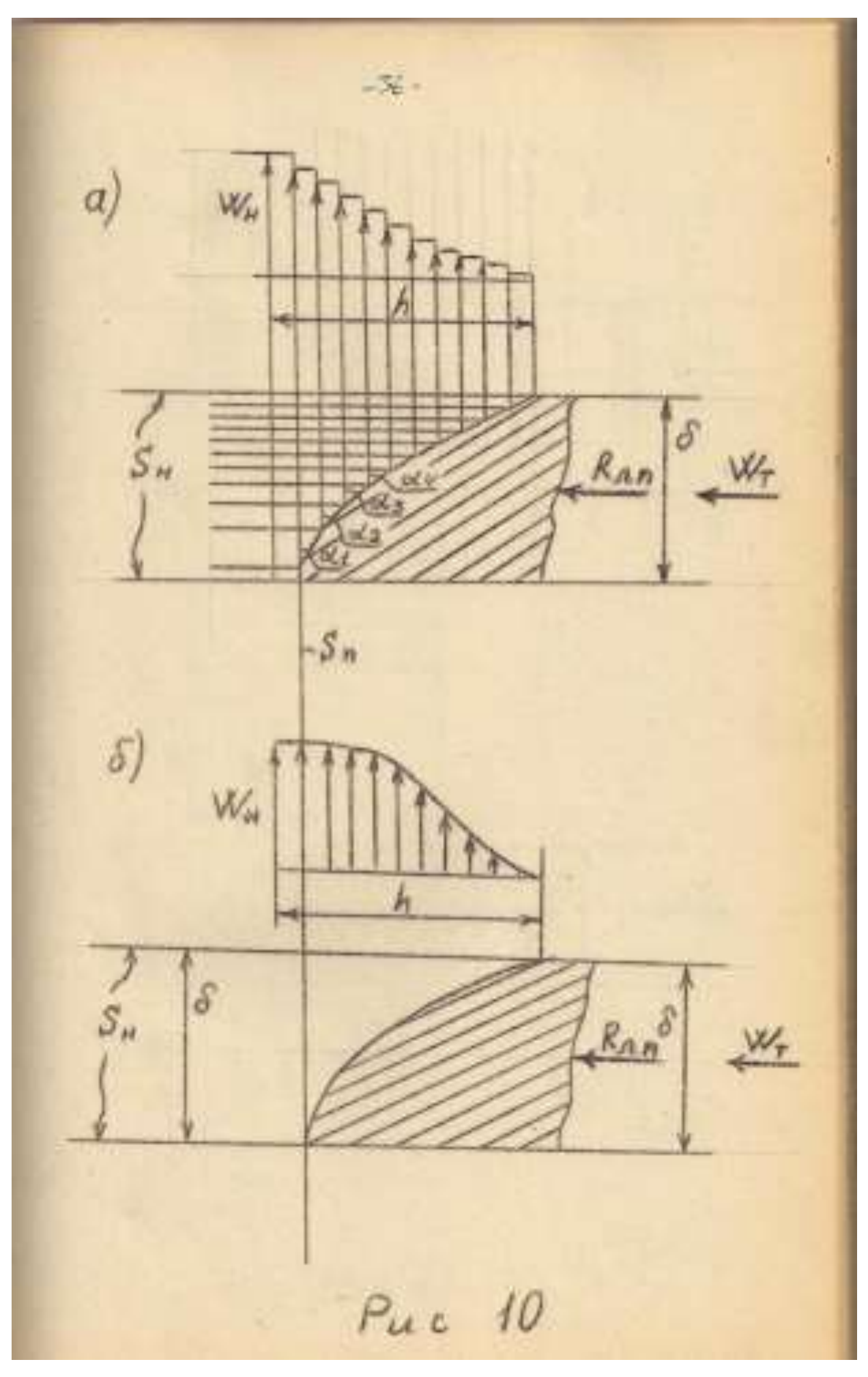

Рис 10

С лобовыми поверхностями второй группы мы познакомились с определённым назначением. В настоящее время существует понятие волнового сопротивления, но что это такое, никто не знает. Просто говорят, что это лучше, а что лучше и как лучше, ещё неизвестно.

Для объяснения этого волнового сопротивления на рис. 11 мы дадим несколько изображений. Посмотрите на них внимательно. Теперь вспомним акустический вид движения. При перемещении пластинки с определённой скоростью из второго положения в первое образуется волна давления, которая состоит из двух потоков: поступательного и нормального. В поступательном потоке скорости не изменяются, если не изменяется скорость движения пластины. В нормальном потоке скорости $W_н$ изменяются. Своим изменением они обязаны скорости звука $C$, которая непрерывно изменяет площадь сечения нормального потока. На рис. 11 показаны эпюры распределения нормальных и поступательных скоростей в акустической волне.

Далее мы должны будем взять площадь сечения вибрирующей пластинки равной площади сечения твёрдого тела в поступательной плоскости исследования $S_{п}$. Затем придать пластинке скорость $W_{т}$ при её перемещении из второго положения в первое, равную скорости движения твёрдого тела. После чего мы с помощью зависимостей акустического вида движения получим эпюры распределения нормальных скоростей на длине волны $\lambda$.

Затем, в соответствии с эпюрой нормальных скоростей в волне мы будем должны спрофилировать лобовую поверхность твёрдого тела. В этом случае длина волны $\lambda$ должна равняться половине длины нормального потока $h$ твёрдого тела. Полученный после такого профилирования профиль лобовой поверхности твёрдого тела будет соответствовать профилям второй группы лобовых поверхностей. После чего мы можем его просчитать в соответствии с вышеизложенными расчётами и получим соответствующую силу сопротивления, которую затем мы сможем сравнить, например, с силами сопротивления лобовой поверхности первой группы, образованной наклонной плоскостью, и выбрать для себя лучший вариант.

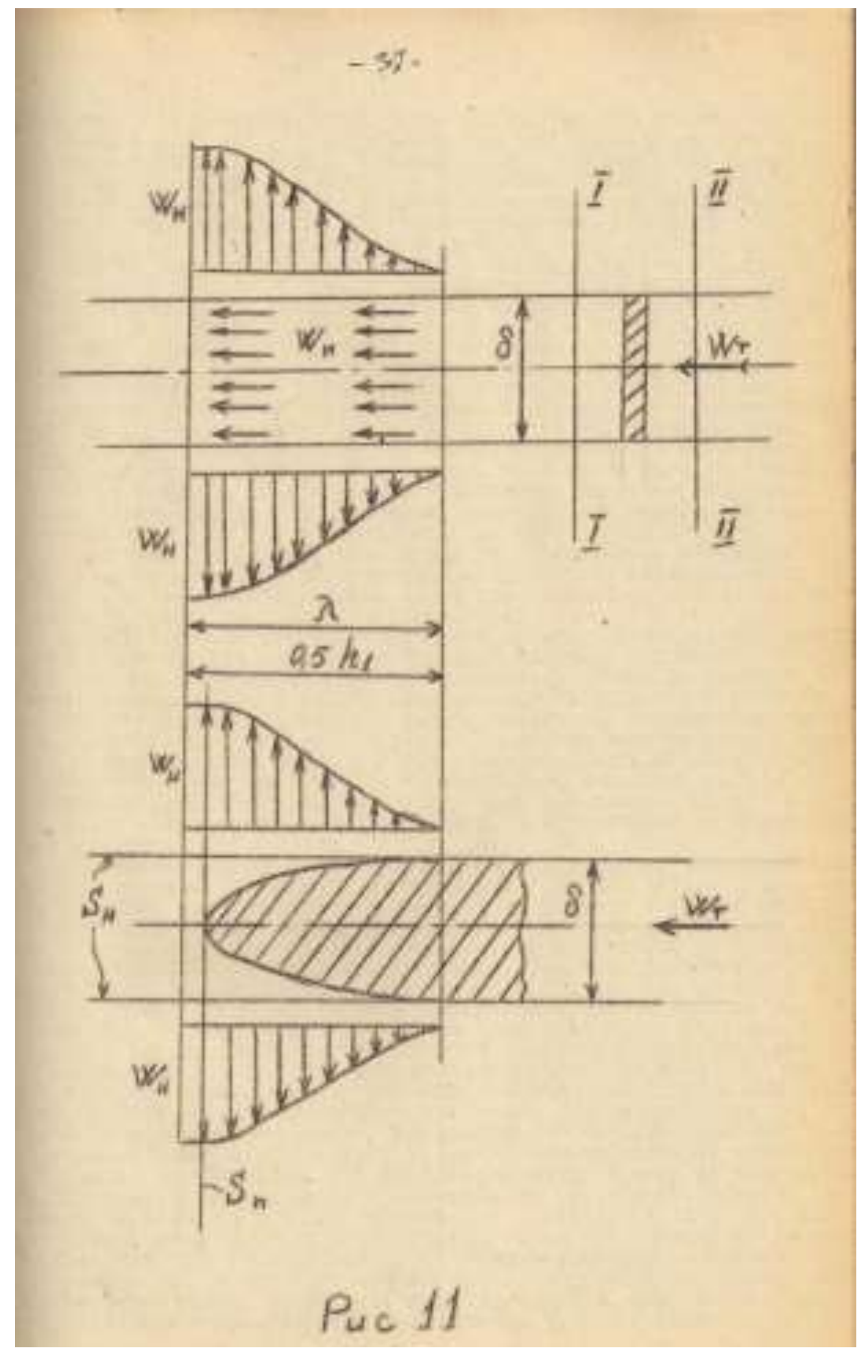

Рис 11

Отметим, что акустическая волна перемещается со скоростью звука, а здесь она будет перемещаться со скоростью твёрдого тела, которое образует поступательный поток для нормальных скоростей. Вот такова сущность волнового сопротивления.

Рассмотрев обтекаемые лобовые поверхности, мы видим, что здесь тоже нет нарушения закона сохранения энергии. Это значит, что при обтекаемой форме лобовой поверхности мы ничего не теряем и ничего не приобретаем в смысле энергетического баланса. Просто при перпендикулярном расположении лобовой поверхности твёрдого тела кинетическая энергия движения тела переходит в полную энергию поступательного потока, а при обтекаемой поверхности кинетическая энергия движения твёрдого тела переходит лишь в потенциальную, или статическую, энергию поступательного потока. В результате чего, мы имеем выигрыш в силе, затрачиваемой на перемещение твёрдого тела в среде.

Полученные выводы можно проверить практически на установившемся потоке, который движется в материальных границах. Лучше будет, если подобные практические исследования вы свяжете с формой лобовой поверхности конкретно интересующего вас движущегося в среде твёрдого тела. Тогда вы получите одновременно конкретные практические результаты, которые вы сможете сверить со своими расчётными результатами. Для этого вам придется взять установившийся поток жидкости, полная энергия которого равнялась бы полной энергии поступательного потока.

Если исследуемое вами движущееся тело очень большое, то вы его можете уменьшить в соответствующем масштабе. При этом вы будете должны сохранить неизменным либо угол наклона $\alpha$, либо характер кривизны лобовой поверхности. Материальные границы вашего установившегося потока должны придавать ему в сечении прямоугольную форму.

Покажем на рис. 12 модельные потоки.

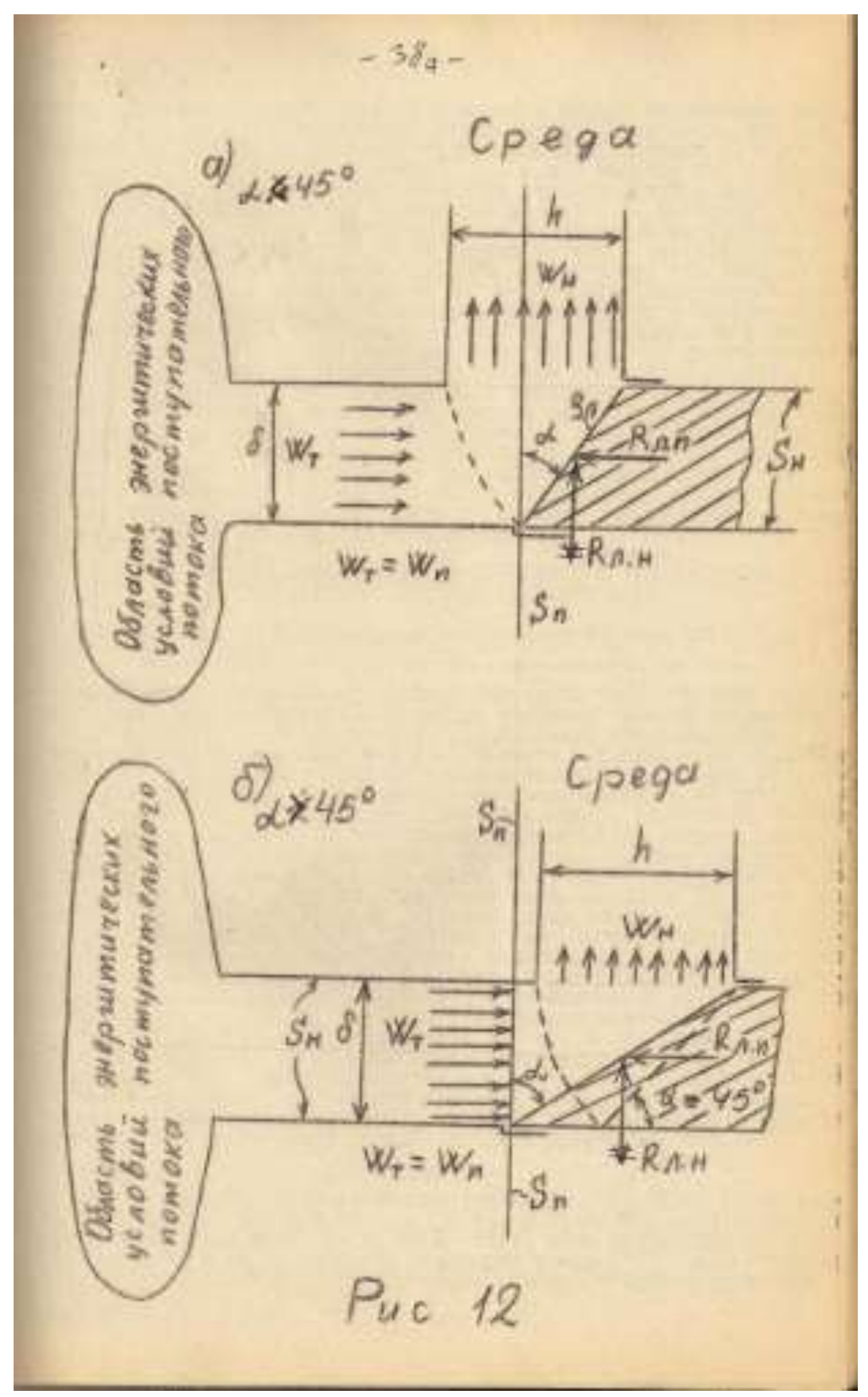

Для моделирования можно брать не только плоские лобовые поверхности, но и криволинейные. Это не принципиально. Просто на наклонных лобовых поверхностях удобнее пояснять.

На рис. 12.*а* мы показали лобовую поверхность с углом наклона $\alpha$ менее 45°, а на рис. 12.*б* – лобовую поверхность с углом наклона более 45°, поскольку мы условились, что при наклоне менее 45° лобовая поверхность образует нормальный поток, а при угле наклона более 45° существует только поступательный поток.

Далее, как видно из рисунка 12.*а*, ширина поступательного потока должна равняться толщине твёрдого тела $\delta$, а ширина нормального потока должна равняться ширине нормального потока $h$. Лобовую поверхность можно сделать подвижной, как это показано на рис. 12, тогда можно будет замерить величины нормальных $R_{л.н}$ и поступательных $R_{л.п}$ величин сил лобового сопротивления. Если все вышеизложенные условия будут соблюдены (рис. 12.*а*), то при замере энергии на входе и на выходе этого потока мы не обнаружим потерь энергии, а во втором случае (рис. 12.*б*) потери полной энергии будут существенными.

Причина различия в результатах замеров в том, что лобовая поверхность с углом наклона более 45° не образует нормального потока, а только лишь поступательный поток. Поэтому поступательный поток, подчиняясь материальным границам, будет вынужден развернуться на 90°, но этот разворот он будет совершать в соответствии с положениями плоского установившегося движения жидкостей и газов. Плоский установившийся поток существует в областях с различными энергетическими уровнями. В результате чего в данном случае (рис. 12.*б*) будут существенные потери энергии. Таким способом вы не только сможете убедиться в правильности своих расчётов, но и получите их количественное подтверждение. Дополнительно вы лишний раз убедитесь в действии закона сохранения энергии, ведь скорости движения в поступательных модельных потоках должны быть предельными.

Далее переходим к исследованию сопротивления тыльной поверхности твёрдого тела.

***Глава II***

# ТЫЛЬНОЕ СОПРОТИВЛЕНИЕ ПРОФИЛЯ ПРИ РАВНОМЕРНОМ И ПРЯМОЛИНЕЙНОМ ДВИЖЕНИИ ТВЁРДОГО ТЕЛА

## II. 1. ПОНЯТИЕ О НЕОБТЕКАЕМОЙ ПОВЕРХНОСТИ ТЫЛЬНОЙ ЧАСТИ ТВЁРДОГО ТЕЛА

Особенностью тыльной необтекаемой поверхности твёрдого тела является то, что по отношению к направлению движения тела она расположена сзади твёрдого тела. В то же время она представляет собой плоскость расположенную перпендикулярно к направлению движения твёрдого тела. Характер расположения тыльной необтекаемой плоскости накладывает свои особенности на её взаимодействие со средой. В этом случае она как бы обгоняет жидкость среды, которая должна непрерывно её догонять. В этот период организуется силовая связь между жидкостью среды и тыльной поверхностью, аналогичная силовой связи между магнитом и железным телом, которые взаимно притягиваются под действием силового поля. Подобное взаимодействие происходит между тыльной поверхностью твёрдого тела и жидкостью среды.

Величину взаимодействия, определяемого силовым полем, как мы выше договорились, выражаем через количественные характеристики жидкостного потока. Используя этот характер связи, мы произведём исследование взаимодействия необтекаемой тыльной поверхности движущегося твёрдого тела со средой. В данном случае среда так же будет реагировать двумя взаимно перпендикулярными потоками. Покажем взаимосвязь необтекаемой тыльной поверхности со средой на рис. 13.

Для исследования поступательного потока мы приняли поступательную плоскость $S_п$, а для исследования нормального потока мы приняли нормальные плоскости исследования $S_н$.

Начнём с поступательного потока, так как его характеристики мы можем определить из условия движения. Согласно этим условиям площадь сечения поступательного потока в поступательной плоскости исследования будет равна площади тыльной поверхности, то есть:

$$f_{п} = s_{тыл}.$$ [10]

Скорость движения поступательного потока $W_{п}$ будет равна скорости движения твёрдого тела $W_{т}$. В противном случае жидкость среды будет либо обгонять твёрдое тело, либо безнадёжно от него отстанет, чего не происходит. Плотность среды $\rho$ нам известна. Поэтому мы можем записать уравнение движения для поступательного потока. Оно будет иметь вид:

$$M_{п} = f_{п}\rho W_{п}. \qquad (63)$$

Затем запишем уравнение сил как произведение расхода массы на скорость:

$$f_{п}P_{п} = M_{п}W_{п} = f_{п}\rho W_{п}^{2}. \qquad (64)$$

Силу потока, воздействующую на тыльную плоскость, мы определили уравнением (64). Эта сила должна уравновешиваться тыльной силой $R_{тыл}$, которая затрачивается на преодоление сил поступательного потока, то есть:

$$R_{тыл} = f_{п}P_{п} = f_{п}\rho W_{п}^{2}. \qquad (65)$$

В уравнении (65) мы определили величину силы, которую нужно приложить к твёрдому телу, чтобы преодолёть тыльное сопротивление среды. Нам остаётся рассмотреть энергию поступательного потока.

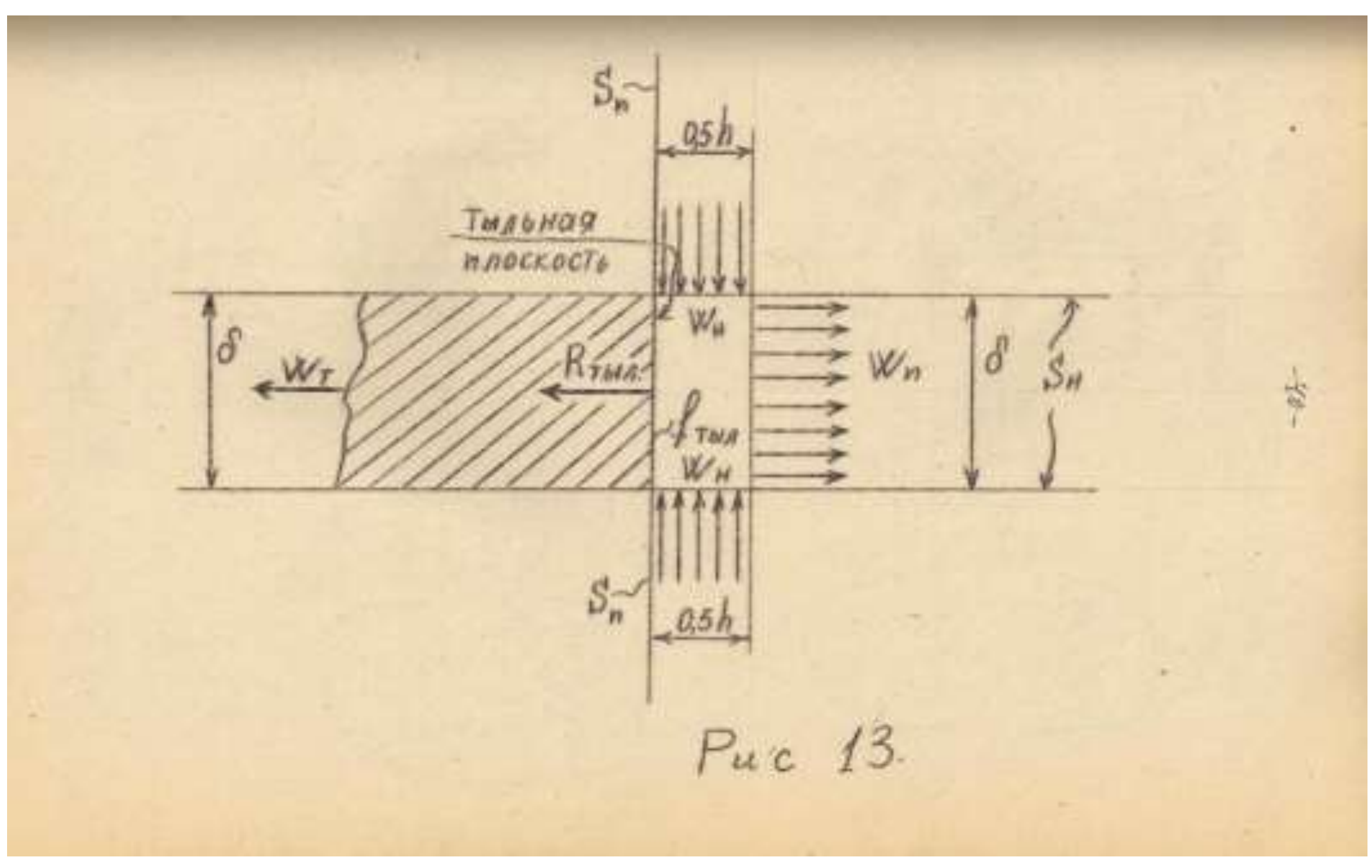

Рис 13.

Судя по силовому взаимодействию поступательного потока с тыльной плоскостью, можно сказать, что кинетическая энергия движения твёрдого тела полностью переходит в полную энергию поступательного потока, то есть:

$$U_{п} = VP_{пол.} = V\rho W_{п}^{2}. \qquad (66)$$

В этом случае скорость поступательного потока $W_{п}$ будет равна скорости твёрдого тела $W_{т}$, то есть $W_{п} = W_{т}$.

[10] На рис. 13 площадь тыльной поверхности обозначена как $f_{тыл.}$, но в данном случае и в следующем параграфе II. 2. площадь тыльной поверхности обозначается прописной буквой «s» как $s_{тыл.}$ (прим.ред)

Теперь разберёмся, что представляет собой энергия поступательного потока. Относительно невозмущенной среды, в которой движется твёрдое тело и которая обладает вполне определённым уровнем энергии, полная энергия поступательного потока определяет убыль энергии в объёме поступательного потока, то есть она является отрицательной величиной. Следовательно, полная энергия выражает величину убыли энергии среды в объёме этого потока. Это значит, чем больше скорость твёрдого тела, тем меньше энергии остается в объёме поступательного потока. Например, при достижении твёрдым телом скорости звука в атмосфере Земли в поступательном потоке энергия полностью отсутствует, то есть она равна нулю. Ибо величина отрицательной энергии этого потока будет равна полной энергии среды.

По отношению к нормальному потоку, площадь сечения которого определяется на нормальной плоскости исследования $S_н$, полная энергия поступательного потока является положительной, или действительной полной энергией, так как она определяет разность энергий среды и объёма потока. Поэтому нам остается принять за полную энергию нормального потока энергию поступательного потока и расписать её в соответствии с уравнением энергии Бернулли. Получим:

$$U_н = VP_{пол.\,н} = V\rho W_п^2 = VP_{ст} + \frac{1}{2} V\rho W_н^2. \qquad (67)$$

Запишем это уравнение в виде удельной энергии как:

$$P_{пол.\,н}(1/м^3) = \rho W_п^2 = P_{ст}\,(1/м^3) + \frac{1}{2}\rho W_н^2. \qquad (68)$$

Используя условия предельного движения в нормальном потоке, мы определим величину скорости нормального потока как:

$$\frac{1}{2}\rho W_п^2 = \frac{1}{2}\rho W_н^2.$$

Отсюда мы видим, что скорости поступательного и нормального потоков равны между собой. Далее по уравнению движения мы можем найти площадь сечения нормального потока, так как расход массы в единицу времени нам известен. Он будет равен расходу массы поступательного потока, то есть:

$$M_п = M_н = 1 \cdot h \cdot \rho W_н. \qquad (69)$$

Так как нормальные и поступательные скорости равны, то и величина $h$ будет равна толщине твёрдого тела $\delta$. Величина $h$ представляет собой общую длину нормального потока, но так как в нашем случае реализуются два нормальных потока (рис. 13), то для каждого из этих потоков она будет равна половине величины $h$, или 0,5 $h$.

Для нормального потока можно ещё записать уравнение сил. Как его записывать, нам уже известно, поэтому мы его записывать не будем.

Как видно, никаких потерь энергии здесь тоже не наблюдается. В этом случае необтекаемость тыльной поверхности тоже определяется полным переходом кинетической энергии движения твёрдого тела в полную отрицательную энергию поступательного потока. Практически в этом можно убедиться с помощью потока установившегося вида движения с материальными границами. Покажем этот поток на рис. 14.

На рисунке 14 показан установившийся поток с прямоугольной площадью сечения. Часть этого потока, которая имитирует нормальный поток, свободно соединена со средой. Другая часть, которая имитирует поступательный поток, соединена с областью вакуума.

Тыльная плоскость твёрдого тела составляет определённую часть материальной границы этого потока.

Нормальный поток можно имитировать либо одним потоком, либо двумя потоками, как это показано на рис. 14. Это не имеет принципиального значения. Затем мы в области вакуума уменьшаем энергию на величину соответствующую полной энергии поступательного потока для соответствующего движения твёрдого тела.

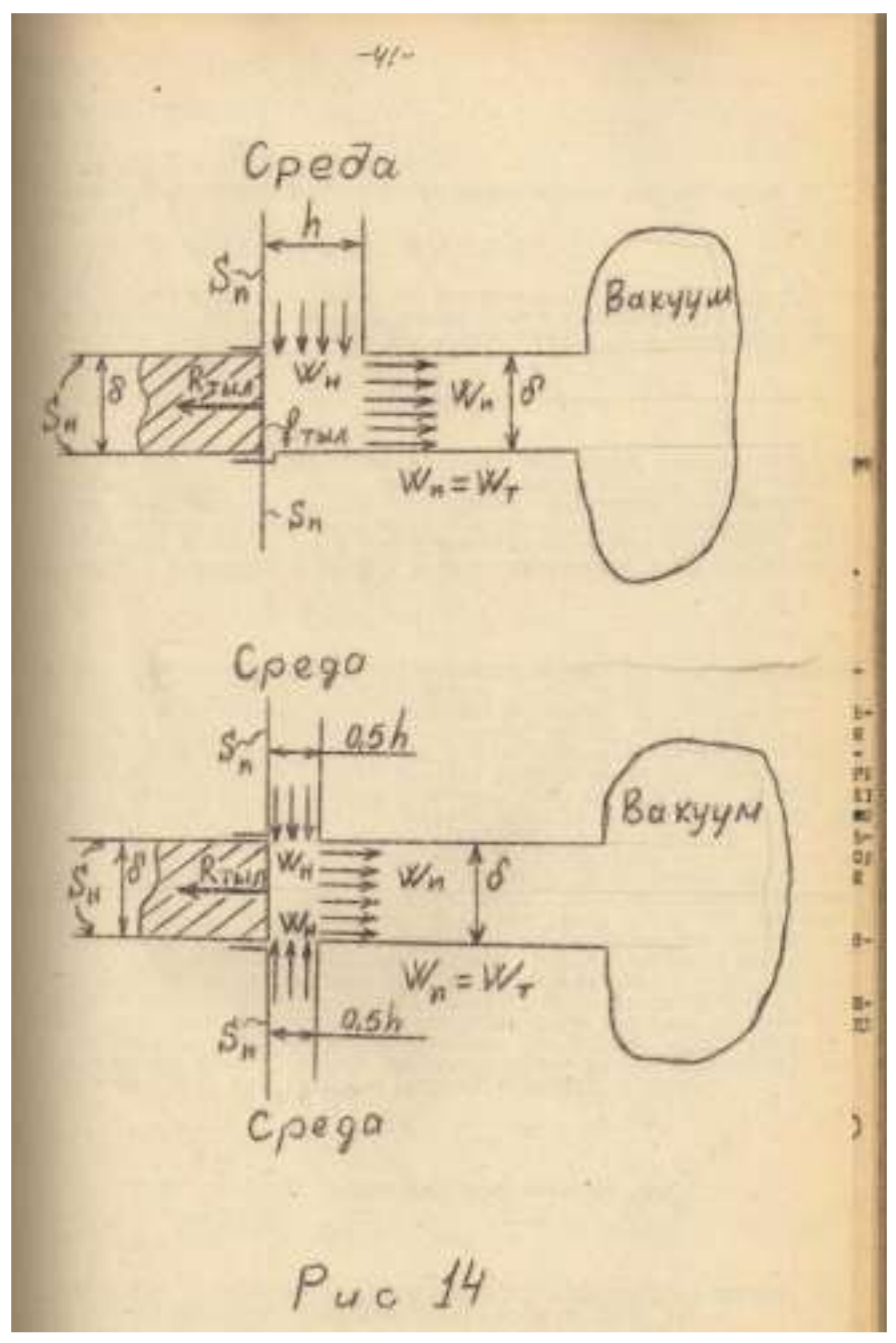

Жидкость в установившемся потоке приходит в движение, и мы начинаем снимать замеры интересующих нас характеристик. Получив замеры энергии на входе и на выходе, мы убедимся, что никаких потерь нет. Величина силы $R_{тыл}$, уравновешивающая взаимодействие потока с тыльной плоскостью твёрдого тела, окончательно убедит вас в правильности расчёта.

## II. 2. ПОНЯТИЕ ОБ ОБТЕКАЕМОЙ ПОВЕРХНОСТИ ТЫЛЬНОЙ ЧАСТИ ТВЁРДОГО ТЕЛА

С количественным определением обтекаемости мы уже познакомились выше. Основой его является величина силы, которая затрачивается на перемещение твёрдого тела в среде. Величина этой силы в свою очередь зависит от того, в каких условиях происходит взаимодействие лобовой поверхности со средой. Решающим обстоятельством в этих условиях является форма лобовой поверхности, от которой зависят условия перехода кинетической энергии твёрдого тела в полную, или потенциальную, энергию поступательного потока, а от этого зависит величина силы для твёрдого тела. Отметим, что понятия полной и потенциальной энергии тождественны. Просто мы здесь подчеркиваем то положение, что в данном случае полная энергия является целой величиной. В другом случае она развернута по уравнению Бернулли, которое характеризует переход потенциальной энергии в кинетическую и наоборот.

Эти же понятия определяют обтекаемость тыльной поверхности, поэтому тыльная поверхность, расположенная перпендикулярно направлению движения твёрдого тела, является необтекаемой поверхностью, ибо в этом случае кинетическая энергия движения твёрдого тела полностью переходит в полную энергию поступательного потока. Любая другая форма тыльной поверхности будет уже обтекаемой, но с различной степенью обтекаемости, которую мы определяем по различию сил, приложенных к твёрдому телу.

Для тыльной поверхности наиболее простой формой является форма, образованная наклонной плоскостью под определённым углом. Наклонные плоскости могут располагаться также симметрично относительно плоскости симметрии твёрдого тела. В общем, подобные формы мы отнесли к первой группе обтекаемых форм, из которых в последующем сможем образовать любую форму тыльной поверхности.

Вторая группа форм обтекаемости тоже присуща тыльной поверхности. По этой причине зависимости поступательных и нормальных потоков будут одинаковы с зависимостями потоков лобовой поверхности. Различие будет заключаться лишь в противоположном направлении движения потоков при тыльных и лобовых поверхностях, что для математики означает противоположность знаков. Поэтому мы дадим зависимости для потоков тыльной поверхности без особых разъяснений. Покажем на рис. 15 взаимодействие наклонной тыльной поверхности со средой.

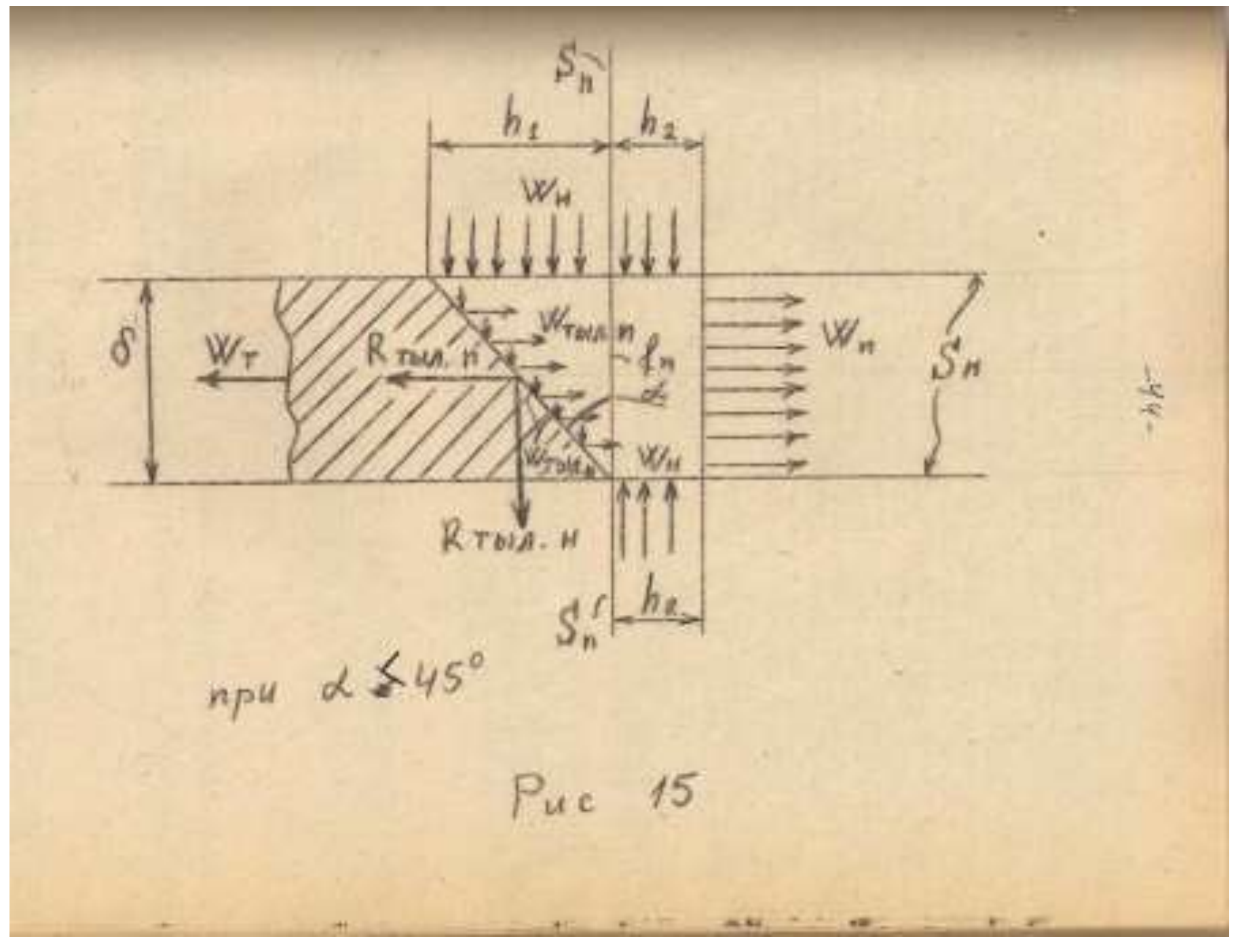

Рис 15

На этом рисунке показано, что твёрдое тело движется с постоянной скоростью $W_т$. Тыльная поверхность этого тела образована плоскостью, наклоненной под углом $\alpha$ относительно поступательной плоскости исследования $S_п$. Полная поверхность имеет площадь равную $s_{тыл}$. Также показаны поступательный и нормальный потоки. Теперь мы должны определить все характеристики для этих потоков.

Для площади сечения поступательного потока $f_п$, расположенной в поступательной плоскости исследования $S_п$, по скорости движения твёрдого тела $W_т$, которая в этом случае равна скорости поступательного потока $W_п$, с помощью уравнения движения мы определим расход массы в единицу времени в поступательном потоке. Для чего нам придется воспользоваться уравнением (63). Запишем его:

$$M_{п} = f_{п}\rho W_{п}^{2}.$$

Далее поступательный поток взаимодействует с тыльной плоскостью на всей её площади $s_{тыл}$. Так как эта площадь больше площади сечения поступательного потока $f_{п}$, то характеристики для этого потока взаимодействия будут иными:

$$M_{п} = s_{тыл}\rho W_{тыл.п}. \qquad (70)$$

Из уравнения движения (70) мы можем определить величину тыльной поступательной скорости. Расход массы мы определили по уравнению (63), а площадь тыльной поверхности будет равна:

$$s_{тыл} = \frac{\delta}{\sin\alpha}\cdot 1_{п.ед.} \qquad (71)$$

Скорость $W_{тыл.п}$ и расход массы $M_{п}$ мы можем определить по уравнению сил силового взаимодействия поступательного потока с твёрдым телом:

$$R_{тыл.\,п} = s_{тыл}\,P_{дин.\,тыл} = s_{тыл}\,\rho W_{тыл.п}^{2}. \qquad (72)$$

По уравнению сил (72) мы получили величину силы $R_{тыл.п}$, которую необходимо приложить к твёрдому телу, чтобы перемещать его в среде с соответствующей скоростью.

Полную энергию поступательного потока в плоскости исследования $S_{п}$ мы определяем по уравнению (66). Запишем его:

$$U_{пол} = VP_{пол} = V\rho W_{п}^{2}.$$

Оно характеризует полный переход кинетической энергии движения твёрдого тела в энергию поступательного потока. На наклонной тыльной плоскости мы получили меньшую величину скоростей взаимодействия $W_{тыл.п}$, что приводит к противоречию. Это противоречие устраняется, если полную энергию поступательного потока мы распишем по уравнению Бернулли:

$$U_{пол} = VP_{пол} = VP_{ст.п} + \frac{1}{2}V\rho W_{п.п}^{2}. \qquad (73)$$

Уравнение (73) является полным уравнением энергии поступательного потока, конечно, отрицательной.

Величину полной энергии поступательного потока мы определили по уравнению (66). Величину статической, или потенциальной, энергии мы можем определить тоже как переход энергии движения твёрдого тела в потенциальную энергию поступательного потока на площади $s_{тыл}$ наклонной тыльной поверхности, или:

$$U_{пол} = VP_{ст.п} = V\rho W_{тыл.п}^{2}. \qquad (74)$$

В уравнении (74) все характеристики нам известны. Поэтому мы можем определить величину потенциальной энергии. Зная теперь величину полной и потенциальной энергии, мы можем по уравнению (73) определить величину кинетической энергии потока и величину скорости поступательного потока $W_{п.п}$ как:

$$VP_{дин} = \frac{1}{2}V\rho W_{п.п}^{2}. \qquad (75)$$

Далее запишем уравнение (73) в виде удельной энергии поступательного потока, получим:

$$P_{\text{пол.п}}\,(1/\text{м}^3) = P_{\text{ст.п}}\,(1/\text{м}^3) + \frac{1}{2}\rho W_{\text{п.п}}^2\,. \qquad (76)$$

Теперь мы можем пользоваться ограничениями уравнения энергии для поступательного потока. Они записываются в таком виде:

$$P_{\text{ст.п}}\,(1/\text{м}^3) \geq \frac{1}{2}\rho W_{\text{п.п}}^2\,. \qquad (77)$$

В то же время мы знаем, что каждое конкретное значение этого неравенства зависит от каждого конкретного значения угла наклона тыльной плоскости $\alpha$. Предельному значению неравенства (77), когда неравенство переходит в равенство, то есть

$$P_{\text{ст.п}}\,(1/\text{м}^3) = \frac{1}{2}\rho W_{\text{п.п}}^2\,,$$

соответствует предельное значение угла $\alpha$. Подставим в это равенство значение статических сил давления $P_{\text{ст.п}}$, получим:

$$\rho W_{\text{тыл.п}}^2 = \frac{1}{2}\rho W_{\text{п.п}}^2\,. \qquad (78)$$

Из уравнения (78) мы определим предельную величину тыльной поступательной скорости $W_{\text{тыл.п}}$, а с её помощью определим уже предельный угол наклона тыльной поверхности $\alpha_{\text{пр}}$. Для тыльной плоскости он тоже равен 45°.

Для лобовой поверхности предельный угол наклона $\alpha_{\text{пр}}$ определяет предел существования нормального потока. Для тыльной плоскости предельное ограничение по углу наклона $\alpha$ выглядит несколько иначе. Это связано с особенностями взаимодействия тыльной поверхности со средой. Ибо объём поступательного потока образуется потому, что твёрдое тело как бы обгоняет жидкость среды в пределах этого объёма. Это значит, что жидкость среды не может организовать движения в объёме поступательного потока в направлении движения твёрдого тела, то есть жидкость не может догнать твёрдое тело в направлении его движения. Характер такого взаимодействия определяет в объёме поступательного потока уменьшение или падение величины полной энергии. Отсюда следует, что пополнение объёма поступательного потока жидкостью среды может происходить в любом случае только за счёт нормального потока. Тогда ограничения по углу наклона тыльной плоскости будут означать, что при изменении угла наклона тыльной плоскости $\alpha$ в пределах меньших, чем величина $\alpha_{\text{пр}}$, будут изменяться скорости нормального $W_{\text{н}}$ и поступательного $W_{\text{п.п}}$ потоков. Скорость поступательного потока $W_{\text{п.п}}$ мы имеем в виду как вторичную скорость, которая образуется за счёт дополнительного переноса потенциальной энергии среды нормальным потоком в объём поступательного потока.

Предельный угол наклона тыльной плоскости $\alpha_{\text{пр}}$ определяет минимальную величину скорости нормального потока $W_{\text{н.мин}}$ и величину максимальной вторичной скорости поступательного потока $W_{\text{п.п.макс}}$. Это значит, что при увеличении угла наклона тыльной плоскости больше предельного значения $\alpha_{\text{пр}}$, значения вышеуказанных скоростей не изменятся, и мы должны ими пользоваться для всех углов наклона, величина которых превышает величину $\alpha_{\text{пр}}$.

Будем считать, что мы выяснили условия нормального и поступательного потоков, которые ограничиваются количественно уравнением энергии.

Теперь мы можем приступить к определению характеристик взаимодействия наклонной тыльной поверхности с нормальным потоком.

Для нормального потока полной энергией будет потенциальная энергия поступательного потока, которая определяется уравнением (74). Распишем эту полную энергию по её составляющим по уравнению Бернулли. Получим:

$$U_{\text{пол. н}} = V\rho W^2_{\text{тыл.п}} = VP_{\text{ст. н}} + \frac{1}{2}V\rho W_{\text{н}}^2. \tag{79}$$

Далее запишем уравнение (79) как удельную энергию нормального потока:

$$P_{\text{пол.н}}(1/\text{м}^3) = \rho W^2_{\text{тыл.п}} = P_{\text{ст. н}}(1/\text{м}^3) + \frac{1}{2}\rho W^2_{\text{н}}. \tag{80}$$

Из уравнения (80) мы сможем определить величину скорости нормального потока $W_{\text{н}}$, используя условия ограничения уравнения энергии. Но здесь у нас могут реализоваться два случая: когда $\alpha < \alpha_{\text{пр}}$ (рис. 15) и когда $\alpha > \alpha_{\text{пр}}$ (рис. 16).

Для первого случая мы определяем нормальную скорость из равенства потенциальной и кинетической энергии уравнения (80), то есть:

$$\frac{1}{2}\rho W^2_{\text{тыл.п}} = \frac{1}{2}\rho W^2_{\text{н}}. \tag{81}$$

Из уравнения (81) мы получим величину скорости нормального потока на площади его сечения, расположенной в нормальной плоскости исследования, равной скорости взаимодействия тыльной поверхности с нормальным потоком $W_{\text{тыл.п}}$.

Для второго случая, когда $\alpha > \alpha_{\text{пр}}$, мы определяем нормальную скорость потока $W_{\text{н.мин}}$, для тыльной плоскости с углом наклона $\alpha$ равному предельному углу наклона $\alpha_{\text{пр}}$ тоже по уравнению (80), получим:

$$\frac{1}{2}\rho W^2_{\text{тыл.п}} = \frac{1}{2}\rho W^2_{\text{н.мин}}. \tag{82}$$

Определив скорость нормального потока, мы затем начинаем определять геометрические характеристики нормального потока. Для чего нам будет необходимо найти величину $h$ для нормального потока.

Для первого случая, когда $\alpha < \alpha_{\text{пр}}$, она определяется как

$$h = \frac{\delta}{\sin\alpha}. \tag{83}$$

Для второго случая, когда $\alpha > \alpha_{\text{пр}}$, она определяется как:

$$h_{\text{пр}} = \frac{\delta}{\sin\alpha_{\text{пр}}}. \tag{84}$$

На рисунках 15 и 16 показаны первый и второй случай движения нормального потока.

Для первого случая, при $\alpha < \alpha_{\text{пр}}$ (рис. 15), величина $h$ равна сумме:

$$h = h_1 + h_2 + h_2.$$

Это значит, что объём поступательного потока распространяется за пределы поступательной плоскости исследования $S_{\text{п}}$.

Для второго случая, при $\alpha > \alpha_{\text{пр}}$ (рис. 16), величина $h$ не доходит до поступательной плоскости исследования. Это значит, что длина поступательного потока находится в зоне профиля.

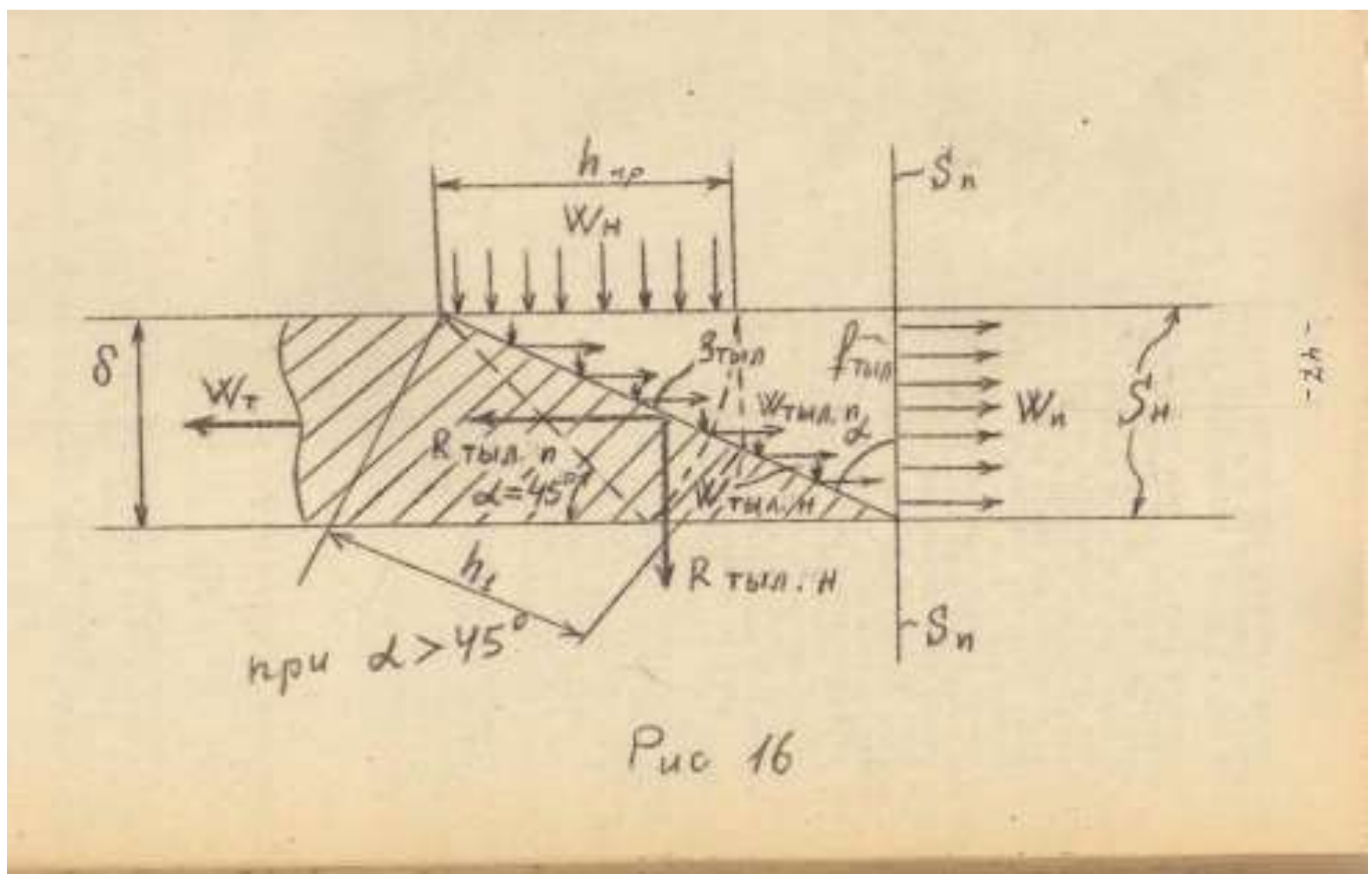

Рис 16

Дальше мы выясним, с какой силой действует нормальный поток на наклонную тыльную поверхность твёрдого тела.

В первом случае, при $\alpha < \alpha_{\text{пр}}$ (рис. 15), нормальный поток располагается по площади тыльной поверхности на длине $h_1$. Используя эту величину, мы можем определить, какая часть нормального потока взаимодействует с наклонной тыльной плоскостью. Это мы можем сделать по уравнению движения:

$$M_{\text{тыл. н}} = h_1 \cdot 1 \cdot \rho W_{\text{н}}. \qquad (85)$$

По уравнению (85) мы найдем величину расхода массы $M_{\text{тыл.н}}$, которая действует на тыльную поверхность. Поэтому мы должны будем записать уравнение движения для всей тыльной площади $s_{\text{тыл}}$:

$$M_{\text{тыл.н}} = s_{\text{тыл}} \cdot \rho W_{\text{тыл.н}}. \qquad (86)$$

Из уравнения (86) мы найдем величину скорости $W_{\text{тыл.н}}$ взаимодействия тыльной плоскости с нормальным потоком. Теперь, зная скорость взаимодействия, мы можем определить и силовое взаимодействие как

$$R_{\text{тыл. н}} = s_{\text{тыл}} \cdot \rho W^2_{\text{тыл.н}}. \qquad (87)$$

По уравнению (87) мы получили величину силы $R_{\text{тыл.н}}$, которая уравновешивает динамические силы давления нормального потока. В общепринятом понимании эта сила называется подъёмной силой.

Для второго случая, при $\alpha > \alpha_{\text{пр}}$ (рис. 16), силовое взаимодействие определяется следующим образом. Расход массы нам известен, так как он равен поступательному расходу массы:

$$M_{\text{п}} = M_{\text{н}}.$$

Силовое взаимодействие нормального потока происходит на части тыльной поверхности, которая ограничивается величиной $h_1$. Эту величину мы можем определить

как

$$h_1 = \frac{h}{\cos\alpha}.$$

Используя эту зависимость, мы можем получить величину площади взаимодействия. Далее с помощью этой величины и уравнения движения мы найдем величину скорости взаимодействия:

$$M_{\text{н}} = h_1 \cdot 1 \cdot \rho W_{\text{тыл.н}}. \qquad (88)$$

Зная скорость взаимодействия, мы можем определить и силовое взаимодействие как:

$$R_{\text{тыл.н}} = h_1 \cdot 1 \cdot \rho W^2_{\text{тыл.н}}. \qquad (89)$$

Сила $R_{\text{тыл.н}}$, приложенная со стороны твёрдого тела, уравновешивает динамические силы давления. Мы получили подъёмную силу для наклонной тыльной плоскости для второго случая, когда $\alpha > \alpha_{\text{пр}}$.

Отметим, что тыльная поверхность твёрдого тела может иметь любую форму. Взаимодействие этих поверхностей со средой можно определить тоже либо приближенным методом, то есть разделением их на множество участков наклонных тыльных поверхностей с различными углами наклона (как мы делали для лобовой поверхности), либо воспользоваться дифференциальным вычислением, но для этого потребуется самостоятельно записать зависимости в дифференциальной форме.

Для тыльной поверхности тоже возможно волновое сопротивление, или волновое взаимодействие. На рисунке 17 покажем, как это происходит.

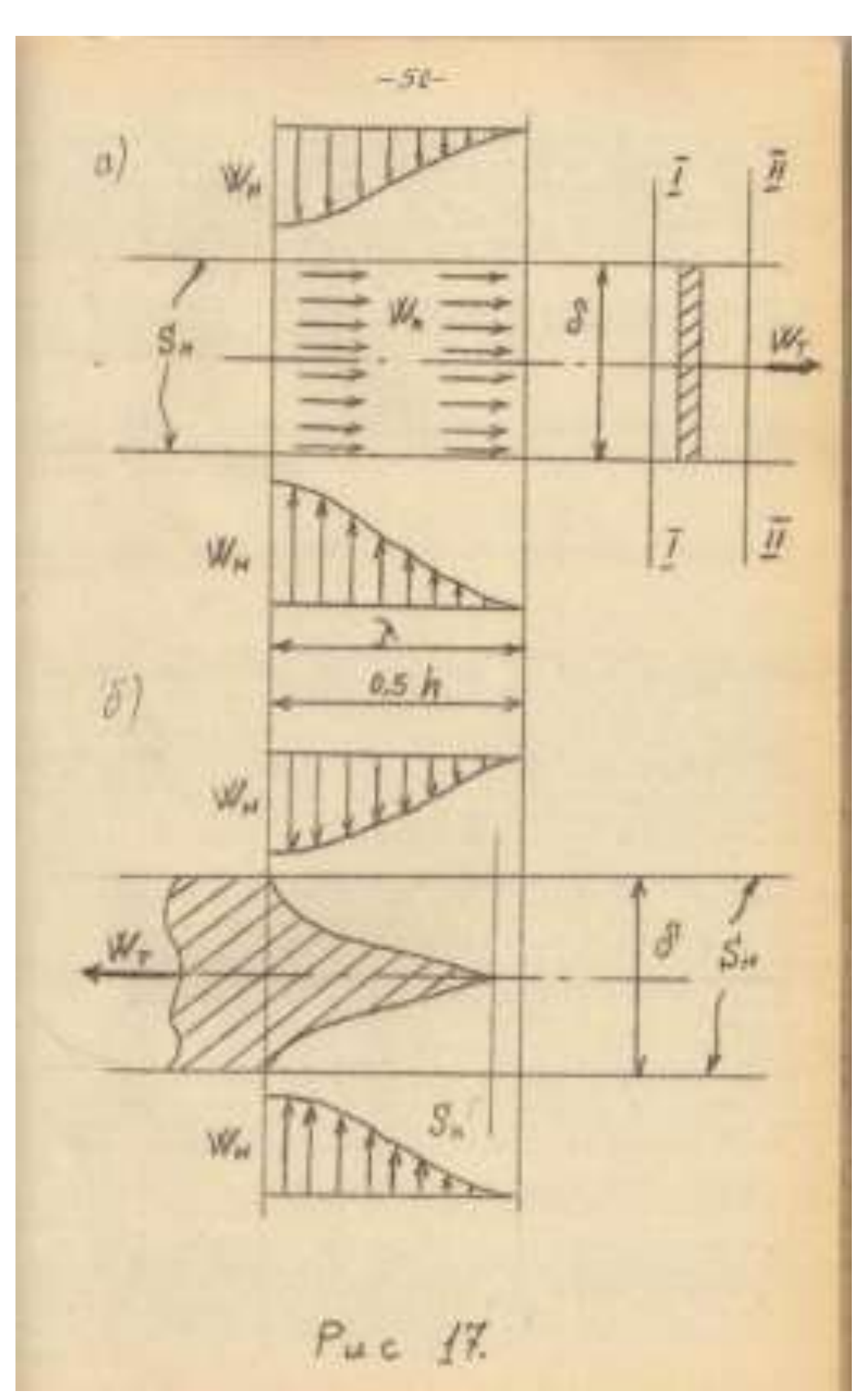

Рис 17.

На рис. 17.*а* показано движение пластинки из первого положения во второе. При этом движении образуется волна разрежения. Мы должны будем взять пластину с площадью равной площади сечения твёрдого тела, и совершить движение этой пластиной со скоростью $W_{пл}$ равной скорости твёрдого тела $W_т$. После чего мы получим профиль нормальных скоростей $W_н$ с помощью зависимостей акустического вида движения. После чего в соответствии с эпюрой распределения нормальных скоростей мы должны будем спрофилировать тыльную поверхность, как показано на рис. 17.*б*. После того, как мы получим профиль тыльной поверхности, мы сможем просчитать величину силового взаимодействия со средой и затем сравнить с силовым взаимодействием тыльных поверхностей других профилей

Будем считать, что мы получили необходимые количественные зависимости и соответствующие пояснения для тыльной поверхности твёрдого тела.

Отметим, что количественные величины мы можем получить практически на модельных потоках. Как это сделать, мы покажем на рисунках. На рис. 18 дадим модельные потоки для тыльной плоскости, у которой угол наклона $\alpha$ меньше предельного угла $\alpha_{пр}$, то есть $\alpha < \alpha_{пр}$.

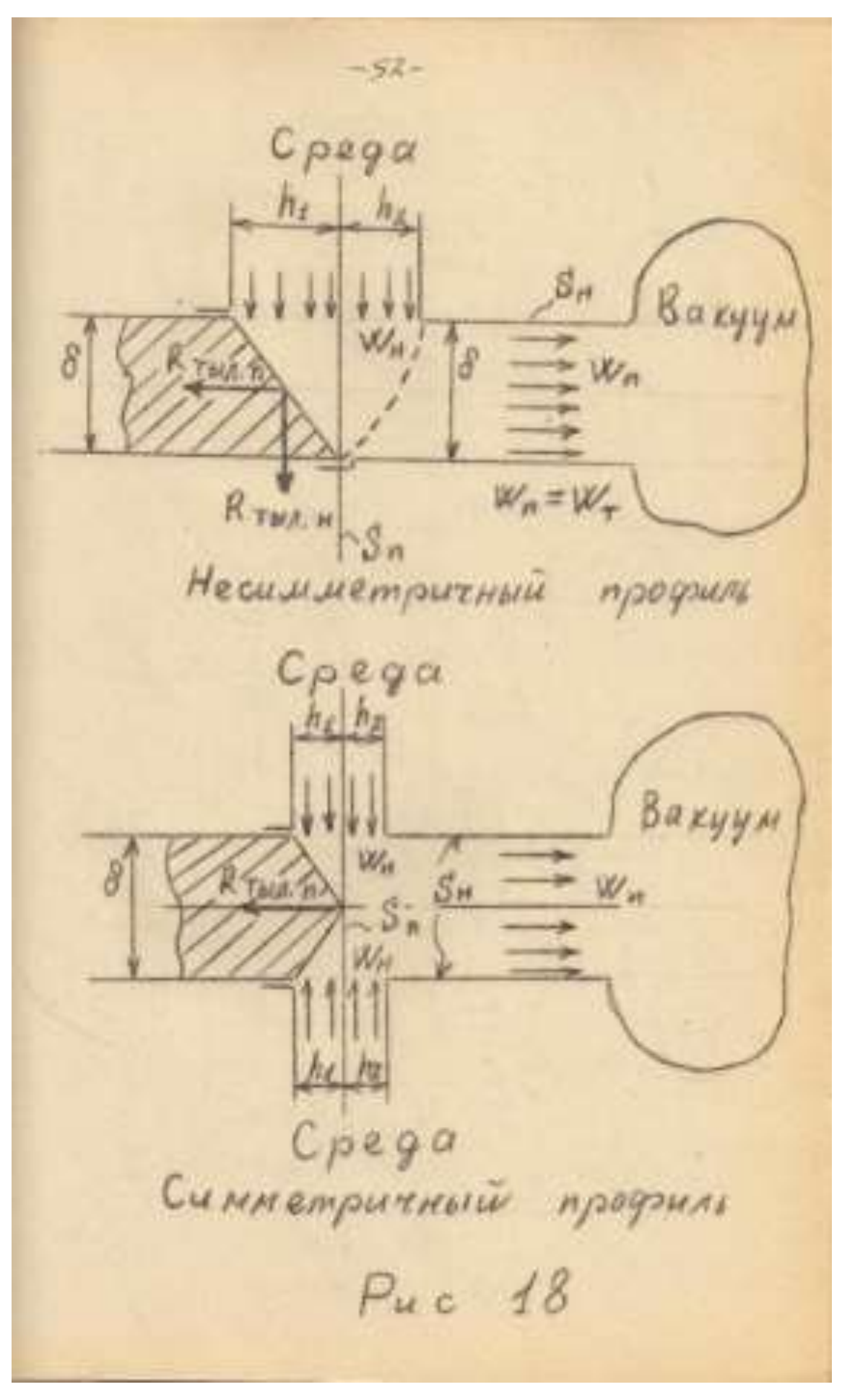

Рис 18

На рис. 19 мы дадим модельные потоки для тыльной плоскости, у которой угол наклона $\alpha$ больше предельного $\alpha_{пр}$, то есть $\alpha > \alpha_{пр}$ . Рисунки 18 и 19 наглядно показывают, каким образом можно получить практические результаты, используя установившийся поток.

Ещё раз напомним, что геометрические размеры потока и твёрдого тела можно уменьшать в определённом масштабе. Неизменными должны оставаться угол $\alpha$, условия среды, и величина вакуума должна соответствовать уменьшению полной энергии, которое образуется при движении твёрдого тела с определённой скоростью $W_т$.

Мы получили необходимые количественные зависимости и качественное описание взаимодействия лобовой и тыльной поверхностей твёрдого тела со средой, в которой движется твёрдое тело

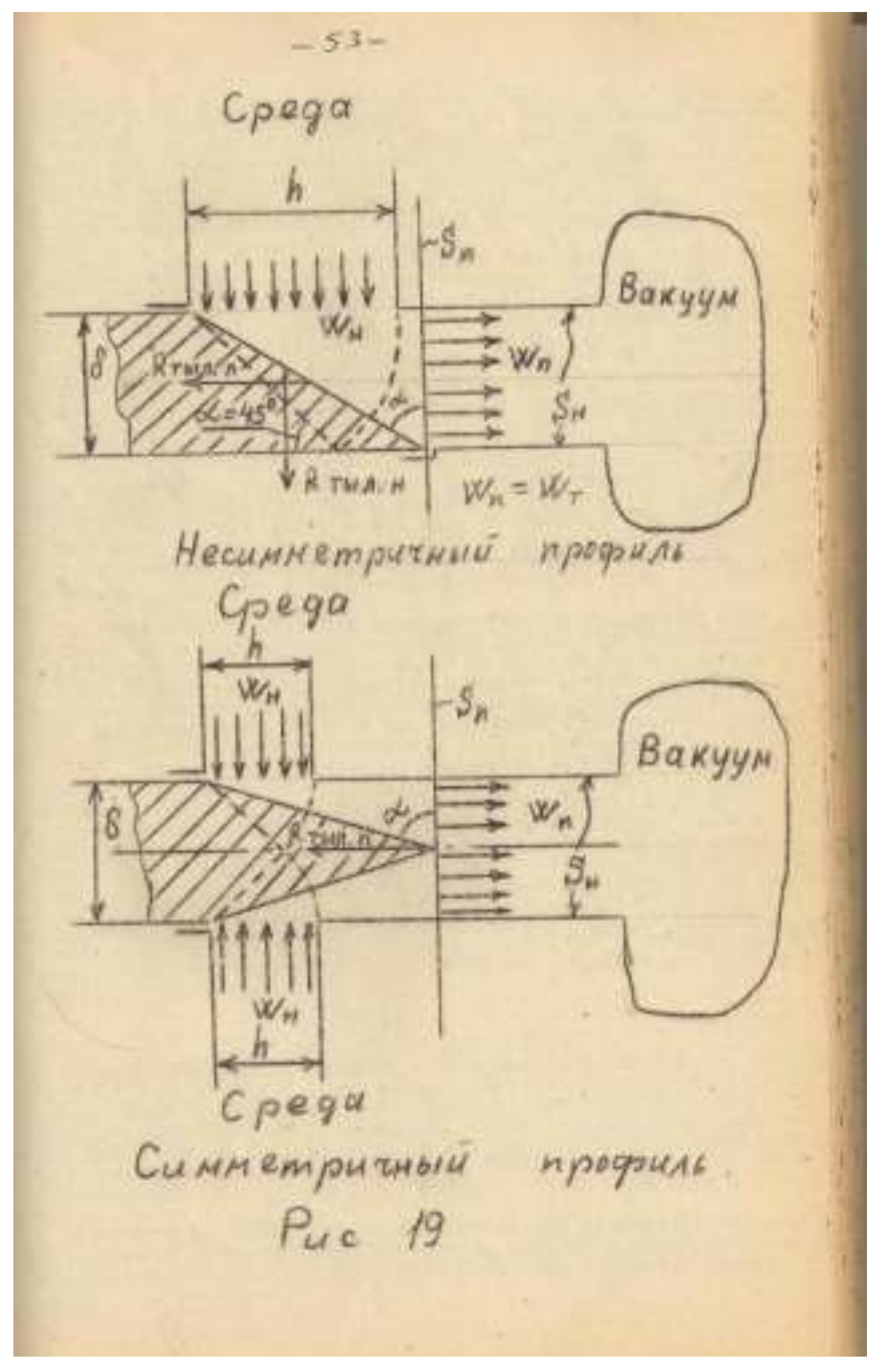

<u>Нам остается подвести итоги.</u>

Для определения необтекаемых форм лобовых и тыльных поверхностей существует одно-единственное понятие – это плоскость, расположенная перпендикулярно направлению движения твёрдого тела. Любые другие формы лобовых и тыльных поверхностей являются в определённой степени обтекаемыми. Показателем обтекаемости является величина силы, которую необходимо приложить к твёрдому телу для обеспечения его перемещения в среде. Поэтому для определения большей обтекаемости лобовых и тыльных поверхностей необходимо построить зависимость, например, в виде кривой, где определить, например, изменение обтекаемости в зависимости от изменения угла наклона поверхности $\alpha$. Затем, уже в соответствии с этой кривой, выбрать обтекаемые поверхности для своих практических целей. Чтобы эта кривая полностью отвечала практическим нуждам, необходимо ещё учесть для неё вязкость и трение, так как эти характеристики и их величины тоже связаны с силовым взаимодействием среды и твёрдого тела.

Во всех современных механиках, динамиках с приставкой “аэро-” существует одно понятие – лобовое сопротивление. Для определения этого понятия брались различные формы твёрдых тел и практическим путем для них получали коэффициенты лобового сопротивления. А затем уже эти коэффициенты использовали для практических целей в качестве коэффициентов пересчёта или характеристики подобия.

Как мы знаем, лобовые и тыльные поверхности в сумме являются неотъёмлемыми геометрическими характеристиками любого движущегося твёрдого тела. Эти поверхности выделены как поверхности силового взаимодействия твёрдого тела со средой, в которой оно движется. На лобовой и тыльной поверхностях происходит принципиально различное взаимодействие. В первом случае оно связано с повышением, или приростом, энергетического уровня в среде, во втором – с убылью энергии в среде. Поэтому мы должны будем сначала вычислить величину сил лобового сопротивления $R_{л.п}$, затем величину сил тыльного сопротивления $R_{тыл.п}$ , и после суммирования этих сил мы получим величину силы $R_{\Sigma}$, которую необходимо приложить к твёрдому телу, чтобы оно перемещалось в среде с соответствующей скоростью:

$$R_{\Sigma} = R_{л.п} + R_{тыл.п}. \qquad (90)$$

Если профили лобовой и тыльной поверхностей несимметричны, то для них таким же способом для каждой в отдельности определяют силы нормального сопротивления ($R_{л.н}$ и $R_{тыл.н}$), или подъёмные силы. Эти силы в зависимости от расположения профиля могут быть направлены либо в одном направлении, либо в противоположных направлениях. В зависимости от этого мы должны будем, руководствуясь положениями механики твёрдого тела, соответственно поступать с этими силами.

Для получения количественных зависимостей взаимодействия мы воспользовались плоскими твёрдыми телами, которые в перпендикулярном направлению своего движения протяжении имеют сколь угодно большую длину. В механике с приставкой “аэро-” твёрдые тела подобной формы имеют определение “крыла”. Мы сохраним это определение для подобных плоских тел.

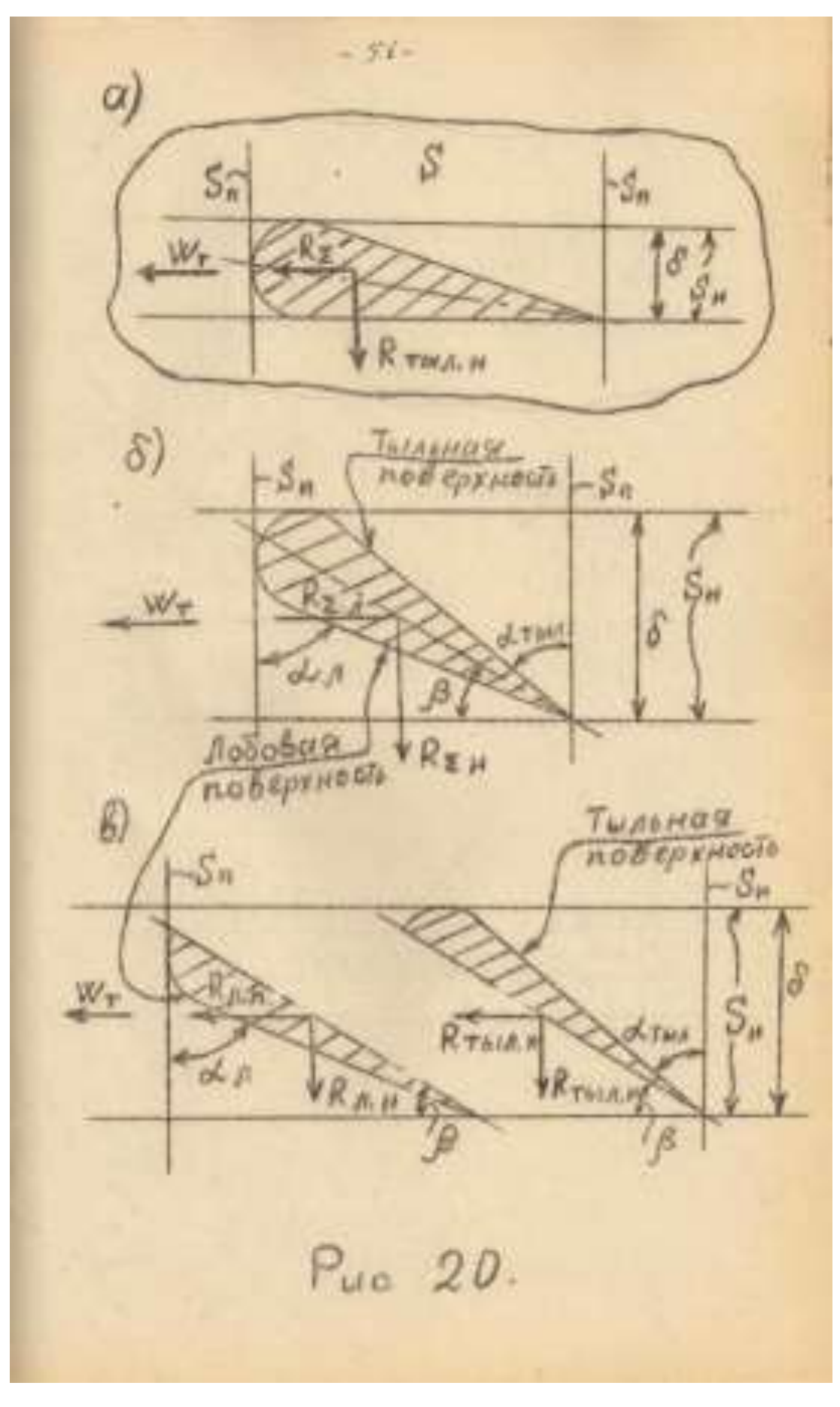

Рис 20.

Геометрическая форма сечения плоского тела, находящаяся в плоскости $S$, которая совпадает с направлением движения твёрдого тела и перпендикулярно расположена относительно плоскости твёрдого тела называется профилем (рис. 20.*а*). Это определение мы тоже сохраним.

В механике с приставкой "аэро-" существует понятие угла атаки $\beta$ для профиля крыла (рис. 20.*б*). Мы сохраним это понятие как символическое понятие, характеризующее разворот профиля по отношению движения крыла. Поэтому в соответствии с новой ориентацией крыла по отношению к направлению его движения мы должны будем выделить для него новые формы лобовой и тыльной поверхностей, как показано на рис. 20.*б*.

Для более удобного понимания этого положения можно представить себе новую лобовую и тыльную поверхности, как показано на рис. 20.*в*. Затем для этих новых лобовых и тыльных поверхностей будет необходимо провести вышеизложенные количественные расчёты и получить результаты в виде определённых сил взаимодействия. В конечном итоге, все эти количественные зависимости нужны человеку, чтобы обеспечить движение твёрдых тел в воде и воздухе с максимальной скоростью при минимальных затратах энергии. Силы лобового и тыльного сопротивления среды в первую очередь зависят от площади сечения твёрдого тела, то есть от толщины этого тела $\delta$.

Коль человек делает летательные и плавающие аппараты из реальных материалов, которые обладают определённой прочностью, то минимальная толщина профиля крыла будет диктоваться прочностными условиями. Получив минимальную толщину профиля из условия прочности, мы должны будем с помощью наших расчётов оформить его лобовой и тыльной поверхностями с минимальной силой взаимодействия.

Для летательных аппаратов большое значение имеет подъёмная сила крыла. Увеличение подъёмной силы для каждого конкретного крыла тоже в первую очередь связано с уменьшением толщины профиля $\delta$. Следовательно, с помощью наших расчётов мы должны будем добиться максимальной подъёмной силы при минимальной толщине профиля.

## II. 3. ИССЛЕДОВАНИЕ ДВИЖЕНИЯ ТВЁРДЫХ ТЕЛ, ИМЕЮЩИХ ЛЮБУЮ КОНКРЕТНУЮ ГЕОМЕТРИЧЕСКУЮ ФОРМУ

Выше мы рассмотрели движение в среде твёрдых тел типа "крыло". Но движущиеся в жидкостях и газах твёрдые тела могут иметь самую разнообразную форму. Наиболее характерными из таких тел являются удлинённые тела, которые в своем сечении имеют либо круглую, либо прямоугольную форму. Эти тела тоже будут иметь лобовую и тыльную поверхности взаимодействия. Необтекаемыми будут плоскости, расположенные перпендикулярно направлению движения. Покажем взаимодействие таких тел со средой на рис. 21.*а*.

На рис. 21.*а* показаны тела прямоугольной и цилиндрической формы. Лобовые и тыльные поверхности этих тел образованы плоскостями, расположенными перпендикулярно относительно направления их движения, то есть данные тела являются необтекаемыми.

Тогда, для количественного расчёта взаимодействия мы должны применить к ним зависимости необтекаемых плоскостей твёрдого тела типа "крыло", которые мы получили выше. При этом мы должны учесть следующие особенности:

что поступательный поток в нормальной плоскости исследования будет иметь прямоугольную или круглую форму, как показано на рис. 21. *а*;

что нормальные плоскости исследования будут располагаться по соответствующему периметру площади сечения поступательного потока. Соответственно, площади сечения нормальных потоков тоже будут располагаться по периметру этих сечений. В одном случае они будут располагаться на прямоугольнике, в другом – на цилиндре (рис. 21.*а*). Значит, длину поступательного потока $h$ мы должны распределять с учётом этих особенностей. Учтя эти особенности, мы легко справимся с расчётом взаимодействия лобовой и тыльной поверхностей.

Отметим, что лобовые и тыльные поступательные потоки всегда будут повторять форму площади сечения твёрдого тела поступательной плоскостью исследования. Соответственно, эту форму будут копировать и нормальные потоки. В то же время периметры площадей сечения, образованные замкнутыми кривыми, которые не имеют симметрии относительно двух взаимно перпендикулярных плоскостей, или образованные кривыми с вогнутыми участками будут вызывать нежелательные эффекты, связанные с увеличением силы сопротивления. Таким кривым надо посвятить специальный раздел исследования, а мы не можем позволить себе подобную роскошь.

На рис. 21.*б* показано вытянутое прямоугольное тело, которое имеет наклонные лобовую и тыльную плоскости. Эти поверхности могут быть также конусными или какой-либо другой формы.

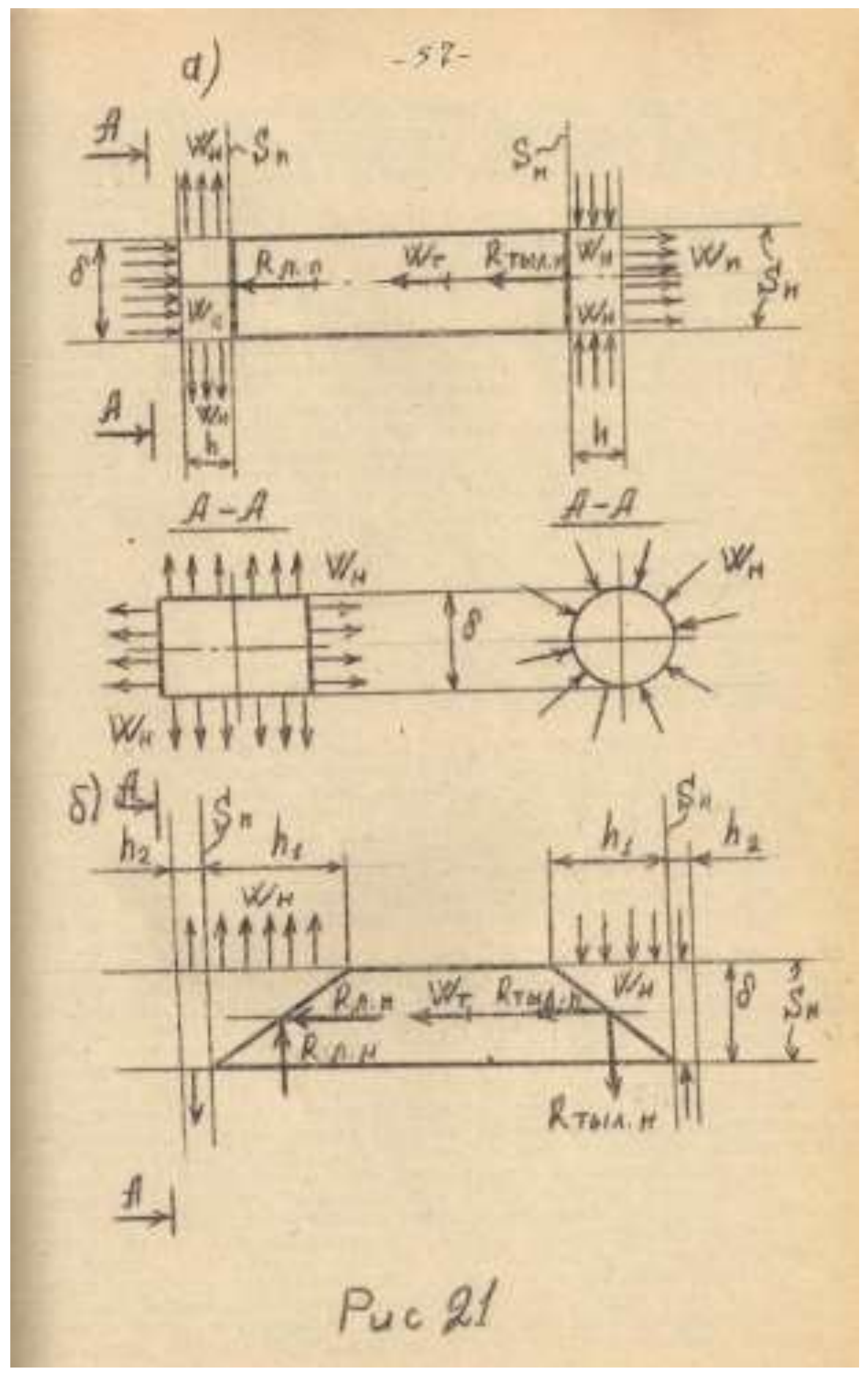

Все эти поверхности будут обладать определённой степенью обтекаемости. Для них мы тоже должны применить расчёты обтекаемых и необтекаемых лобовых и тыльных поверхностей. Мы также должны будем определить предельный угол наклона $\alpha_{\text{пр}}$ и связанные с ним ограничения. При применении данных расчётов мы должны будем учесть особенности распределения поступательных и нормальных потоков по периметру площади сечения твёрдого тела. Практические результаты для взаимодействия лобовой и тыльной поверхностей можно получить тоже с помощью модельных потоков установившегося вида движения, в которых тоже надо учесть особенности распределения поступательного и нормального потоков по периметру площади сечения твёрдого тела.

Так как мы не имеем возможности провести экспериментальные исследования, то обратимся за подтверждающими результатами к живой природе. Ведь она является самым совершенным и самым мудрым творцом. Посмотрим на самые совершенные образцы плавающих и летающих животных.

Наиболее совершенными из плавающих являются рыбы, а наиболее совершенной из них можно считать рыбу-меч, которая может развивать скорость до 130-140 км/час. Лобовые поверхности почти всех рыб образованы наклонными плоскостями с углом наклона больше предельного. Тыльные поверхности образованы наклонными плоскостями тоже с предельно большими углами наклона. Такие морские животные, как дельфины, киты, занимают второе место по плаванию. Они имеют округлые формы лобовой поверхности и тыльные поверхности с предельно большими углами наклона. Дельфин удивляет весь мир быстротой своего плавания. Люди ищут секреты его быстроты почему-то в загадочном устройстве его кожи. Что касается птиц, то они почти все имеют округлые формы лобовой поверхности и тыльные поверхности, образованные наклонными плоскостями.

Теперь сопоставим все природные формы поверхностей взаимодействия. Тыльные поверхности у всех плавающих и летающих будем считать одинаковыми. Следовательно, совершенство в плавании должно определяться совершенством формы лобовой поверхности. Но природа наделила своих лучших пловцов различными лобовыми поверхностями взаимодействия: одних – образованными наклонными плоскостями, других – округлыми формами. Это значит, что эти поверхности при взаимодействии со средой создают приблизительно одинаковое сопротивление, то есть они обладают одинаковой обтекаемостью. Если даже это так, то округлые формы лобовой поверхности имеют существенное преимущество. Они дают возможность заключить в области лобовой поверхности больший объём твёрдого тела. Это важное свойство, если учесть, что подобные твёрдые тела также являются корпусами летательных и плавающих аппаратов.

Далее, наблюдая за дельфинами, мы видим, как они выпрыгивают из воды и, ныряя в неё снова, не разбрызгивают ни одной капли. Это говорит о том, что округлые формы лобовой поверхности у дельфинов соответствуют формам волнового сопротивления для этой поверхности. Мы теперь знаем, как вычисляются и строятся поверхности волнового сопротивления. Если, например, носовой и кормовой частям корабля, как лобовой и тыльной поверхностям взаимодействия, придать форму волнового сопротивления, то при движении подобного корабля не будет возникать волны, которая для обычного корабля тянется, как усы, от его носовой и кормовой частей. Это тоже очень важное свойство корпусов летательных и плавающих аппаратов.

Секреты быстроходности дельфина ищут в его коже, но рыба-меч обладает чешуйчатой твёрдой кожей, а скорость развивает в два с лишним раза большую, чем дельфин. Следовательно, секрет быстроходности дельфина заключается в совершенной по обтекаемости лобовой поверхности как поверхности волнового сопротивления. Мы знаем, что в природе передача энергии идет по линии наименьшего сопротивления, а акустический, или волновой, способ ее передачи свидетельствует об этой природной линии наименьшего сопротивления для передачи энергии. По этой причине лобовая поверхность волнового сопротивления должна обладать минимальным сопротивлением при взаимодействии по сравнению с другими формами лобовой поверхности.

Будем считать, что мы выяснили принцип силового взаимодействия твёрдых тел различной формы со средой, в которой они движутся. Переходим к следующей теме.

## *Глава III*

## ИССЛЕДОВАНИЕ ДВИЖЕНИЯ ОЧЕНЬ ТОНКОЙ ПЛАСТИНКИ, КОТОРАЯ ОБЛАДАЕТ ВПОЛНЕ ОПРЕДЕЛЁННЫМ И КОНКРЕТНЫМ ВЕСОМ

Определение понятия очень тонкой пластинки здесь надо понимать так, что при движении пластинки в среде в направлении её торцов образуются поверхности взаимодействия – лобовая и тыльная - незначительной площади по величине близкой к нулю. Это значит, что их величинами мы можем пренебречь при количественных расчётах.

Вес очень тонкой пластинки означает, что на неё действует сила определённой величины. Поэтому в атмосфере или гидросфере Земли пластинка будет перемещаться под действием этой силы, если ей предоставить свободу движения. Она будет либо всплывать по закону Архимеда, если её удельный вес окажется меньше удельного веса жидкости среды, либо будет падать на поверхность Земли, если её удельный вес будет больше удельного веса вещества среды. Под средой мы здесь понимаем либо атмосферу Земли, либо её водные бассейны.

Твёрдые тела, если им предоставить возможность к свободному падению, движутся с постоянным ускорением, равным $g$ = 9,81 м/сек$^2$. Жидкости и газы, если им предоставить такую же возможность, будут падать с постоянной скоростью, равной $w$ = 3,132 м/сек [1].[11] Зависимость между постоянным ускорением и постоянной скоростью квадратичная, то есть

$$w = \sqrt{g} \quad [1].$$

Образно вы можете представить себе такую пластинку в виде листа, например жести, поверхность которого соответствует поверхности плоскости. Вы также могли наблюдать движение подобного листа в атмосфере. Если ему предоставляли возможность свободно падать, то он падал под действием силы собственного веса, но падал не с постоянным ускорением $g$, а несколько замедлённо. Если вы придавали этому листу определённую начальную скорость в направлении его торцов, то лист стремился сохранить эту направлённость движения независимо от направления действия сил его собственного веса. Приступим к исследованию движения очень тонкой пластины.

Будем считать, что вес пластины равномерно распределен по её площади. Поскольку она является твёрдым телом, то мы можем привести её распределённый вес к сосредоточенному весу, расположенному в центре тяжести пластинки. Тогда вектор сил веса пластинки независимо от её положения будет всегда направлен к центру Земли. Поэтому мы можем рассматривать положение пластинки в её движении относительно вектора силы веса.

Теперь рассмотрим движение пластинки, когда вектор силы её веса будет расположен перпендикулярно плоскости пластины.

При таком расположении пластинки относительно вектора силы веса она может совершать движение в среде только в виде свободного падения. Определимся с этим движением относительно положений механики безынертной массы. Для чего покажем движение пластины на рис. 22.*а*.

Поскольку вектор силы веса расположен перпендикулярно к плоскости пластинки, то направление взаимодействующих потоков тоже будет перпендикулярным по отношению к плоскости пластинки. По этой причине плоскости пластины будут поверхностями взаимодействия твёрдого тела со средой. Тогда одна из этих плоскостей будет лобовой, а другая – тыльной плоскостью. Так как эти плоскости расположены перпендикулярно к направлению движения, то они будут необтекаемыми поверхностями взаимодействия. Затем, уже относительно этих плоскостей, расположим свои плоскости исследования: поступательные $S_{п}$ и нормальные $S_{н}$ (рис. 22.*а*). Отметим, что в плане пластинка может иметь самую разнообразную конфигурацию.

[11] «падение», т.е. движение жидкостей и газов под действием силы земного (любого планетарного) притяжения, означает, что данное движение жидкостей и газов происходит в среде, которую они образуют. Это движение как бы «жидкости в жидкости», «газа в газе» рассмотрено в работе « II.«Строение Солнца и планет солнечной системы с точки зрения механики безынертной массы» на примере строения и динамики атмосфер Земли и Венеры. Там же дано описание простого опыта, демонстрирующего постоянство и величину скорости падения. Без учёта даже существования скорости свободного падения вещества среды невозможно будет рассчитывать, например, парашюты для приземления на других небесных телах.

Центр тяжести *ц.т.* и вектор силы веса пластины $G_т$ тоже показаны на рис. 22.*а*. Поскольку мы проводим исследование свободного падения пластинки, то мы его должны делать не просто так, а в зависимости от чего-то. Поэтому мы будем рассматривать изменение скорости свободного падения пластинки в зависимости от её площади $F$ и силы её веса $G_т$. В движении пластинки нас будут интересовать два крайних предела её свободного падения. В первом случае это будет минимальный предел скорости свободного падения $W_{т.мин}$ при одновременном увеличении её площади $F$ и уменьшении её веса $G_т$. Наглядно можно представить себе это положение, как если бы пластина становилась всё тоньше, при этом часть её массы отбрасывалась. Во втором случае нас будет интересовать максимальный предел скорости свободного падения пластинки $W_{т.макс}$ при одновременном уменьшении её площади $F$ и увеличении её веса, что можно представить себе, как непрерывное превращение пластины в пруток.

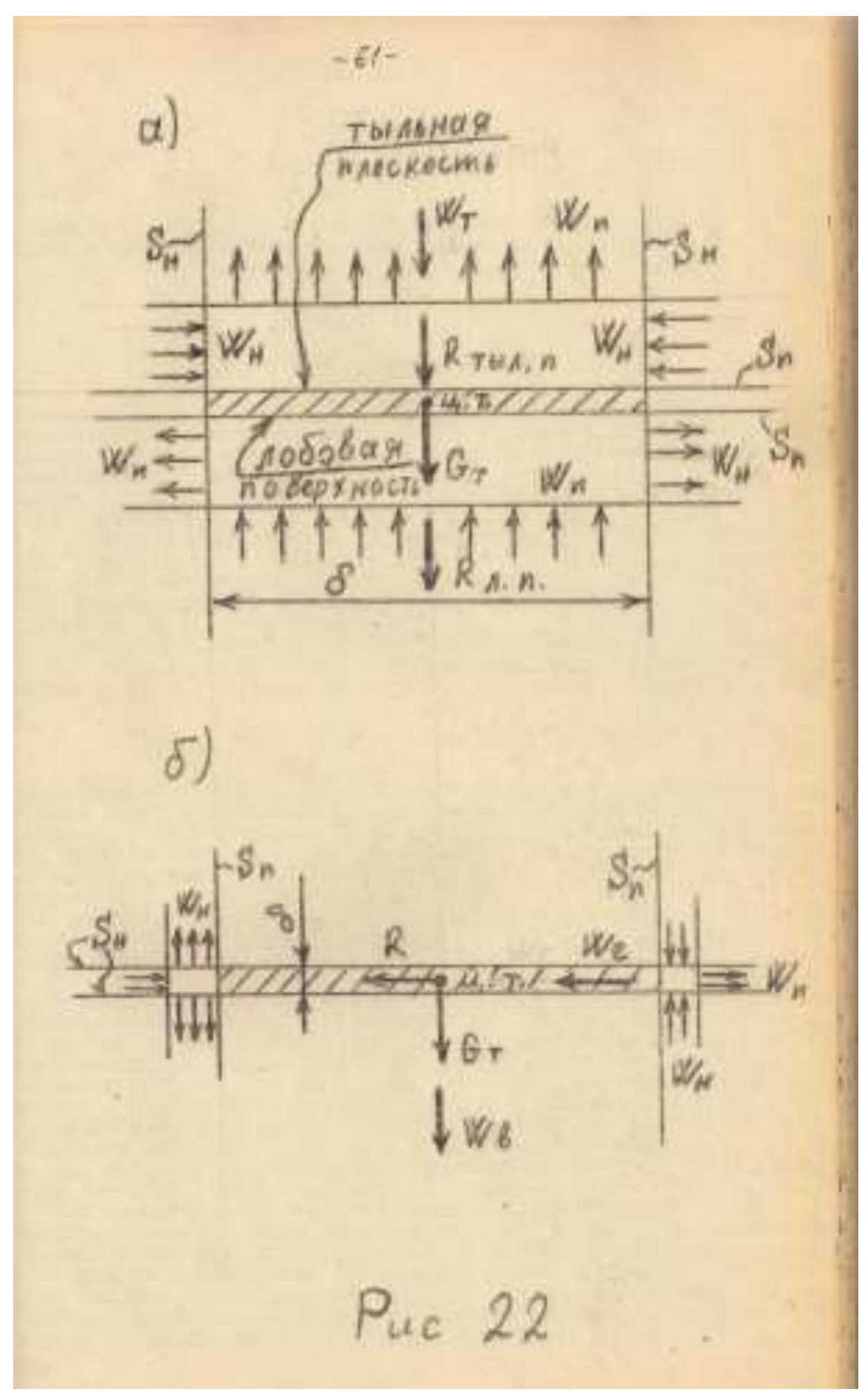

Рис 22

Рассмотрим первый предельный случай.

Коль мы определили поверхности пластины необтекаемыми, то для количественной характеристики взаимодействия мы должны будем пользоваться зависимостями для необтекаемых лобовых и тыльных плоскостей, полученными выше. Получая эти зависимости, мы как бы задавались величиной скорости движения тела и взависимости от неё получали силы взаимодействия. Теперь нам придётся решать обратную задачу. Мы имеем определённую величину силы, которая по величине равна весу пластинки $G_т$, и в зависимости от величины этой силы нам необходимо определить величину скорости $W_т$. Поскольку мы имеем дело с силами, то для количественной характеристики возьмём уравнения сил лобового и тыльного поступательных потоков. Они имеют вид:

$$R_{л.п} = F_л \cdot \rho W_{л.п}^2 \ .$$

$$R_{тыл.п} = F_{тыл} \cdot \rho W_{тыл.п}^2 \ .$$

Эти две силы лобового и тыльного нормальных поступательных потоков в сумме должны равняться величине силы веса пластинки $G_т$, то есть:

$$G_т = R_{л.п} + R_{тыл.п} = F_л \cdot \rho W_{л.п}^2 + F_{тыл} \cdot \rho W_{тыл.п}^2 \ . \qquad (90)$$

Уравнением (90) мы записали силовое равновесие пластинки в свободном падении. Теперь определим, какие доли в этом равновесии занимают лобовые и тыльные силы. В нашем конкретном случае лобовая и тыльная поверхности равны по площади, то есть $F_л = F_{тыл}$ .

В связи с тем, что мы имеем дело с необтекаемыми плоскостями, то поступательные скорости взаимодействия лобовой $W_{л.п}$ и тыльной $W_{тыл.п}$ поверхностей будут равны скорости падения пластинки $W_т$, то есть $W_т = W_{л.п} = W_{тыл.п}$. Плотность среды тоже будет одинаковой для обоих потоков. Поэтому величина лобового сопротивления $R_{л.п}$ будет равна величине тыльного сопротивления $R_{тыл.п}$, то есть $R_{л.п} = R_{тыл.п}$. Это равенство означает, что каждая из этих сил уравновешивает ровно половину веса пластинки, то есть их величина равна $0{,}5G_т$. Для лучшего понимания изменения скорости пластины $W_т$ запишем уравнение (90) через её величину, получим:

$$G_т = R_{л.п} + R_{тыл.п} = 2F\rho W_т^2. \qquad (91)$$

Мы разобрались с силами, а теперь посмотрим, как будут изменяться скорости падения пластинки $W_т$ при увеличении её площади $F$ и уменьшении веса $G_т$.

Из уравнения (91) мы видим, что скорость падения пластины будет уменьшаться. При этом мы будем иметь дело с квадратичной зависимостью уменьшения скорости, так как скорость в этом уравнении стоит в квадрате.

Для этой зависимости мы можем раздельно изменять вес и площадь. Например, уменьшать вес $G_т$ пластины и оставлять её площадь $F$ неизменной или наоборот. От этого суть дела не изменится. Просто комплексным изменением мы подчеркиваем одностороннюю зависимость скорости падения пластины от этих характеристик. Зависимость (91) проста, и мы легко можем представить себе изменение скорости и, естественно, согласимся с реальностью такого изменения.

Далее нас будет интересовать предельное уменьшение скорости падения, то есть величина $W_{т.\ мин}$. Теперь нам придётся вспомнить, что скорость свободного падения жидкостей и газов в поле земного тяготения равна $w = 3{,}132$ м/сек. Это значит, что когда мы уменьшим вес пластинки $G_т$ и увеличим её площадь $F$ настолько, что скорость свободного падения пластинки $W_т$ станет равной скорости падения жидкостей и газов $w$, то есть $W_т = w = 3{,}132$ м/сек, то сила сопротивления тыльного поступательного потока $R_{тыл.п}$ исчезнет, поскольку масса газа или жидкости тыльного поступательного потока будет находится в свободном падении и не будет вызывать силового взаимодействия с тыльной поверхностью пластины. Остаётся только силовое взаимодействие лобового потока. Это значит, что при достижении пластинкой скорости свободного падения жидкостей и газов $w$, силовое сопротивление среды уменьшается ровно наполовину.

Практически мы наблюдали бы падение такой пластины в виде неравномерного изменения её скорости. То есть сначала мы бы увидели, что пластина в момент начала свободного падения как бы увеличивает свою скорость падения. Когда её скорость станет равной 3,132 м/сек, то нам показалось бы, что она остановилась на какое-то мгновение. Затем её скорость падения снова начала бы возрастать, и через короткий промежуток времени пластина снова как бы остановилась на мгновение, и т.д. Такой характер движения будет сохраняться на всём пути пластинки. Это означает, что когда пластинка достигает скорости падения 3,132 м/сек, нам кажется, что она останавливается. В этот момент силовое взаимодействие тыльного потока и пластинки прекращается. В силу этой причины она начинает ускорять своё падение до тех пор, пока снова не включится в работу её тыльный поток. Всё это в сумме мы наблюдаем как неравномерное падение пластины.

Отсюда для себя мы должны будем сделать вывод, что минимальной скоростью падения пластины $W_{т.\ мин}$ является скорость свободного падения жидкостей и газов в поле земного тяготения $w$ = 3,132 м/сек. Минимальная скорость свободного падения пластинки не зависит от дальнейшего увеличения площади пластинки $F$ и уменьшения её веса.

С одной стороны, вы со мной здесь согласитесь, а с другой стороны, скажете, что, например, пух может падать с меньшей скоростью, чем скорость $w$. Вы тут тоже будете правы. Мы живем в мире реальных вещей, поэтому нам придётся разобраться, что означает ваш пример. Ведь мы имеем в виду, в конечном итоге, не идеальные пластины, а вполне реальные, сделанные из реальных материалов.

Для дальнейшего нашего пояснения отнесём вес пластины и силы её взаимодействия со средой к единице площади, то есть разделим уравнение (91) на площадь $F$ пластины, получим:

$$G_т(1/м^2) = P_{л.п} + P_{тыл.п} = 2\rho W_т^2. \qquad (92)$$

Так вот, изготовляя пластину из реальных материалов, мы не сможем сделать её настолько тонкой, чтобы она не теряла своей формы плоскости по всей своей площади или же не разрушилась на мелкие частицы, которые мы называем пылью. Ведь удельный вес любых твёрдых веществ намного больше удельного веса воздуха атмосферы. Так, для того, чтобы получить скорость падения нашей пластинки $W_т \leq$ 3,132 м/сек, мы должны сделать её очень тонкой. А сделать этого не позволяют прочностные характеристики вещества. Поэтому в обыденной практике мы пользуемся пластинками, вес единицы площади которых лежит в пределах зависимости (92), то есть скорость падения такой пластины будет больше или равна 3,132 м/сек. Это значит, что невозможно, например, изготовить парашют, чтобы скорость спуска на нём была меньше, чем скорость падения жидкостей и газов в среде, то есть в атмосфере Земли в данном случае. Но мы можем сделать такой парашют, что скорость спуска на нём будет постоянной независимо от высоты, то есть от изменения плотности атмосферы по высоте. Вот почему мы утверждаем, что скорость $w$ = 3,132 м/сек является минимальной предельной скоростью для скорости свободного падения пластины.

В природе также существуют чешуйчатые пластинки типа пыли, площадь которых измеряется в микронах, а толщина – в миллимикронах. Вот такие пластинки способны падать со скоростью $W_т \leq$ 3,132 м/сек. Для подобных пластинок силовое взаимодействие происходит только по лобовой плоскости. Тогда уравнение (96) принимает вид:

$$G_т(1/м^2) = P_{л.п} = \rho W_т^2. \qquad (97)$$

В уравнении (97) скорость падения пластинки будет зависеть в основном от физических возможностей конкретного вещества к утончению, то есть минимальная скорость падения $W_{т.\ мин}$ здесь будет определяться физическими возможностями каждого конкретного вещества. По этой причине пух тоже падает со скоростью меньше предельной. Различие здесь заключается в том, что его площадь представлена в виде нитевидных переплетений.

В то же время мы можем наблюдать падение реальных пластинок со скоростью меньшей минимальной. Ибо плотность воды имеет сравнительно большую величину, соизмеримую с плотностью твёрдых тел. Следовательно, минимальную скорость падения $W_{т.\ мин}$ надо понимать как минимальную скорость, при достижении которой кончается силовое взаимодействие твёрдых тел со средой по одной из двух поверхностей.

Рассмотрим второй предельный случай, когда вес пластинки увеличивается, а её площадь уменьшается.

При свободном падении подобной пластинки силовая взаимосвязь по её лобовой и тыльной поверхностям будет осуществляться в соответствии с уравнениями (91) и (92). Согласно этим зависимостям, скорость свободного падения пластинки $W_т$ будет увеличиваться при уменьшении её площади $F$ и увеличении её веса $G_т$. В конечном итоге подобное увеличение скорости свободного падения пластины будет стремиться к форме падения твёрдых тел в безвоздушном пространстве, то есть она будет стремиться падать с ускорением $g$.

Отметим, что замена необтекаемых плоскостей пластины обтекаемыми будет приводить к подобному эффекту, то есть к увеличению скорости свободного падения, поскольку обтекаемые поверхности уменьшают величину сил сопротивления лобового и тыльного поступательных потоков. Отсюда следует, что предельные скорости свободного падения пластинки надо понимать как её стремление к форме падения твёрдых тел в соответствии с законами механики твёрдого тела.

## III. 1. ДВИЖЕНИЕ ПЛАСТИНКИ В НАПРАВЛЕНИИ ЕЁ ТОРЦОВ ПОД ДЕЙСТВИЕМ ПРИЛОЖЕННЫХ СИЛ

В этом разделе мы продолжим исследование движения очень тонкой пластинки, которая имеет собственный вес. Напомним ещё раз, что очень тонкая пластинка означает, что при движении пластинки в направлении её торцов величины сил сопротивления лобовой и тыльной поверхностей будут незначительными, и мы можем ими пренебречь (рис. 22.*б*).

Отметим, что понятие “идеализация” имеет в исследовании вполне определённое назначение. Так как мы ищем общее в движении пластинки, которое присуще пластинкам, находящимся в подобном движении, то понятие идеализации означает выделение этого общего в движении любых пластинок. При движении пластинок в торцевом направлении именно это их движение будет общим для них. Лобовое и тыльное сопротивления мы рассмотрели в других задачах, где общим были они сами, и нашли для них решение. Следовательно, понятие идеализации в любых исследованиях означает выделение наглядного общего в исследуемой группе предметов.

Мы здесь делаем подобное примечание потому, что в настоящее время принято понимать идеализацию как искусственный прием для упрощения решения какой-либо задачи, который дает возможность получить её решение с определённой степенью точности и с меньшими физическими затратами для человека. Подобное мнение в отношении идеализации будет неверным.

Продолжим наши исследования. Расположение пластинки относительно направления сосредоточенной силы веса принимаем таким же, как для пластинки в случае её свободного падения, то есть перпендикулярным. Сосредоточенная сила веса всегда направлена к центру Земли. Раньше мы размещали пластинку в определённом положении, затем отпускали её, и она совершала свободное падение. Теперь же мы тоже разместим её в определённом положении, затем в торцевом направлении приложим к ней определённую силу $R$ и заставим её двигаться под действием сразу двух сил: силы собственного веса $G_т$ и приложенной нами силы $R$.

Внешне движение пластинки под действием двух взаимно перпендикулярных сил будет выглядеть следующим образом: если мы в торцевом направлении приложим небольшую силу, то пластинка начнет двигаться в торцевом, или горизонтальном, направлении тоже с небольшой скоростью.

Дополнительно к горизонтальной скорости пластинка получит ещё и вертикальную скорость от действия силы собственного веса.

Затем вернём пластинку в исходное положение и приложим к ней большую силу. Тогда горизонтальная скорость пластинки увеличится, но вертикальная скорость тоже останется.

Так, последовательно увеличивая величину горизонтальной силы, мы, наконец, достигнем такой величины горизонтальной скорости, при которой будет существовать только данная горизонтальная скорость движения пластинки, а вертикальная скорость движения от силы собственного веса будет отсутствовать. Если мы увеличим данную горизонтальную скорость движения пластинки ещё больше, то она будет продолжать движение в этом направлении без вертикальной скорости движения.

Тут мы наблюдаем определённый эффект в движении пластинки, который выражается в том, что при достижении пластинкой какой-то минимально необходимой горизонтальной скорости движения, вертикальная скорость от воздействия силы её собственного веса исчезает. Мы можем здесь сказать, что вертикальная сила веса уравновешивается каким-то способом.

Далее мы должны объяснить этот эффект с помощью законов механики безынертной массы и определить его количественными зависимостями. В то же время мы отказались от сопротивления трения и пренебрегли лобовым и тыльным сопротивлениями торцевых поверхностей пластинки. Получается, что мы, таким образом, исключили силовое взаимодействие пластинки со средой.

Если такую пластинку разогнать до минимально допустимых горизонтальных скоростей, когда исчезает её вертикальная скорость, то она будет сохранять её сколь угодно долго согласно первому закону механики твёрдого тела. Ведь среда в данном случае не будет оказывать тормозящего сопротивления. Но здесь надо правильно понимать инертность твёрдого тела. Действительно, снаряд или камень летят по инерции, когда им сообщают соответствующую скорость горизонтального движения. Самолёт или птица тут же падают, как только исчезает движущая сила. Это говорит о том, что для чисто инерциального горизонтального движения тел требуются бОльшие минимальные скорости, чем для пластинки, совершающей горизонтальное движение в среде. Отсюда следует, что пластинку в её горизонтальном движении поддерживают не инерциальные силы, а иные, которые возникают при её взаимодействии со средой. Поэтому для выявления силового взаимодействия пластинки со средой в её горизонтальном движении мы должны будем применить законы механики безынертной массы. Для этого нам придётся выяснить относительность движения пластинки.

Мы привыкли к тому, что подобные пластинки являются материальными границами различных потоков. В таких случаях границы потока остаются без движения, а жидкость совершает свое движение относительно этих границ. Если мы теперь возьмем, например, материальные границы определённого установившегося потока, поместим их в среду и начнём перемещать их, сохраняя направление и скорость того потока, который совершал движение в этих границах, то в объёме этих границ снова организуется установившийся поток со всеми ему присущими характеристиками. Но с той лишь разницей, что в этом случае жидкость потока будет находиться в состоянии покоя, а её границы будут совершать движение относительно этой неподвижной жидкости.

Конечно, с границами потока подобный эксперимент провести невозможно, например, из-за различия площадей сечения, поскольку мы границы потока можем перемещать с одинаковой скоростью для всех её участков. С отдельными участками подобного потока мы можем провести подобный эксперимент. Поэтому мы нашу очень тонкую пластинку можем считать границей части установившегося потока. Это значит, что пластинка, совершая горизонтальное движение относительно неподвижной жидкости среды, одновременно является границей потока установившегося вида движения. По этой причине она должна будет взаимодействовать со средой в соответствии с теми законами, которые присущи взаимодействию границ потока установившегося вида движения.

В отличие от границы обычного установившегося потока, движущаяся в горизонтальном направлении пластинка является одновременно границей двух одинаковых симметрично расположенных относительно этой пластинки установившихся потоков. Вот эти два расположенных симметрично потока определяют энергетическое и силовое взаимодействие пластинки с окружающей её средой. Зная величину горизонтальной скорости пластинки $W_{т.г}$, мы можем определить величину динамических сил $P_{дин.г}$ относительно установившихся потоков пластинки.

Согласно второму закону механики безынертной массы, эта величина для единицы площади будет равна:

$$P_{дин.г} = \rho W_г^2. \qquad (94)$$

Эта динамическая горизонтальная сила $P_{дин.г}$ для данных потоков является одновременно силой взаимодействия. Это значит, что она образует полную энергию этих относительных установившихся потоков, величина которой будет равна произведению динамических сил давления на объём потока:

$$U_{пол.г} = V_г P_{дин.г} = V_г \rho W_г^2. \qquad (95)$$

Полная энергия горизонтальных потоков, полученная по уравнению (95), является дополнительной, или энергией прироста, по отношению к полной энергии среды, в которой пластинка совершает своё движение.

Характеристики вертикальных потоков пластинки при её свободном падении мы получили выше, то есть мы записали уравнения сил давления (91) и (93), которые при свободном падении пластинки уравновешивают её вес. Это значит, что при горизонтальном движении пластинки мы имеем дело с двумя потоками взаимодействия: горизонтальным относительным потоком и вертикальным потоком взаимодействия с весом пластинки. Отсюда напрашивается вывод, что при горизонтальном движении пластинки с полным отсутствием её вертикального движения силы давления вертикального потока должны будут уравновешиваться силами давления горизонтального установившегося относительного потока.

Количественно мы можем сказать, что только горизонтальное движение пластинки в среде возможно лишь в том случае, когда динамические силы давления горизонтального установившегося относительного потока будут больше динамических сил давления вертикального потока. Используя уравнения сил (92) и (94), мы можем записать количественное силовое неравенство в таком виде:

$$2P_{дин.г} > 2P_{дин.в}, \text{ или}$$

$$2\rho W_г^2 > 2\rho W_в^2. \qquad (96)$$

Неравенством (96) мы записали силовое неравенство для единицы площади пластины. Если мы в это неравенство подставим площадь пластины $F$, то оно будет выражать силовое взаимодействие по всей площади пластины. Цифра 2 перед динамическими силами давления означает, что пластина имеет два горизонтальных потока и два вертикальных потока: лобовой и тыльный.

Затем вы спросите, почему мы записали силовое неравенство по всей площади $F$ пластины, когда для горизонтального потока лобовой и тыльной поверхностями служат торцевые поверхности пластины. Здесь нам придётся вспомнить, что полная энергия горизонтального потока является одновременно полной энергией его нормального потока, а направление этого нормального потока совпадает с направлением вертикального потока.

По этой причине силовое взаимодействие между этими потоками осуществляется по площади $F$ пластины.

Получив неравенство сил давления (96), мы можем отсюда получить минимальную горизонтальную скорость $W_{г.мин}$ для любой конкретной пластинки, при которой начинает осуществляться только горизонтальное движение. Для чего нам сначала придётся определить для этой пластинки скорость её свободного падения $W_{т.в}$. Затем величину этой скорости подставим в неравенство (96) и из этого неравенства получим минимальную горизонтальную скорость $W_{г.мин}$ как чуть бОльшую, чем вертикальная скорость, которая в своем пределе стремится к этой вертикальной скорости, то есть:

$$W_{г.мин} \cong W_{в}.$$

Будем считать, что мы получили необходимые характеристики для горизонтального полёта пластины.

Мы рассматривали движение очень тонкой пластины, когда её плоскость располагалась перпендикулярно направлению сосредоточенной силы её веса. Но пластинка может располагаться под любым углом $\beta$ относительно этой силы. Покажем такую пластинку на рис. 23.

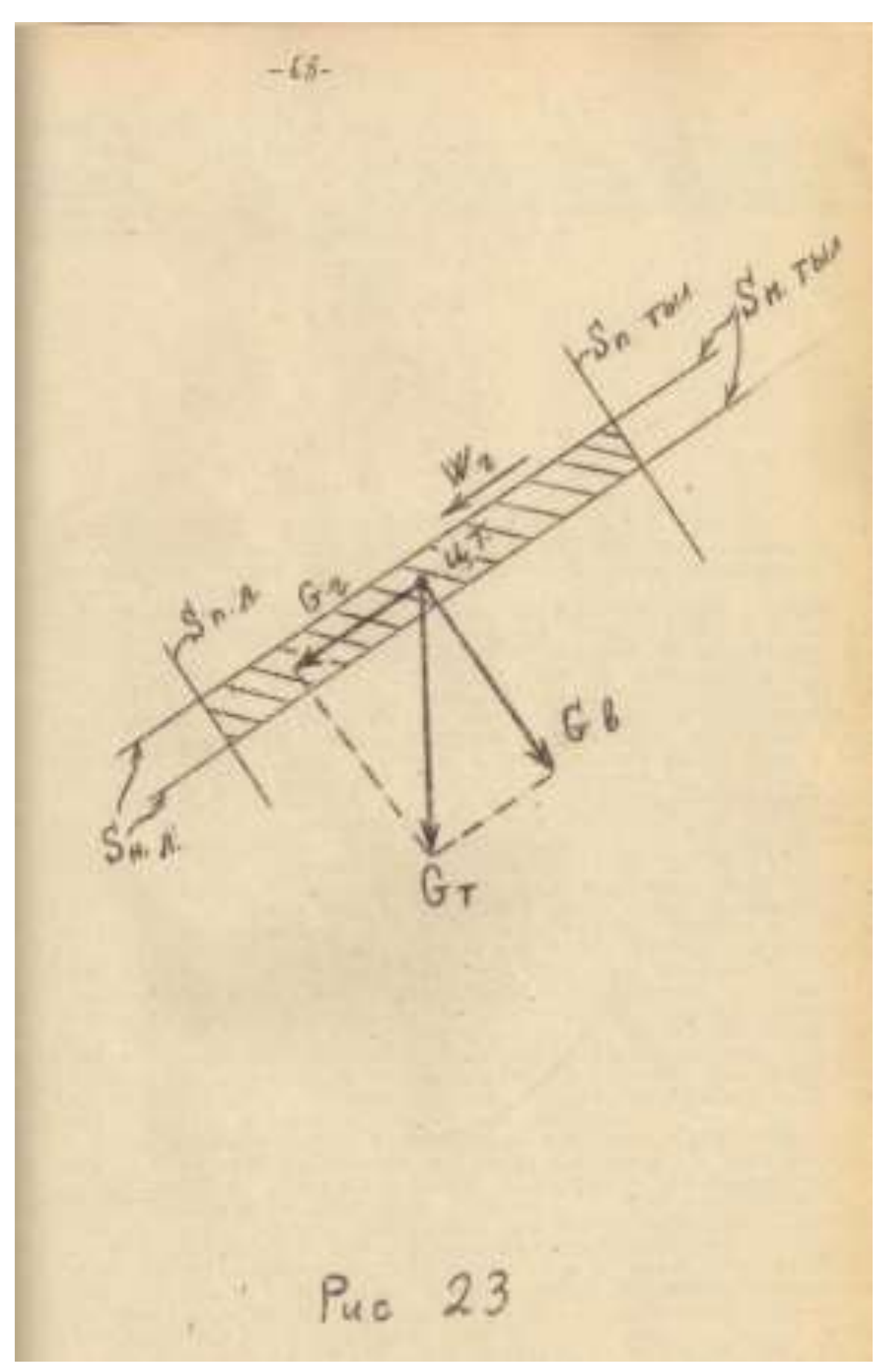

Рис 23

При наклонном положении пластины её сосредоточенный вес $G_т$ не изменяет своего направления. Тогда, согласно законам и положениям механики твёрдого тела, мы эту силу $G_т$ можем разложить на две составляющие, одна из которых $G_г$ будет действовать в направлении торцов пластинки. Назовем её горизонтальной составляющей $G_г$. Другая составляющая силы веса будет действовать перпендикулярно плоскости пластины. Назовем эту составляющую вертикальной составляющей $G_в$ силы веса. Составляющие силы веса говорят нам о том, что при наклонном положении пластинки на неё действуют сразу две неуравновешенные силы в виде взаимно перпендикулярных составляющих сил её веса.

Если горизонтальная составляющая силы веса пластины $G_г$ будет в состоянии обеспечить минимальную скорость $W_{г.мин}$ горизонтального движения, то пластина начнёт двигаться в направлении своих торцов, при этом сохраняя угол своего наклона. Для подобной минимальной скорости горизонтального полёта пластины вертикальную скорость мы будем определять не по полной величине силы тяжести $G_т$, а по её вертикальной составляющей $G_в$. Для этой силы лобовой и тыльной поверхностями сопротивления будут плоскости пластины, перпендикулярно расположенные к этой силе. Следовательно, силовое взаимодействие для вертикальной составляющей силы веса будет определяться зависимостями свободного падения пластины с той лишь разницей, что мы должны будем заменить в них силу веса $G_т$ на её вертикальную составляющую $G_в$.

Начало отсчёта угла $\beta$ мы будем вести от направления сил тяжести. Тогда для горизонтального полёта угол $\beta$ будет максимальным, то есть $\beta = 90°$. При этой величине угла горизонтальная составляющая равна нулю, а вертикальная составляющая равна величине сосредоточенной силы веса пластины $G_т$, то есть

при $\beta = 90°$

$$G_г = 0, \quad G_в = G_т;$$

а при $\beta = 0°$

$$G_г = G_т, \; G_в = 0.$$

Мы определили предельные значения угла $\beta$ и связанное с ним количественное распределение составляющих сил веса пластинки, которое определяет нам, в конечном итоге, движение пластины в наклонном положении.

Мы рассмотрели идеализированное движение пластины в среде. Теперь нам остаётся снова вернуться на бренную землю и вспомнить, что всякие пластины изготавливаются из конкретных материалов. Поэтому реальные пластины всегда имеют конкретную толщину, определённую либо прочностными характеристиками материала, либо какими-то другими. По этой причине подобные пластины имеют сравнительно большую величину сопротивления в своем торцевом направлении, да и сила трения по плоскостям этой пластины будет иметь немалую долю в сопротивлении, которой уже нельзя будет пренебречь.

Движение пластины под углом наклона от действия составляющей силы её веса называют либо планирующим полётом, либо пикирующим. Относительно угла атаки $\beta$ пластинки к планирующим полётам мы будем относить такое её движение, при котором угол $\beta$ имеет предельно большую величину и в своём пределе стремится к величине 90°. К пикирующим полётам мы отнесем такое движение пластинки, при котором угол наклона $\beta$ имеет минимальную величину и в своем пределе стремится к нулю.

Пикирующий полёт в своем пределе (при $\beta = 0°$) означает, что сила веса пластины $G_т$ уравновешивается лобовыми и тыльными поверхностями, которые образуют потоки взаимодействия на её торцевых поверхностях. Это значит, что для пикирующего полёта мы должны будем применить зависимости свободного падения пластины, заменив в них плоскостные поверхности взаимодействия торцевыми. При пикирующем полёте обычно интересуются тем, какую максимальную скорость может развить пластинка в свободном падении.

В планирующем полёте обычно интересуются тем, какое максимальное расстояние она может преодолёть в этом полёте. Максимальное же расстояние полёта пластинки будет находиться в прямой зависимости от величины угла её наклона $\beta$. Чем больше будет этот угол, тем большее расстояние она сможет преодолеть.

Максимальный угол наклона для каждой пластинки определяется её геометрическими характеристиками и весом. Исходя из этих условий, мы постараемся определить количественно максимальный угол наклона каждой конкретной пластины. Это делается в следующей последовательности:

1. определяют общую площадь $F$ пластины и её вес $G_т$;
2. затем определяют величину скорости её свободного падения по зависимостям (91) и (92);
3. в зависимости от величины скорости свободного падения определяют минимальную скорость $W_{г.мин}$ горизонтального полёта по неравенству (96);
4. пользуясь величиной горизонтальной минимальной скорости, определяют величину сил взаимодействия лобовой и тыльной поверхностей в торцевом направлении пластины, так как здесь минимальная горизонтальная скорость будет выражать скорость движения твёрдого тела $W_т$.
5. Проделав эти действия, мы получим минимально необходимую величину силы для обеспечения горизонтального полёта. В планирующем полёте мы должны будем обеспечить её углом наклона пластины $\beta$, который определяет нам горизонтальную составляющую силы веса $G_г$. Поэтому мы должны взять угол наклона таким, чтобы величина горизонтальной составляющей силы веса пластинки равнялась бы величине минимальной силы $R_{т.мин}$, которая необходима для обеспечения горизонтального полёта. Из этого условия мы получим величину угла наклона пластинки $\beta$. Но эта его величина будет не точной, а приблизительной, то есть мы получим величину угла наклона пластинки в первом приближении. В практических расчётах, если найденный приближенный угол наклона по величине будет близок к 90°, то этот приближенный угол можно оставить как расчётный угол наклона.
6. Получив угол наклона пластины $\beta$ в первом приближении, мы должны будем в соответствии с его величиной разложить сосредоточенную силу веса пластины $G_т$ на две её составляющие: горизонтальную $G_г$ и вертикальную $G_в$. Затем, пользуясь величиной вертикальной составляющей, мы должны снова проделать расчёты по пунктам 2, 3, 4 и 5. После чего мы снова получим угол наклона во втором приближении, то есть он тоже будет приближенным. Поэтому мы должны будем снова повторить расчёты по пунктам 2, 3, 4, 5, 6.

Подобные приближенные расчёты мы должны будем проделывать столько раз, пока не добьемся необходимой для нас степени точности значения угла наклона пластины. Найденный таким способом угол наклона пластины будет являться максимальным углом планирования $\beta_{макс}$, при котором каждая пластина может преодолеть максимальное для себя расстояние.

Для каждой конкретной пластинки её максимальный угол планирования будет означать, что если мы сделаем наклон пластинки больше максимального, то горизонтальная составляющая силы веса не обеспечит ей в этом случае планирующего полёта и пластинка перейдет в свободное падение. Под любым углом наклона меньше максимального пластинка будет совершать планирующий полёт, но чем меньше будет угол её наклона, тем больше будет скорость её планирования, тем меньшее расстояние она сможет преодолёть. При предельном уменьшении угла наклона её планирующий полёт перейдет в пикирующий полёт. В этом заключается практическое назначение полёта пластинки под различными углами наклона.

Вернёмся немножко назад и рассмотрим практическое назначение минимальной скорости горизонтального полёта пластины, когда её сосредоточенная сила тяжести расположена перпендикулярно её плоскости.

Знание этой величины дает нам возможность с помощью неравенства (96) определить минимальное значение величин стартовой и посадочной скоростей для каждой конкретной пластинки. Незнание величины минимальной скорости горизонтального полёта привело к гибели многих людей, особенно на заре развития авиации. Незнание стартовой минимальной скорости компенсировалось для них тем, что если они её не достигали, то их летательный аппарат просто не взлетал. При посадке, не зная ограничительного минимума горизонтальной скорости, люди стремились посадить летательный аппарат как можно с меньшей скоростью и в этом стремлении переходили запретный минимум. Как только этот минимум нарушался, летательный аппарат переходил из планирующего полёта в свободное падение, и катастрофа была неизбежной. Поскольку подобные казусы случались на сравнительно небольшой высоте, то подобные явления связывали с влиянием поверхности земли, а не с нарушением ограничения по минимальной горизонтальной скорости.

Неравенство (96) также определяет количественно максимальную высоту полёта, или потолок полёта пластинки в атмосфере Земли. Это происходит потому, что с ростом высоты происходит уменьшение плотности атмосферы. Уменьшение её плотности влечёт за собой увеличение скорости свободного падения пластинки, или вертикальной скорости, что в свою очередь, согласно неравенству (96), приводит к увеличению минимума горизонтальной скорости. Поэтому на максимальной высоте, или потолке, предельно возможная горизонтальная скорость конкретной пластины стремится в своём пределе к величине скорости свободного падения пластины, то есть неравенство (96) будет стремиться к равенству. Отсюда следует, что все предельные ограничения горизонтального полёта, в том числе и подъёмная сила, связаны непосредственно только с величиной площади пластинки и её горизонтальной скоростью.

Реально существующие пластинки имеют различную толщину, которая в их горизонтальном полёте организует силовое взаимодействие со средой через лобовую и тыльную поверхности торцов пластинки. Это значит, что при горизонтальном полёте пластинки её силовая установка затрачивает энергию на преодоление торцевого сопротивления, которое является излишней роскошью в этом полёте, поскольку для обеспечения горизонтального полёта необходимо только разогнать пластинку до соответствующей скорости. Практически силовое взаимодействие торцевых поверхностей пластины определяет для каждой конкретной пластины её максимальную горизонтальную скорость, а эта скорость в свою очередь определяет потолок полёта пластинки в соответствии с неравенством (96).

В настоящее время понятие подъёмной силы пластинки сводят к одному понятию подъёмной силы, которая создается за счёт асимметрии лобовой и тыльной поверхностей в торцевом направлении пластинки. Как вы теперь понимаете, это неверное понимание. Как таковая подъёмная сила, или несущая способность пластинки, количественно и качественно определяется её площадью и горизонтальной скоростью, которые ограничивают пластинку в её горизонтальном полёте по минимальной скорости и высоте полёта. Асимметричность лобовых и тыльных поверхностей в торцевом направлении действительно создает подъёмную силу, как мы определили в своих расчётах выше, но эта сила обеспечивает лишь скороподъёмность пластинки в пределах её полёта, ограничения для которого дает площадь пластинки и её горизонтальная скорость.

Это значит, что подъёмная сила асимметрии профиля не уменьшает величины минимальной горизонтальной скорости и не увеличивает высоты полёта, или потолок полёта пластины. Это положение очевидно, так как мы при определении минимума горизонтальной скорости и максимума высоты полёта использовали предельные энергетические и силовые возможности относительных потоков, которые для лобовых и тыльных асимметричных поверхностей являются взаимодействующими потоками. Поэтому мы должны понимать действие подъёмных сил асимметричных поверхностей взаимодействия как способ обеспечения набора высоты пластинкой при минимальных силовых возможностях обеспечения горизонтального полёта. Ведь при наклонном положении пластинки с набором высоты в её торцевом направлении будут действовать не только силы лобового и тыльного сопротивлений, но и горизонтальная составляющая силы веса пластинки $G_{г}$. Если силовая установка пластинки не обеспечивает равновесие этой суммарной величины, то наклонный полёт пластинки с набором высоты будет для неё невозможен. Если силовая установка пластинки уравновешивает лобовое и тыльное сопротивления асимметричных торцов пластины, то горизонтальный полёт для пластины будет возможен, но в этом случае набор высоты обеспечивается подъёмными силами асимметричных поверхностей взаимодействия. Так определяется силовой минимум для пластинки в её горизонтальном полёте.

На этом мы закончим исследование движения очень тонкой пластинки и перейдём к следующей теме.

## *ЧАСТЬ 2*

## ДВИЖЕНИЕ ТВЁРДЫХ ТЕЛ В СРЕДЕ, ПРИ КОТОРОМ ВНОСИМАЯ ИХ ДВИЖЕНИЕМ УДЕЛЬНАЯ ЭНЕРГИЯ ВОЗМУЩЕНИЯ ПРЕВЫШАЕТ ПО ВЕЛИЧИНЕ УДЕЛЬНУЮ ЭНЕРГИЮ СРЕДЫ

В первой части мы познакомились с движением твёрдых тел в среде и определились с их количественными и качественными характеристиками. Узнали, что силовое взаимодействие твёрдых тел со средой происходит по их лобовым и тыльным поверхностям. Все эти определения и характеристики остаются верными для твёрдых тел, которые в поступательном потоке взаимодействия со средой в зоне лобовой поверхности создают удельную энергию, превышающую удельную энергию среды. Любая среда, какая бы она ни была, в том числе атмосфера Земли и её водные бассейны, обладает определённым запасом энергии. В зависимости от величины объёма среды её энергия может иметь очень большую величину, например, такова энергия земной атмосферы. С такой величиной энергии, как говорится, мы не можем тягаться, а вот по удельной энергии мы можем её превзойти. Такое происходит, когда твёрдое тело движется в атмосфере Земли со сверхзвуковой скоростью. Для водных бассейнов такой показательной и шумной характеристики нет. Ибо в воде превышение удельной энергии среды при движении твёрдых тел происходит на меньших скоростях их движения.

Как мы выше сказали, количественные и качественные характеристики, полученные для движения твёрдого тела в среде, которое создаёт удельную энергию меньшую, чем удельная энергия среды, верны и для данного движения твёрдого тела в среде, в том числе верны понятия обтекаемости и необтекаемости, движения очень тонкой пластинки в среде. В то же время подобное движение имеет определённые качественные и количественные различия, которые должны быть учтены. Для данной темы мы определим эти различия в общем плане и учтём их качественно и количественно.

### *Глава I*

### ДВИЖЕНИЕ ТВЁРДЫХ ТЕЛ, ИМЕЮЩИХ ОБТЕКАЕМЫЕ И НЕОБТЕКАЕМЫЕ ПОВЕРХНОСТИ, В ГАЗОВОЙ СРЕДЕ

Выше мы определили среду как пространство, заполненное идеальной жидкостью. При движении твёрдых тел и жидкости, и газы среды вели себя действительно одинаково – как идеальная жидкость. При данном движении твёрдого тела нам придется учитывать сжимаемость газов в потоках взаимодействия. Это различие является одним из основных различий, которое мы должны учесть для данного вида движения твёрдого тела в среде. Поэтому теперь мы будем иметь дело со средой как с пространством, заполненным идеальным газом.

Здесь понятие идеальности газа определено тем, что идеальный газ должен подчиняться термодинамическим процессам, типа адиабатического, при своём сжатии и расширении. Для лучшего понимания в качестве вещественного ощущения подобной среды мы должны иметь в виду атмосферу Земли.

Поскольку повышение удельной энергии происходит в потоках взаимодействия по лобовой поверхности твёрдого тела, то мы начнём свое исследование именно с этих потоков и их поверхностей взаимодействия. Далее принимаем твёрдое тело типа “крыло”, которое имеет сколь угодно большую длину. Лобовую поверхность взаимодействия принимаем необтекаемой, то есть плоскостью, расположенной перпендикулярно направлению движения тела. Также мы знаем, что в потоке взаимодействия прирост энергии по величине большей, чем величина удельной энергии среды, создается в том случае, если тело движется со скоростью большей, чем скорость звука в данной среде. Поэтому считаем, что твёрдое тело движется в газовой среде со скоростью $W_т$ большей, чем скорость звука $C$ в этой среде. Покажем движущееся твёрдое тело с его потоками взаимодействия на рис. 24.

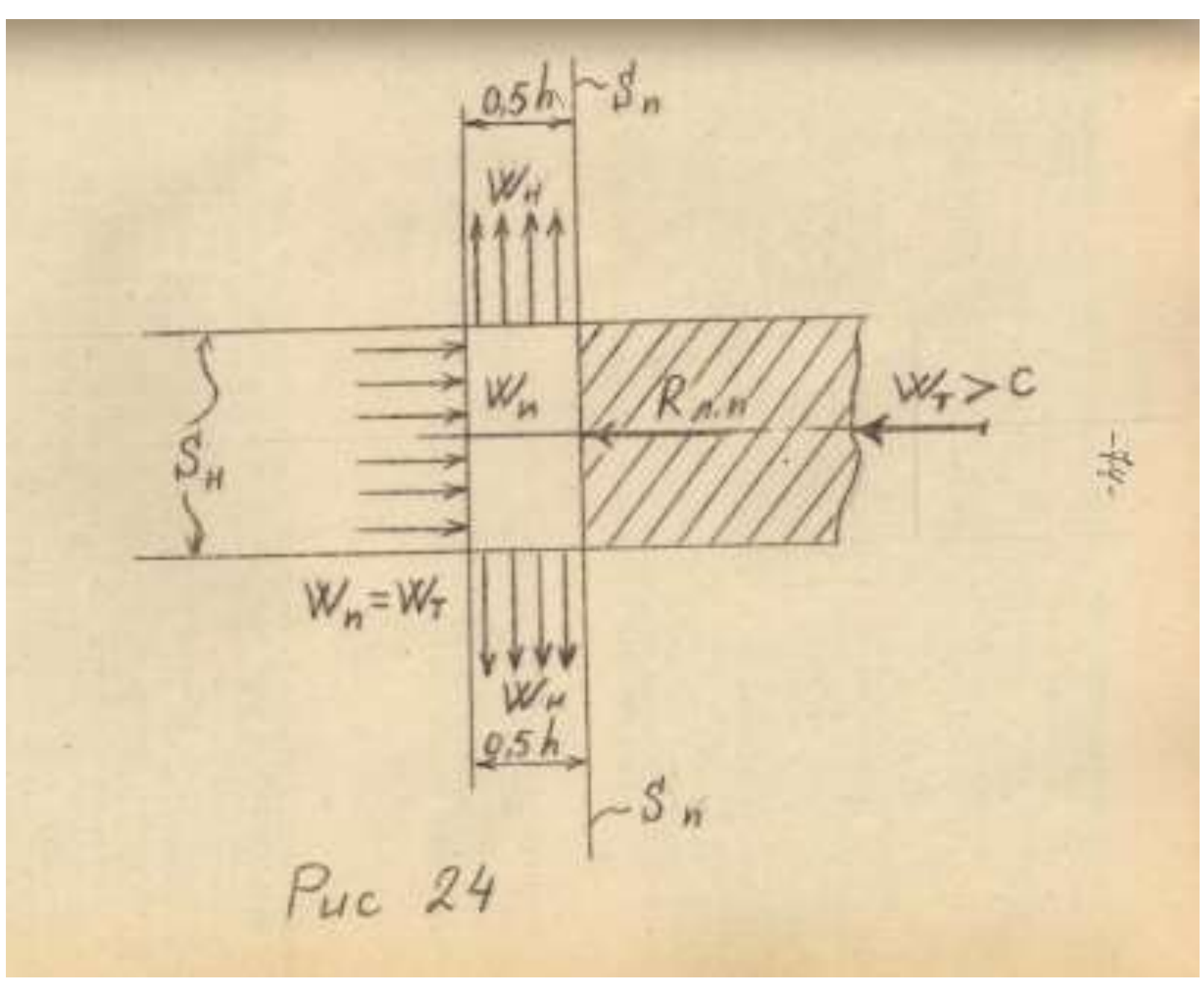

Рис 24

Здесь мы видим обычную и даже привычную для нас картину: необтекаемую лобовую поверхность, нормальные $S_н$ и поступательную $S_п$ плоскости исследования, а также нормальный и поступательный потоки взаимодействия, то есть принцип взаимно перпендикулярного движения здесь сохраняется.

Начнём с поступательного потока. Коль мы сказали, что ранее полученные выкладки верны и для этого потока, то запишем их для него.

Для поступательного потока принимаем площадь сечения $f_п$, равную площади лобовой плоскости $s_л$, расположенной на поступательной плоскости исследования $S_п$. Тогда ***уравнение движения*** для этого потока запишется в таком виде:

$$M_п = s_л\, \rho W_т.$$

***уравнение сил*** будет иметь вид:

$$R_л = s_л\, \rho W_т^2.$$

***уравнение энергии*** запишем в таком виде:

$$U_л = V_п P_{дин} = V_п \rho W_т^2.$$

Здесь уравнение энергии тоже выражает полный переход энергии движения твёрдого тела в полную энергию поступательного потока.

Все эти уравнения правильные, но не полные, так как они не учитывают сжимаемости газа. Ведь тело движется со скоростью большей, чем скорость звука. Разберёмся, что это означает.

Скорость звука в газовой среде $C$ означает скорость распространения возмущения. Она является физической величиной и зависит от давления и температуры. Кроме того, для каждого конкретного газа она является определённой величиной. В связи с тем, что наше тело движется со скоростью большей скорости звука, то возмущения, вносимые лобовой поверхностью тела, не должны распространяться вперёд по движению. Это значит, что поступательный поток должен быть плоским, а не объёмным, то есть он должен располагаться по лобовой плоскости, а перед плоскостью будет находиться невозмущенная среда, чего в действительности не происходит.

Поступательный поток имеет объём, так как тело движется и вытесняет вполне определённый объём газа, который мы записали выше количественными уравнениями движения, сил и энергии. Поэтому мы должны будем рассмотреть, каким образом образуется поступательный поток. При скорости тела $W_т > C$ в среде, где газ может развивать максимальную скорость, равную скорости звука[12], мы должны были записать уравнения движения и сил в таком виде:

уравнение движения: $M_C = s_л \rho C$ ;

уравнение сил: $R_{л.C} = s_л \rho C^2$ .

Эти зависимости выражают характеристики потока для движущегося тела, скорость которого равна скорости звука, то есть $W_т = C$.

Вкличины этих уравнений будут меньше, чем вышезаписанных для полной скорости тела $W_т$, поскольку площади потоков остаются неизменными и плотность тоже одинакова Она равна плотности невозмущённой среды. Коль твёрдое тело движется с постоянной скоростью и через равные промежутки времени проходит равные расстояния в среде, то единственным правильным уравнением остаётся уравнение движения, записанное нами для скорости движения $W_т$ твёрдого тела.

В то же время, мы знаем, что при движении твёрдого тела в газе среды со сверхзвуковой скоростью происходит разогрев газа в поступательном потоке. По этой причине в настоящее время бытует мнение, что в поступательном потоке газ сжимается в соответствии с адиабатическим процессом, в результате которого происходит разогрев газов. Данное явление называют даже скачком уплотнения. Проверим это общепринятое мнение.

При адиабатическом сжатии газа происходит не только разогрев, но и сжатие газа, что выражается в увеличении его плотности. Допустим, что в нашем случае тоже произошло адиабатическое сжатие газа в поступательном потоке и плотность его увеличилась $\rho_п$, то есть стала больше, чем плотность газа в невозмущенной среде: $\rho_п > \rho$. Тогда уравнение движения для поступательного потока мы должны будем записать в таком виде:

$$M_{п1} = s_л \rho_п W_т .$$

Отсюда мы видим, что в отличие от нашего правильного уравнения движения в данном уравнении увеличилась плотность газа, а другие характеристики должны при этом оставаться неизменными. Чего быть не может, поскольку расход массы в единицу

[12] см. [2] – *ред.*

времени $M_{п1}$ для данного потока будет больше расхода массы $M_{п}$ нашего правильного уравнения движения, то есть $M_{п1} > M_{п}$ , так как $\rho_{п} > \rho$.

В связи с тем, что из невозмущенной среды в поступательный поток расход массы может поступать в соответствии с количественными величинами скорости движения твёрдого тела $W_{т}$ и плотности газа невозмущенной среды $\rho$, то отсюда следует, что увеличение расхода массы за счёт увеличения плотности газа в поступательном потоке невозможно потому, что там невозможен больший расход массы в единицу времени, чем тот расход массы, который пополняет этот поток со стороны невозмущенной среды. Отсюда следует, что плотность газа в поступательном потоке не изменяется, с какой бы скоростью не перемещалось твёрдое тело. Плотность остается постоянной и равной плотности газа невозмущенной среды. Что уже указывает на невозможность адиабатического сжатия газа в поступательном потоке.

В то же время мы знаем, что в поступательном потоке происходит разогрев газа, хотя плотность его не изменяется. С термодинамической точки зрения, подобное явление может происходить в качестве изохорного процесса, то есть когда в замкнутом объёме постоянное количество газа нагревают или охлаждают. При этом давление и температура газа изменяются, а его плотность остаётся неизменной. Следовательно, мы выяснили, что разогрев газа в поступательном потоке проходит в качестве изохорного процесса. С этой точки зрения мы должны будем уточнить количественные зависимости для поступательного потока. Для этого мы должны хорошо представлять себе взаимодействие поступательного потока с лобовой плоскостью, которое осуществляется по изохорному процессу.

При сверхзвуковом движении твёрдого тела поступление массы газа через поступательный поток остается неизменным, плотность которого тоже не изменяется. Поэтому уравнение движения поступательного потока взаимодействия с необтекаемой лобовой плоскостью остается неизменным для любой скорости движения твёрдого тела.

Уравнение энергии должно измениться, поскольку в объёме поступательного потока происходит разогрев газа. Что указывает на прирост тепловой энергии. Коль эта энергия не возникает из ничего, то это означает, что часть энергии движения твёрдого тела переходит в тепловую. Обозначим эту часть тепловой энергии для единицы объёма как $P$ $(1/м^3)_{теп}$.

Отметим, что мы этой же буквой $P$ обозначаем полные и статические давления. В то же время нам приходится обозначать этой буквой удельную энергию жидкостей и газов, так как величины их совпадают. Чтобы в дальнейшем не путаться в обозначениях сил давления и удельной энергии, мы к обозначению удельной энергии будем добавлять в скобках такое обозначение: $(1/м^3)$. Тогда $P(1/м^3)$ будет означать удельную энергию, а $P$ – силы давления.

Вернёмся к прерванному тексту. Мы обозначили тепловую часть энергии поступательного потока. Чтобы получить полное уравнение энергии для поступательного потока, мы должны будем добавить эту часть энергии к уже имеющемуся уравнению энергии. Тогда получим:

$$U_{л.п} = V_{п}\rho W_{т}^2 + V_{п}P\ (1/м^3)_{теп}. \qquad (97)$$

Уравнение (97) будет уравнением энергии поступательного потока взаимодействия при движении твёрдого тела в газовой среде со сверхзвуковой скоростью, или с возбуждаемой удельной энергией больше, чем удельная энергия среды.

Тепловая энергия является физической величиной, которая изучается другими науками. Поэтому без помощи этих других наук мы не сможем представить себе полную картину перехода энергии движения твёрдого тела в тепловую. Поэтому опишем её только с точки зрения механики безынертной массы.

При движении твёрдого тела в газовой среде со сверхзвуковой скоростью в поступательном потоке взаимодействия возникает удельная энергия, превышающая удельную энергию среды. Так как скорость тела превышает скорость звука в газовой среде, то поступательный поток не может организоваться. Ведь скорость звука является физической величиной, от которой зависит организация этого потока.

Поэтому для организации потока в данных условиях движения необходима другая скорость звука, которая бы превосходила скорость движения твёрдого тела. Эта новая скорость звука для газа может существовать в поступательном потоке за счёт дополнительного поступления тепловой энергии, которая образуется за счёт энергии движения твёрдого тела. В результате чего в поступательном потоке увеличивается не только температура, но и давление. В общем, изменяются те физические характеристики, которые в свою очередь изменяют скорость звука в поступательном потоке взаимодействия. В одностороннем порядке мы можем дать только такое объяснение для теплового взаимодействия. Дальше нам остается ожидать, когда другие науки дополнят объяснение этого теплового взаимодействия.

Даже не зная полной качественной картины перехода энергии движения твёрдого тела в тепловую, мы всё же можем определить её количественно по уравнению энергии (97). Но одного этого уравнения не достаточно, поэтому нам придется получить ещё ряд необходимых зависимостей. На данный момент мы имеем уравнение движения и уравнение энергии поступательного потока взаимодействия. В существующем виде уравнение сил поступательного потока определяется скоростью взаимодействия, или скоростью движения твёрдого тела. При сверхзвуковом движении твёрдого тела эта количественная величина сил давления сохраняется, так как скорость и плотность поступательного потока остаются неизменными. К этой величине мы должны будем добавить ещё силы давления, которые затрачиваются на прирост тепловой энергии в поступательном потоке. Обозначим эти силы давления как $P_{теп}$. Тогда уравнение сил будет иметь такой вид:

$$R_л = s_л \rho W_т^2 + s_л P_{теп}. \qquad (98)$$

Уравнение (98) дает нам величину сил лобового сопротивления $R_л$, которая создается суммой динамических ($s_л \rho W_т^2$) и тепловых сил давления ($s_л P_{теп}$).

С точки зрения механики безынертной массы мы получили все необходимые зависимости для поступательного потока взаимодействия. Но в этих уравнениях появились силы и энергии, связанные с разогревом поступательного потока, которые определяются физическими величинами. Связь между физическими величинами газа выражается термодинамическими зависимостями. В нашем случае в потоке взаимодействия реализуется изохорный процесс. Например, для изохорного процесса мы можем записать связь между удельным объёмом $v$ и давлением $P$ с помощью зависимостей политропного процесса как:

$$Pv^n = P_1 v_1^n = \text{const}. \qquad (99)$$

Для изохорного процесса показатель политропы $n$ будет равен $n = \pm\infty$. В общем, с помощью изохорного процесса термодинамики мы сможем определить связь между температурой и давлением газа поступательного потока взаимодействия. Но эту связь мы можем определить относительно неподвижных, или статических, условий газового объёма. Это значит, что количественных зависимостей изохорного процесса будет ещё недостаточно для полной количественной характеристики поступательного потока взаимодействия. Ведь мы ещё не имеем количественной зависимости, связывающей скорость полёта твёрдого тела с температурой газа поступательного потока и связь этой скорости с силой, затрачиваемой на тепловое преобразование.

Поскольку эта связь ещё не определена законами природы, то в этом случае пользуются экспериментальными данными. Для чего будет необходимо замерить температуру поступательного потока при различных скоростях движения твёрдого тела для каждого конкретного газа. Затем полученные результаты свести в таблицу или представить в виде графика.

Аналогичным способом необходимо произвести замеры сил давления, которые переходят или создают соответствующие температуры поступательного потока, то есть произвести замеры сил давления при различных скоростях движения твёрдого тела. Результаты тоже можно представить либо в виде табличных данных, либо в виде графиков. Тогда мы будем иметь экспериментальную зависимость температуры и сил давления от скорости движения твёрдого тела. С помощью этих зависимостей, уравнений движения, сил, энергии поступательного потока взаимодействия и зависимостей изохорного процесса мы сможем определить все необходимые характеристики любого поступательного потока взаимодействия для различных сверхзвуковых скоростей движения твёрдого тела. В настоящее время мы можем получить величину сил лобового сопротивления только таким способом.

Для нормального потока мы определим характеристики, исходя из уравнения энергии. Энергия движения твёрдого тела преобразуется в полную энергию поступательного потока. Величину этой полной энергии составляет энергия движения и тепловая энергия. Полная энергия поступательного потока у нас записана уравнением (97). Она одновременно будет являться полной энергией нормального потока, то есть:

$$U_{\text{л.п}} = U_{\text{л.н}}.$$

Запишем уравнение (97) в виде удельной энергии как:

$$U_{\text{л.н}} = \rho W_{\text{т}}^2 + P(1/\text{м}^3)_{\text{теп.}}. \qquad (100)$$

Для нормального потока его скорости $W_{\text{н}}$ будут ограничиваться только величиной полной энергии, то есть предельным случаем, когда потенциальная энергия равна кинетической и равна половине полной энергии. Из этого ограничения мы получим скорость нормального потока как:

$$\frac{1}{2}U_{\text{л.н}} = \frac{1}{2}(\rho W_{\text{т}}^2 + P(1/\text{м}^3)_{\text{теп}}) = \frac{1}{2}\rho W_{\text{н}}^2. \qquad (101)$$

Из уравнения (101) мы видим, что нормальная скорость $W_{\text{н}}$ по величине будет больше скорости поступательного потока, то есть больше скорости движения твёрдого тела $W_{\text{т}}$:

$$W_{\text{н}} > W_{\text{п}} = W_{\text{т}}.$$

Это произошло потому, что в данном случае полная энергия стала больше на величину тепловой энергии.

Затем по уравнению движения нормального потока мы определим величину его площади сечения, а вернее, величину $h$, которая является одной стороной площади сечения, а вторую её сторону составляет единица длины, согласно нашим условиям. Запишем уравнение движения нормального потока:

$$M_{\text{н}} = h\rho W_{\text{н}}.$$

В этом уравнении все характеристики нам известны, кроме $h$. Нормальный расход массы $M_{\text{н}}$ будет равен поступательному расходу массы $M_{\text{п}}$, величина которого нам известна. Плотность по условию движения равна плотности невозмущенной среды, а нормальную скорость $W_{\text{н}}$ мы получим из уравнения энергии (101). Здесь мы можем отметить, что величина $h$ будет меньше толщины твёрдого тела $\delta$, так как $W_н > W_п$. Раньше мы получили, что $h$ равнялась $\delta$. В данном случае этого не происходит.

Получив величину $h$, мы тем самым определили не только площадь сечения нормального потока, но и длину объёма поступательного потока. Вот таким способом получаются характеристики поступательного и нормального потоков взаимодействия лобовой необтекаемой плоскости при движении твёрдого тела в газовой среде со сверхзвуковой скоростью.

Здесь мы определили не только характеристики потоков взаимодействия, но и установили, что газы ведут себя в данном случае как несжимаемые жидкости. Состояние газа в поступательном потоке определяется зависимостями изохорного процесса.

С помощью потока установившегося вида движения мы можем практически определить силу лобового сопротивления необтекаемой поверхности. Для этого лобовую плоскость нам придётся разместить, как показано на рис. 25, и организовать установившийся поток в соответствии с изохорным процессом. В изохорической камере для газов организуются условия, соответствующие изохорному процессу, то есть плотность газа $\rho$ в этой камере должна равняться плотности газа среды, в которой имитируется движение. Это значит, что давление в этой камере повышается за счёт температуры, то есть подогрева. После подогрева давление в изохорической камере должно равняться давлению окружающей среды $P_{ср}$ плюс давление поступательного потока $P_п$, которое должно соответствовать полётным условиям. Газ в изохорической камере должен подогреваться от температуры среды до полётной температуры. Далее из изохорической камеры установившийся поток газа направляется на лобовую поверхность, как показано на рис. 25.

Площадь сечения этого потока равна площади лобовой плоскости. Скорость движения газа в установившемся потоке $W_п$ должна равняться скорости нормального потока $W_н$. Как мы знаем, скорость нормального потока $W_н$ больше скорости движения твёрдого тела $W_т$ и скорости поступательного потока.

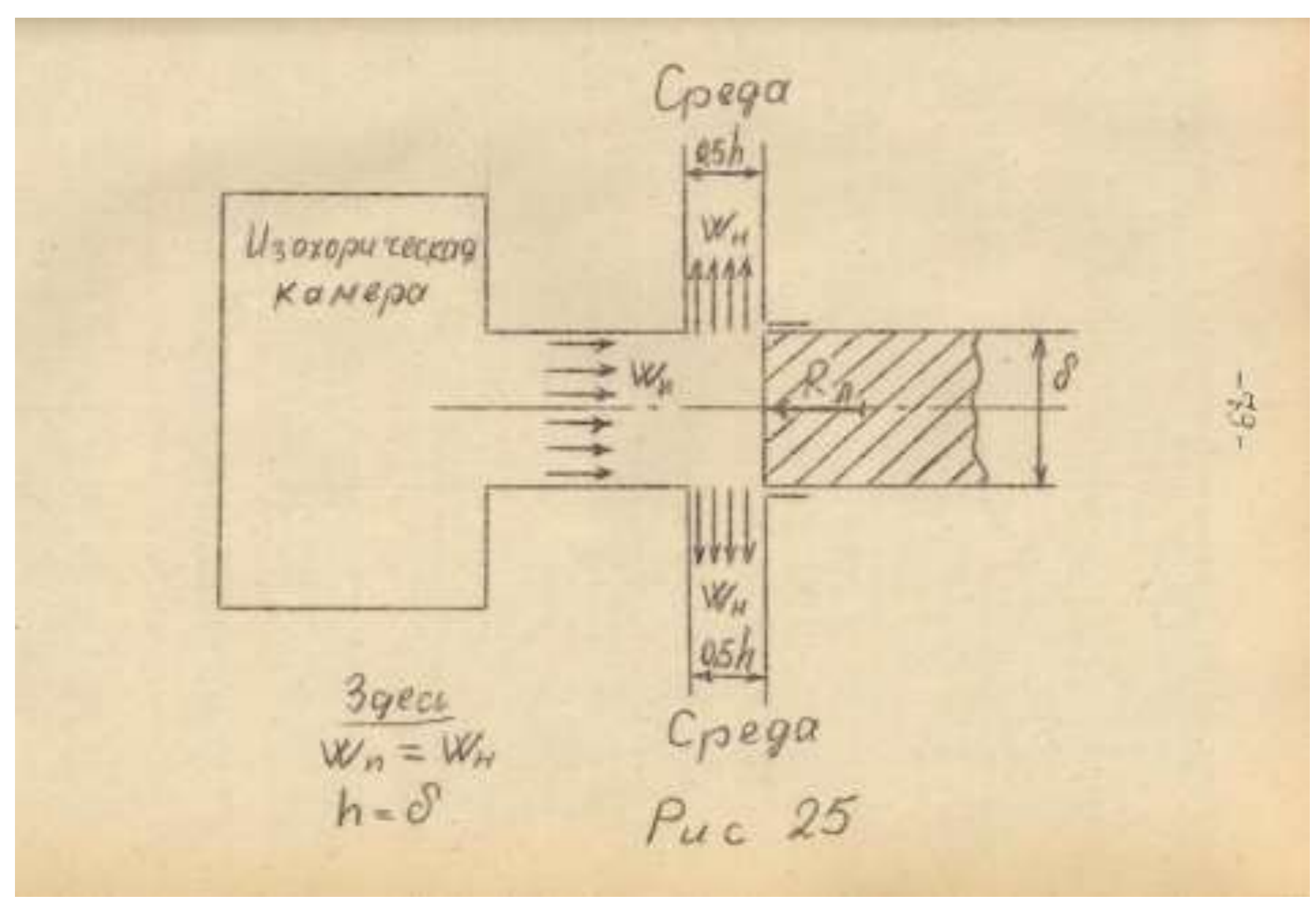

Рис 25

Эти условия влекут за собой увеличение расхода массы в единицу времени в установившемся изохорном потоке, поскольку мы в этом потоке увеличили скорость движения газа от величины скорости движения твёрдого тела до величины скорости нормального потока. Плотность газа при этом остается неизменной. В то же время площадь установившегося потока будет больше действительной площади сечения нормального потока. В действительном потоке $h<\delta$, а в имитируемом потоке $h = \delta$. Сохранив все вышеизложенные условия для установившегося потока и сделав правильное расположение лобовой плоскости относительно этого потока, при обдувке этим потоком необтекаемой лобовой плоскости мы получим истинную величину лобового сопротивления $R_л$, которая по величине соответствует силе лобового сопротивления в полётных условиях твёрдого тела.

Мы договорились рассматривать движение твёрдого тела в среде с возбуждаемой удельной энергией больше удельной энергии среды в общем плане. Поэтому будем придерживаться этого общего плана. Мы рассмотрели взаимодействие необтекаемой лобовой поверхности твёрдого тела, и, как мы знаем, необтекаемой бывает только одна-единственная форма поверхности. Все остальные формы лобовой поверхности обладают различной степенью обтекаемости, которая определяется величиной сил лобового сопротивления относительно лобового сопротивления необтекаемой поверхности.

В данном случае - в простейшей форме - обтекаемые лобовые поверхности определяются наклонной плоскостью с различным углом наклона $\alpha$ . Предельный угол наклона $\alpha_{пр}$ этих поверхностей тоже определяется ограничениями уравнения энергии с учётом изохорного состояния газа в поступательном потоке. Каким образом учитывается изохорное состояние газа в поступательном потоке, мы дали на примере необтекаемой лобовой поверхности.

Для тыльных необтекаемых и обтекаемых поверхностей предельным уменьшением энергии в поступательном потоке является полная энергия среды, в которой движется твёрдое тело. По этой причине сверхзвуковое движение будет характеризоваться для тыльных поверхностей величиной силы тыльного сопротивления и ростом объёма поступательного потока.

Для твёрдых тел конечных размеров, типа “корпус”, все полученные нами расчёты будут верны, конечно, с учётом их характерных размеров. Движение очень тонкой пластинки будет определяться теми же условиями.

***

Нам остается отметить особенности движения твёрдого тела, с возбуждаемой этим движением удельной энергией большей удельной энергии среды, при его движении в жидкостной среде.

Эти особенности движения твёрдого тела в жидкостной среде мы будем отмечать относительно водных бассейнов нашей планеты. Водная среда бассейнов Земли обладает широким диапазоном энергетических уровней. У поверхности бассейнов их энергетический уровень почти одинаков с энергетическим уровнем атмосферы, а в глубинах энергия может достигать величин порядка десятков и сотен атмосфер. Скорость распространения звука в воде больше скорости звука в воздухе атмосферы. С учётом этих особенностей мы в общем плане подчеркнём отличие взаимодействия твёрдых тел с жидкостной средой от взаимодействия при их движении в газовой среде. Это различие мы рассмотрим для потока взаимодействия необтекаемой лобовой поверхности твёрдого тела.

Теперь предположим, что твёрдое тело движется на небольшой глубине, близко к поверхности водного бассейна. В этом случае энергетический уровень жидкостной среды будет почти равен энергетическому уровню газовой среды. Тогда при скорости движения твёрдого тела порядка 10м/сек или 40 км/час твёрдое тело будет возбуждать энергию большую, чем энергия среды. По сравнению с величиной скорости звука в воде – это очень маленькая величина скорости. В этом случае величина удельной энергии поступательного потока будет превышать удельную энергию среды. Скорость распространения возмущения, или скорость звука, имеет намного большую величину по отношению к скорости движения твёрдого тела.

Повышенная энергия поступательного потока и повышенная скорость распространения возмущения должны были бы привести к тому, что поступательный поток должен был бы непрерывно увеличивать свой объём в направлении движения. Подобное увеличение должно было бы происходить за счёт увеличения массового расхода в поступательном потоке и его уменьшения в нормальном потоке. Согласно уравнениям движения, эти расходы должны быть одинаковыми. Это значит, что увеличение объёма поступательного потока невозможно. В действительности соблюдается одинаковость расходов. Одновременно происходит прирост температуры в объёме поступательного потока. Прирост тепловой энергии в поступательном потоке свидетельствует о том, что жидкость этого потока находится в состоянии изохорического процесса. Правда, в термодинамике для жидкостей не дано никаких процессов. В то же время практика показывает, что в объёме поступательного потока реализуется изохорический процесс. Поэтому мы будем считать, что для жидкостей существует изохорический процесс.

В газовом поступательном потоке его объём как бы проталкивается вперед по ходу движения с помощью изохорного процесса, который дает возможность увеличить скорость звука в объёме поступательного потока. По этой причине он требует дополнительных сил сопротивления. В жидкостях действие изохорного процесса является обратным по отношению к объёму поступательного потока. Он как бы удерживает этот объём от его распространения вперед по ходу движения, то есть с точки зрения механического движения его действие является противоположным действием по отношению к изохорному процессу в поступательном газовом потоке. По этой причине изохорный процесс поступательного потока жидкости должен был бы снижать величину сил лобового сопротивления. В практике такой случай не был отмечен. В то же время в газовом поступательном потоке увеличение температур наблюдается ещё при приближении скорости твёрдого тела к скорости звука, но дополнительного прироста сил сопротивления не происходит. В общем, будем считать, что подобные физические явления требуют дополнительного исследования в таких науках, как термодинамика и физика, так как с помощью одних только законов механики безынертной массы решить эту проблему невозможно. Но мы будем считать, что силовое взаимодействие в поступательном потоке организуется только по законам механики безынертной массы. Зависимость температуры в поступательном потоке от скорости движения твёрдого тела мы можем получить только опытным путем и представить её в виде таблицы или графиков.

При дальнейшем увеличении скорости движения твёрдого тела на глубине, близкой к поверхности водного бассейна, происходит непрерывный рост температур в поступательном потоке. При скорости порядка 150 км/час температура в объёме поступательного потока достигает температуры кипения воды. Начинает реализовываться процесс кипения. Процесс кипения жидкости в объёме поступательного потока называют кавитацией. Ибо образование паровых пузырьков происходит под большим давлением, и по этой причине они вызывают эрозию поверхности взаимодействия твёрдого тела.

При дальнейшем увеличении скорости движения твёрдого тела жидкость в объёме поступательного потока будет переходить из жидкого состояния сразу в парообразное, вернее, газообразное. Как движутся твёрдые тела в газах, мы рассмотрели выше. На скорости звука происходит «скачок уплотнения», то есть к механическим силам сопротивления добавляются термодинамические силы сопротивления. По этой причине силы сопротивления вырастают скачкообразно. Для газов существует одна замечательная точка, а для жидкостей их – три.

О физическом или термодинамическом приросте сил сопротивления при подобных скоростях движения твёрдого тела мы можем судить относительно, то есть по отдельным практическим примерам. Приведем один из них. Несколько лет тому назад в Англии на одном из озёр была предпринята попытка установить абсолютный рекорд скорости на воде. Она осуществлялась на лодке с реактивным двигателем. Результат этой попытки был трагическим. На скорости 800 км/час лодка взмыла вверх, описала мертвую петлю и скрылась в водах озера. В журналах были приведены фотографии этого печального конца. Всех удивило такое поведение лодки, но причину до сих пор не выяснили.

Мы можем сказать, что лодка побывала в трех замечательных точках жидкости. На скорости 40 км/час, 150 км/час и выше происходит либо плавный прирост термодинамического сопротивления, либо этого прироста не происходит, но разогрев жидкости существует. На скорости 800 км/час произошел резкий скачкообразный прирост термодинамических сил сопротивления, и лодку подбросил вверх нормальный поток газа, то есть лодка при скорости 800 км/час достигла скорости звука в водном газе. Это значит, что трагически погибший англичанин был первым человеком, который штурмовал звуковой барьер водного газа.

Одним из интересных созданий природы является рыба-меч, которая развивает в воде скорость до 130-140 км/час. По скорости она почти достигает второй замечательной точки. Это значит, что поступательный поток лобового взаимодействия имеет температуру порядка 80°- 90°. Вот такую температуру приходится выдерживать рыбе - меч в её скоростном плаванье. Да и форма её лобовой поверхности более совершенна по отношению к взаимодействию поступательного потока. Так что, рыбу-меч можно сделать эмблемой покорителей скоростных замечательных точек жидкостей и газов.

С другой стороны, меч этой рыбы не только придает её лобовой поверхности идеальную обтекаемую форму, но он является в прямом смысле мечём. Эта рыба питается мелкими морскими организмами. Коль природа всё делает целесообразно, то, надо полагать, что, проносясь через стаю таких мелких организмов, она убивает их, а затем спокойно поедает их.

Мы для себя должны здесь отметить, что скорость движения твёрдого тела более 100 км/час в водных бассейнах Земли гибельно сказывается на мелких морских животных, которые обитают в теплых поверхностных водах. Поэтому большое количество кораблей, имеющих скорость более 100 км/час, в короткое время могут превратить реки, моря и даже океаны в безжизненные пустыни. Об этом надо хорошо помнить при создании и использовании быстроходной плавающей техники.

Мы разобрали все необходимые моменты движения твёрдого тела в жидкостях и газах и выяснили, что это движение может существовать лишь под действием сил, которые необходимы для преодоления лобовых и тыльных сил сопротивления среды. Коль твёрдые тела движутся в пространстве среды, то они должны иметь автономные способы для создания толкающих сил. Обычно на твёрдых телах мускульная сила или сила двигателей преобразуется в толкающую силу с помощью движителей. Эта толкающая сила уравновешивает силы лобового и тыльного сопротивления и тем самым создает соответствующую скорость движения твёрдого тела. В данной работе мы не будем рассматривать двигатели, а вот несколько типов движителей мы рассмотрим такие, как винт и крыло.

В настоящее время движитель, преобразующий вращательное движение двигателя в толкающую силу поступательного движения твёрдого тела, называют винтом, поскольку принцип преобразования сил этим движителем основан на принципе действия винта. Мы же рассмотрим лопастной движитель, принцип работы которого будет основан на законах механики безынертной массы. Чтобы отличить подобный движитель от винта, назовем этот лопастной движитель “ТОЛИК”[13]. Мы сделали различие по принципу работы этих двух лопастных движителей, а по назначению они могут служить и вентиляторами и осевыми компрессорами, и авиационными или корабельными винтами.

[13] назван в честь сына (прим.ред.)

## *ЧАСТЬ 3*

### *Глава I*
### ПРИНЦИП РАБОТЫ ДВИЖИТЕЛЯ “ТОЛИК”

Мы разберём принцип работы этого движителя относительно его назначения как корабельного винта. Такое назначение нам необходимо лишь для последовательности изложения, а сам принцип работы лопастного движителя “ТОЛИК” будет общим, так как он реализуется при любом назначении этого движителя.

При проектировании корабля, прежде всего, задается его грузоподъёмность и скорость. После чего в работу вступают проектировщики. По заданной грузоподъёмности они выбирают объём корпуса корабля в соответствии с законом Архимеда. После выбора объёма корпуса они начинают профилировать его носовую и кормовую поверхности как лобовую и тыльную поверхности твёрдого тела. Придавая этим поверхностям всё более и более обтекаемую форму, они тем самым добиваются минимальной величины силы, необходимой для обеспечения заданной скорости корабля.

Когда сделан выбор по обтекаемости формы носовой и кормовой поверхности корпуса корабля, тогда в соответствии с заданной скоростью движения корабля $W_т$ и профилем носовой и кормовой поверхностей вычисляют величину сил сопротивления, или величину силы $R_\Sigma$. Проектировщик должен будет в соответствии с этой силой выбрать силовую установку, например, типа двигателя внутреннего сгорания. Серийные двигатели различаются по величине мощности и числу оборотов. В соответствии с мощностью каждый двигатель развивает определённую тягу. Величина силы тяги должна будет равняться величине силы $R_\Sigma$. В этом случае выбор двигателя будет сделан правильно. Подбор двигателя выглядит просто, но, в то же время, это является проблемой из-за того, что не знают, как перевести мощность в величину силы тяги.

С точки зрения механики безынертной массы, это делается следующим образом. Мы получили величину силы $R_\Sigma$. Согласно второму закону механики безынертной массы, её можно выразить через характеристики потока в таком виде:

$$R_\Sigma = FP_{дин} = F_д \rho W_д^2. \qquad (1)$$

В уравнении (1) нам не известны ни скорость, ни площадь. Здесь площадь $F_д$ выражает площадь сечения установившегося потока лопастного движителя. Так как движители тоже создают в среде избыточную энергию, избыточное давление, мы поступаем следующим образом. В нашем примере мы будем рассматривать тот случай, когда величина удельной избыточной энергии движителя не будет превышать удельную энергию среды. Исходя из этого условия и величины сил давления среды, мы можем задаться скоростью установившегося потока движителя как:

$$P_{ср} > \rho W_д^2, \qquad (2)$$

где $P_{ср}$ – статическое давление среды. У поверхности воды оно имеет величину порядка 1 кг/см$^2$.

Поэтому, следуя неравенству (2), мы можем задаться для этих условий величиной скорости $W_д$ и по уравнению (1) определить площадь сечения потока движителя $F_д$, так как величина силы $R_\Sigma$ нам известна и плотность жидкости нам тоже известна.

Мощность $N$ определяется величиной работы в единицу времени. Работа для жидкостей определяется как произведение объёма на давление, то есть

$$L = VP_д. \qquad (3)$$

Тогда мощность будет равна работе $L$, делённой на время $t$, то есть

$$N = \frac{L}{t} = \frac{V}{t} P_д. \quad (4)$$

Для нашего конкретного случая величину сил давления для единицы площади мы можем получить как:

$$P_д = \rho W_д^2. \quad (5)$$

Объём жидкости $V$ мы можем представить в виде площади $F_д$, умноженной на длину $A$, как

$$V = F_д \cdot A. \quad (6)$$

Подставим в уравнение (4) значения уравнений (5) и (6), получим:

$$N = F_д \frac{A}{t} \cdot \rho W_д^2. \quad (7)$$

В уравнении (7) длина $A$ делённая на время даст нам скорость $W_д$. Подставим её в уравнение (7), получим:

$$N = F_д \cdot \rho W_д^3. \quad (8)$$

Уравнение (8) является уравнением мощности, выраженной через характеристики потока. Подставим в уравнение (8) значение сил, получим:

$$N = R_\Sigma \cdot W_д. \quad (9)$$

Уравнение (9) тоже является разновидностью уравнения для мощности. Получив уравнения (8) и (9) для мощности, мы можем теперь осознанно выбирать силовую установку. Для этого нам необходимо задаться величиной скорости $W_д$ или величиной площади сечения потока $F_д$. Тогда силу $R_\Sigma$ мы сможем преобразовать в мощность, а по величине мощности выбрать необходимый двигатель.

В последующем выбор скорости потока $W_д$ или площади его сечения $F_д$ будет связан с вашим личным опытом, но лучше для этих целей построить графики. Тогда без всякого труда можно будет задаваться, например, величиной скорости.

Отметим ещё один момент. Здесь мы, задаваясь скоростью или площадью сечения, полагаем, что энергия двигателя полностью преобразуется в тягу движителя, то есть мы полагаем, что коэффициент полезного действия движителя равен единице. Практически такого случая не может быть. Поэтому для практических целей подобное преобразование надо делать с учётом КПД движителя. Вот таким образом делается подбор силовой установки в первом приближении. Подобный подбор делается не только для плавающих технических устройств, но и летающих таких, как самолёт, ракета и т.п.

Подбирая силовую установку, мы одновременно получим некоторые общие характеристики и для движителя, и для его потока. Далее при расчёте геометрических величин движителя мы должны будем исходить из этих общих характеристик. Принцип расчёта геометрических характеристик движителя в определённых моментах совпадает с принципом расчёта центробежных насосов, который мы получили в работе [2][14] и которым мы воспользуемся в данном случае, а также воспользуемся зависимостями для плоского установившегося движения жидкости. Поэтому далее поступаем с лопастями

---

[14] Редактор переместил расчёт центробежного насоса в качестве Приложения к тексту «I. Механика жидкости и газа, или механика безынертной массы».

движителя “ТОЛИК” так же, как с крыльчаткой центробежного насоса.

В колесе центробежного насоса поток движется одновременно в двух взаимно перпендикулярных направлениях: в тангенциальном и радиальном. Для колеса движителя мы тоже полагаем подобное движение жидкости. Тогда прирост энергии потока жидкости в объёме этого колеса будет создаваться лопастями движителя в тангенциальном потоке.

При выборе силовой установки для полученной силы $R_\Sigma$ мы задавались скоростью установившегося потока движителя $W_д$. Соответственно в зависимости от этой скорости получили силы давления (5) и работу (3). Так как мы определили, что при вращательном движении колеса движителя прирост энергии образуется в тангенциальном потоке, то мы должны будем приравнять выбранную нами скорость $W_д$ к скорости тангенциального потока $W_{tg\ д}$ , то есть $W_д = W_{tg\ д}$. Таким способом мы получим величину тангенциальной скорости. Исходя из найденной величины скорости, мы можем определить величину сил давления для единицы площади тангенциального потока как:

$$P_{tg\ д} = \rho W_{tg\ д}^2,$$

а также величину удельной энергии, или единицы объёма потока, как:

$$P_{tg\ д}\,(1/м^3) = \rho W_{tg\ д}^2.$$

Исходя из полученных величин и имея понятие об обтекаемости, мы можем представить себе тангенциальный поток в виде поступательного потока взаимодействия необтекаемой лобовой плоскости твёрдого тела с площадью, равной единице площади, когда это тело движется со скоростью, равной тангенциальной скорости.

Всё это было бы так, если бы тангенциальный поток имел прямоугольный, а не цилиндрический объём. Но всё равно, полученные таким образом величины сил давления и удельной энергии, согласно второму закону механики безынертной массы будут равны по величине тем силам и энергии, которые мы получили при подборе силовой установки. Без рисунка, наверное, всё это трудно понять. Поэтому на рис. 26 мы покажем то, о чём идёт речь.

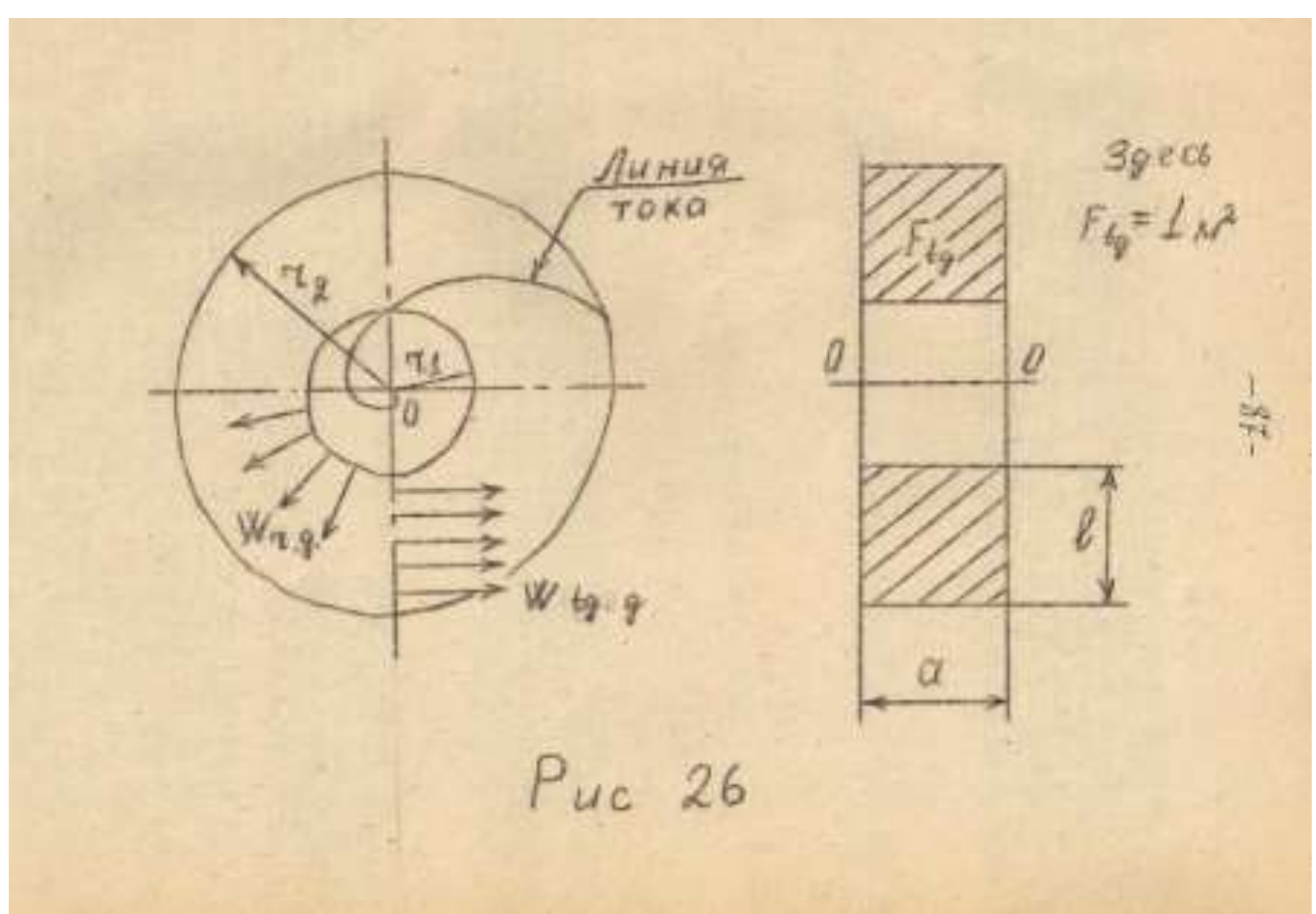

Рис 26

На рисунке 26 мы показали поток плоского установившегося движения. Как вы знаете, он имеет цилиндрическую форму. Также показали направление тангенциальных скоростей $W_{tg\ д}$ и радиальных скоростей $W_{r\ д}$. Площадь сечения тангенциального потока имеет форму прямоугольника со сторонами $a$ и $l$. Величина этой площади постоянна для данного потока. Площадь сечения радиального потока имеет цилиндрическую форму, и величина её в объёме потока зависит от величины радиуса цилиндра.

Для единицы площади сечения тангенциального потока мы выше получили величины сил давления, энергии как для потока взаимодействия с необтекаемой лобовой поверхностью. Фактически, мы таким способом получили величину удельной энергии тангенциального потока, которая возникает в этом потоке как энергия его взаимодействия с лопастью движителя.

При выборе силовой установки мы также нашли площадь движителя $F_{д}$. Зная эту площадь, а плотность жидкости нам тоже известна, мы можем найти величину расхода массы в единицу времени $M_{д}$ по уравнению движения как:

$$M_{д} = F_{д} \cdot \rho W_{д}.$$

Вот этот расход массы в единицу времени должен равняться расходу массы в единицу времени тангенциального потока $M_{tg\ д}$, то есть $M_{д} = M_{tg\ д}$. Это значит, что площадь сечения тангенциального потока $F_{tg\ д}$ должна равняться площади сечения движителя, то есть $F_{д} = F_{tg\ д}$. В основе всех этих равенств, как вы поняли, лежит закон сохранения энергии и второй закон механики безынертной массы, которыми мы руководствовались.

К настоящему времени мы нашли характеристики тангенциального потока. Вернее, известные или выбранные характеристики установившегося потока движителя, полученные при подборе силовой установки, мы перевели в характеристики потока из условия равенства энергии. В установившемся потоке жидкости мы довольно просто определяем величину скорости движения жидкости.

В тангенциальном потоке плоского установившегося вида движения насосного типа постоянная величина тангенциальной скорости $W_{tg\ д}$ по его площади сечения создается вращающейся лопастью, а вращающаяся лопасть имеет различные величины линейной скорости, которые зависят от величины радиуса вращения. Согласно работе [2], величина постоянной скорости тангенциального потока определяется по окружности равных скоростей $r_{р.с.}$

При выборе тангенциальных скоростей нам придётся поступить следующим образом. Выбранная нами силовая установка является вполне конкретным двигателем, который имеет не только определённую мощность, но и определённое число оборотов в единицу времени $n$. Величина линейной скорости будет зависеть от числа оборотов $n$ и от величины радиуса вращения $r$ как:

$$W_{окр} = 2\pi r \cdot n. \qquad (10)$$

В нашем случае величина окружной скорости $W_{окр}$ должна равняться величине, полученной нами скорости тангенциального потока $W_{tg\ д}$, то есть $W_{окр} = W_{tg\ д}$. Уже известную величину тангенциальной скорости мы подставим в уравнение (10), получим:

$$W_{tg\ д} = 2\pi r \cdot n. \qquad (11)$$

В уравнении (11) нам известны все величины, кроме радиуса. Поэтому, решив его относительно радиуса, мы найдем его величину. Найденная нами величина будет радиусом окружности равных скоростей $r_{р.с.}$ Этот радиус является предельной величиной для внутренней границы плоского установившегося потока насосного типа. Зная это условие, мы можем определить величину $a$, то есть высоту плоского установившегося потока, так как на цилиндрической поверхности с радиусом окружности равных скоростей радиальные и тангенциальные скорости равны по величине. Тогда мы можем записать:

$$F_{д} = F_{tg\,д} = 2\pi r_{р.с} \cdot a.$$

Решив это уравнение относительно высоты потока $a$ (рис. 26), получим её величину:

$$a = \frac{F_{tgд}}{2\pi r_{р.с}}$$

Зная высоту потока, мы сможем определить его ширину:

$$F_{tg\,д} = a \cdot l, \text{ или } l = \frac{F_{tgд}}{a}.$$

Вот таким способом мы определили геометрические размеры площади сечения тангенциального потока и одновременно геометрические размеры всего плоского установившегося потока.

Затем мы определяем геометрию линии тока, которая создает нам поверхность лопасти движителя. Форма линии тока записывается уравнением логарифмической спирали. В полярных координатах оно имеет такой вид:

$$r = ae^{\varphi}.$$

Величина $a$ в уравнении логарифмической спирали принимается из начальных условий, то есть при $\varphi = 0$, она будет равна радиусу окружности равных скоростей: $a = r_{р.с}$. Затем, задаваясь величиной угла $\varphi$ в определённой последовательности его интервалов, мы получим искомую геометрию линии тока. Ибо основой для построения искомой линии тока, или начальными условиями построения, является окружность равных скоростей.

После проведенных расчётов мы получим геометрические размеры плоского установившегося потока движителя и поверхности его лопасти. Сразу же отметим, что, как и для центробежных насосов, мы можем увеличивать радиусы граничных поверхностей плоского установившегося потока движителя, но при этом величина площади сечения тангенциального потока должна оставаться неизменной, то есть ширина потока $l$ и высота $a$ должны оставаться неизменными. Для этих изменений построение линии тока тоже из начальных условий окружности равных скоростей. Уменьшать радиус внутренней поверхности потока меньше радиуса окружности равных скоростей мы не можем. Для данных условий изменить величину радиуса окружности радиуса равных скоростей мы можем только за счёт изменения числа оборотов двигателя, то есть мы можем взять другой двигатель такой же мощности, но с другой величиной числа оборотов, или применить редуктор для изменения числа оборотов.

Вот все те величины, которые мы должны получить, используя расчёт центробежного насоса. Теперь мы будем вести расчёт движителя с учётом его особенностей.

Начнём с того, что при расчёте центробежного насоса мы, получив площадь и профиль лопасти, должны будем оставить эти величины неизменными, какое бы число лопастей мы не размещали в объёме потока. Для движителя, который работает в среде, мы не можем этого позволить. Силовая установка способна вращать только одну лопасть, которую мы получили для неё. Если к этой лопасти мы добавим ещё одну такую же лопасть, то мы должны будем увеличить мощность двигателя вдвое. Это значит, что если мы хотим применить для движителя две лопасти, то мы должны разделить площадь подобранной лопасти пополам, то есть высоту лопасти $a$ нужно разделить на две части. Если мы хотим сделать три лопасти в нашем движителе, то эту высоту $a$ нужно разделить на три части и т.д. Таким способом мы оставляем площадь подобранной лопасти неизменной.

Полученную выше логарифмическую спираль мы должны применить в качестве профиля лопасти как зеркально отображенную (рис. 27). Это делается потому, что жидкость в насосе движется одновременно в радиальном и тангенциальном направлениях, а в движителе она будет двигаться в тангенциальном и нормальном направлениях, то есть мы не только сохраняем взаимно перпендикулярное движение жидкости, но и изменяем его перпендикулярность в радиальном направлении.

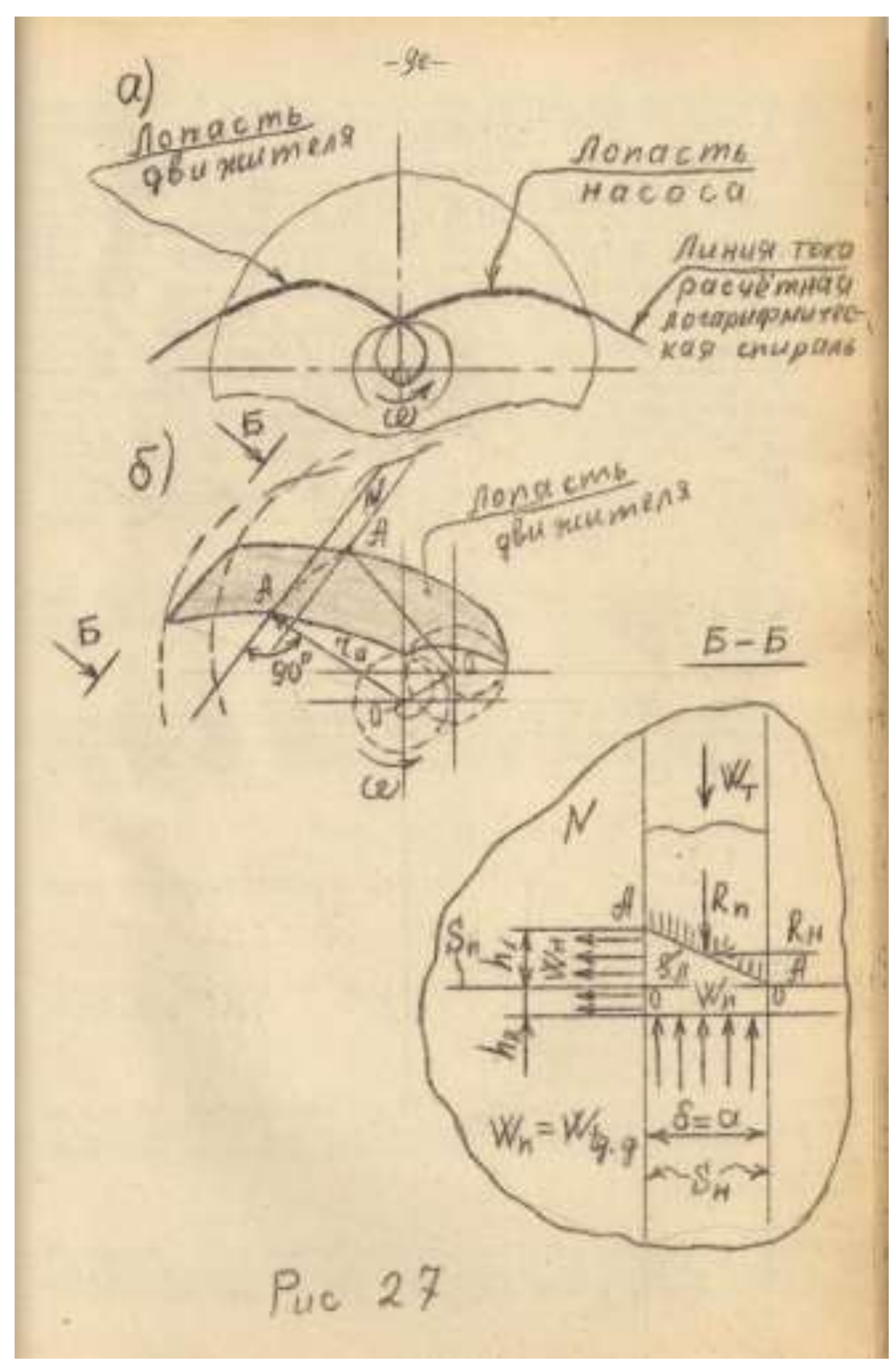

Рис 27

Но одного такого мероприятия ещё недостаточно, чтобы изменить направление движения потока жидкости. Мы должны ещё создать направлённость нормальному потоку.

С точки зрения движения твёрдых тел в среде, которые взаимодействуют с ней через лобовую и тыльную поверхности, лопасть движителя тоже является твёрдым телом, которое совершает движение в среде. Но здесь различие будет заключаться в том, что при вращательном движении лопасти движителя и при профилировании её поверхности в виде логарифмической спирали, лопасть движителя будет взаимодействовать со средой только через лобовую поверхность, то есть в этом случае мы как бы имеем движение твёрдого тела только с одной лобовой поверхностью взаимодействия. В этом и заключается весь фокус плоского установившегося вида движения жидкостей и газов.

Так как поверхности лопастей движителя являются лобовыми поверхностями взаимодействия, то мы и должны будем поступить с ними, как с лобовыми поверхностями. Тогда, чтобы организовать нормальный поток жидкости, мы должны будем развернуть лопасти под определённым углом $\alpha$, как показано на рис. 27.*б*. На величину этого угла $\alpha$ поворачивается вся поверхность лопасти.

Угол $\alpha$ является нашим старым знакомым. Он превращает необтекаемую лобовую поверхность твёрдого тела в обтекаемую. Здесь это превращение выглядит несколько особенно. Ведь поверхность лопасти имеет форму логарифмической спирали. Поэтому поступаем следующим образом: проводим из центра окружности *O* радиус *r* в точку *A*. Затем перпендикулярно к этому радиусу располагаем плоскость *N*. После этого в плоскости *N* разворачиваем поверхность лопасти движителя на угол $\alpha$ (рис. 27.*б*). В этом случае мы получаем картину взаимодействия обтекаемой лобовой плоскости с углом наклона $\alpha$ с жидкостью среды.

Далее в этой плоскости *N* располагаем наши плоскости исследования: поступательную $S_п$ и нормальную $S_н$. Тогда тангенциальный поток движителя будет играть роль поступательного потока взаимодействия, то есть $W_п = W_{tg\,д}$.

Разворот лопасти на угол $\alpha$ мы проводим аналогичным образом для всех её точек. Тогда передняя кромка лопасти сохраняет форму логарифмической спирали, а задняя удаляется относительно неё на угол $\alpha$.

Развернув, таким образом, лопасть движителя на угол $\alpha$, мы должны будем оставить неизменной площадь сечения тангенциального потока, то есть её высоту *a* и ширину *l*. Это значит, что при переводе необтекаемой лобовой поверхности в обтекаемую мы должны увеличить её площадь в соответствии с её углом наклона $\alpha$ (рис. 27.*б*). Сохраняя переднюю кромку лопасти в виде логарифмической спирали и разворачивая относительно неё заднюю кромку лопасти на угол $\alpha$, мы тем самым сохраняем на всей её поверхности скорость тангенциального потока постоянной, то есть не только постоянной, но и неизменной по величине. Поэтому в лобовом сечении лопасти плоскостью *N* эта скорость будет сохранять свою величину неизменной и будет служить в нашем случае скоростью поступательного потока. Следовательно, сечение лопасти в точке *A* плоскостью *N* будет не частным случаем, а общим, и мы будем исследовать силовое взаимодействие в этой точке не как для определённой точки, а как для всей поверхности лопасти.

Теперь мы можем определить обтекаемость лопасти движителя. Здесь нас будет интересовать величина угла наклона $\alpha$, при которой сила взаимодействия $R_н$ нормального потока с наклонной лобовой поверхностью будет максимальна. Ведь эта сила создает толкающее усилие, которое необходимо для движения корабля.

Энергию поступательного, или тангенциального, потока мы знаем. Она равна:

$$P_{tg\,д}\,(1/м^3) = \rho W_{tg}{}^2{}_{д}. \qquad (12)$$

Эта энергия является полной энергией поступательного потока. Для обтекаемой лобовой поверхности она расписывается по уравнению Бернулли:

$$P_{tg\,д}\,(1/м^3) = P_{ст}(1/м^3) + \frac{1}{2}\rho W_п^2. \qquad (13)$$

Статическая, или потенциальная, энергия этого поступательного потока определяется при непосредственном взаимодействии поступательного потока с площадью лобовой поверхности. Площадь лобовой поверхности мы можем найти как:

$$s_{л.д} = \frac{F_{tg.д}}{\sin\alpha}.$$

Тогда уравнение движения примет вид:

$$M_{\text{п.д}} = s_{\text{л.д}} \cdot \rho W_{\text{л.п}} = \frac{F_{tg.\text{д}}}{\sin \alpha} \cdot \rho W_{\text{л.п}}. \qquad (14)$$

Уравнение движения для поступательного потока полной энергии будет иметь такой вид:

$$M_{tg\ \text{д}} = F_{tg\ \text{д}} \cdot \rho W_{tg\ \text{д}}. \qquad (15)$$

Так как расходы масс для этих двух потоков равны между собой, то есть $M_{\text{п}} = M_{tg\ \text{д}}$, то из этого равенства мы найдем соотношение между скоростями этих потоков как:

$$\frac{F_{tg.\text{д}}}{\sin \alpha} .\cdot \rho W_{\text{л.п}} = F_{tg\ \text{д}} \cdot \rho W_{tg\ \text{д}}. \qquad (16)$$

Уравнение (13) полной энергии поступательного потока накладывает предельные ограничения на распределение между потенциальной и кинетической энергией потока. Это ограничение будет выражаться как:

$$\frac{1}{2} P_{tg}\ (1/\text{м}^3) = P_{\text{ст}}\ (1/\text{м}^3).$$

Распишем его через характеристики потоков как:

$$\frac{1}{2} \rho W_{tg\ \text{д}}^2 = \rho W_{\text{л}}{}^2{}_{\text{п}}$$

По уравнению (16) заменим лобовую поступательную скорость, получим:

$$\frac{1}{2}\ W_{tg\ \text{д}}^2 = W_{tg\ \text{д}}^2 \cdot \sin^2\alpha.$$

Из этого равенства мы определим величину угла наклона $\alpha$ при предельном распределении энергии в поступательном потоке как:

$$\sin\alpha = \sqrt{0{,}5}.$$

Отсюда мы получаем величину угла $\alpha$ равной 45°. Эту величину угла $\alpha$ выше мы приняли равной 45° и назвали предельным углом наклона лобовой поверхности $\alpha_{\text{пр}}$. Теперь мы вычислили величину предельного угла наклона лобовой поверхности.

Для лобовой поверхности лопасти движителя этот предельный угол наклона $\alpha_{\text{пр}} = 45°$ означает, что при наклоне лобовой поверхности на угол меньше предельного угла будет образовываться нормальный поток. Если угол наклона превысит предельную величину, то нормальный поток образовываться не будет. В нашем случае данное положение означает, что мы можем наклонять лобовую поверхность лопасти движителя на угол не более чем 45°, так как в этом случае образуется необходимый для нас поступательный поток. Таким способом мы определили предел наклона лобовой поверхности.

При исследовании обтекаемой лобовой плоскости твёрдого тела мы получили, что скорости нормального потока $W_{\text{н}}$ равны скоростям поступательного потока $W_{\text{п}}$, который взаимодействует непосредственно по обтекаемой лобовой плоскости лопасти, то есть $W_{\text{п}} = W_{\text{н}}$, а лобовая поступательная скорость $W_{\text{л.п}}$ зависит от величины площади лобовой поверхности и вычисляется по равенству (16) в зависимости от скорости тангенциального потока, как:

$$W_{\text{л.п}} = W_{tg\,\text{д}} \cdot \sin\alpha.$$

Тогда:

$$W_{\text{н}} = W_{\text{л.п}} = W_{tg\,\text{д}} \cdot \sin\alpha. \qquad (17)$$

Из равенства (17) мы видим, что скорости нормального потока уменьшаются в зависимости от величины угла наклона $\alpha$ лобовой поверхности. Это будет только одной стороной уменьшения скорости нормального потока.

Подобное уменьшение нормальных скоростей приводит к росту площади сечения нормального потока, которая выходит за пределы лобовой поверхности, что влечёт за собой уменьшение расхода массы в части нормального потока, которая непосредственно взаимодействует с лобовой поверхностью. Эту часть нормального потока мы выше определили уравнением движения (60) как:

$$M_{\text{н.л}} = h_1 \cdot 1 \cdot \rho W_{\text{н}}.$$

В свою очередь, эта часть нормального потока будет взаимодействовать со всей площадью лобовой поверхности. Мы это взаимодействие записали уравнением движения (61) как:

$$M_{\text{н.л}} = s_{\text{л}} \cdot \rho W_{\text{н.л}}.$$

Из этого уравнения мы получили величину скорости $W_{\text{н.л}}$ взаимодействия нормального потока с лобовой поверхностью. Величина этой скорости получается даже меньше величины скорости нормального потока полного сечения, то есть $W_{\text{н.л}} < W_{\text{н}}$. Величину нормальных сил взаимодействия $R_{\text{н.л}}$ мы записали выше уравнением (62) как:

$$R_{\text{н.л}} = s_{\text{л}} \cdot \rho W_{\text{н.л}}^2.$$

По этому уравнению мы получаем тягу движителя, то есть мы таким способом получили величину тяги движителя “ТОЛИК” при подобранной нами выше силовой установке. Как видим, величина тяги находится в прямой зависимости от величины площади лобовой поверхности и величины нормальной скорости взаимодействия. А величины площади лобовой поверхности и нормальной скорости в свою очередь зависят от угла наклона $\alpha$ лобовой плоскости. Эти зависимости мы определили выше.

Из них следует, что при величине угла наклона лобовой плоскости равной нулю ($\alpha = 0°$) площадь $s_{\text{л}}$ взаимодействия тоже будет равной нулю, то есть $s_{\text{л}} = 0$. При предельном угле наклона ($\alpha = 45°$) площадь взаимодействия будет иметь максимальное значение.

Для нормальных скоростей всё будет наоборот. При величине угла наклона равной нулю ($\alpha = 0°$) они будут иметь максимальное значение, а при предельном угле ($\alpha = 45°$) они будут иметь минимальное значение.

Вот эти два условия являются руководством при выборе оптимального значения угла наклона лобовой поверхности $\alpha$. Руководствуясь ими, мы должны будем выбрать соответствующий угол наклона лобовой поверхности для лопастей движителя. Для чего нам придется сделать несколько вариантов числового просчёта. В связи с тем, что величина тяги находится в квадратичной зависимости от скорости, то просчёт надо начинать с минимальных углов наклона $\alpha$. После проведения вариантов расчёта для угла наклона мы окончательно выбираем величину угла наклона лобовой поверхности и величину тяги $R_{\text{н.л}}$, которая для просчитанных вариантов должна иметь максимальное значение. Если мы не имеем возможности варьировать мощностью силовой установки, то этот расчёт движителя будет окончательным.

Для нашей постановки задачи данный расчёт тяги движителя является предварительным. Ведь мы получили определённую величину силы $R_{\Sigma}$, которая должна обеспечивать движение твёрдого тела с заданной скоростью. По величине этой силы мы подбираем силовую установку соответствующей мощности. После расчёта движителя для этой силовой установки мы получили силу тяги $R_{\text{н.л}}$, которая меньше по величине заданной силы тяги, то есть $R_{\text{н.л}} < R_{\Sigma}$.

Как мы видим, уменьшение силы тяги в движителе произошло не за счёт потерь энергии, а за счёт особенностей преобразования движения лопастей движителя в движение нормального потока. В настоящее время подобное явление относят к потерям энергии и количественно выражают через коэффициент полезного действия (КПД), который характеризует полноту использования располагаемой энергии.

В нашем случае мы не можем воспользоваться этим коэффициентом. Ибо при движении твёрдого тела в среде обтекаемость его лобовой и тыльной поверхностей приводит к снижению потребной величины силы для обеспечения заданной скорости. В движителе подобная обтекаемость снижает величину силы тяги при преобразовании движения тангенциального потока в движение нормального потока. Во всех этих случаях мы не наблюдали потери энергии. Поэтому не можем применить коэффициент полезного действия.

При данном преобразовании сил взаимодействия речь идет об эффективности использования сил. Поэтому мы должны оценивать количественное преобразование сил взаимодействия коэффициентом эффективного использования сил (КЭИС), например, при движении такого твёрдого тела в среде, как корпус корабля.

Обтекаемость лобовых и тыльных поверхностей приводит к снижению потребных сил, то есть к положительному коэффициенту эффективного использования сил, а для движителя подобное снижение приводит к отрицательному КЭИС.

Для характеристики технических возможностей коэффициент эффективного использования сил (КЭИС) является новым понятием. Мы его запишем тоже в виде отношения сил. Для лобовой и тыльной поверхности твёрдого тела его надо записывать раздельно, как отношение величины силы лобового сопротивления обтекаемой поверхности к величине силы сопротивления необтекаемой поверхности:

***для лобовой поверхности***: $\text{КЭИС}_{\text{л}} = \dfrac{R_{\text{л.обт}}}{R_{\text{л.необт}}}$;

***для тыльной поверхности***: $\text{КЭИС}_{\text{тыл}} = \dfrac{R_{\text{тыл.обт}}}{R_{\text{тыл.необт}}}$.

Мы назвали их положительными коэффициентами. Чем меньше будет величина КЭИС, тем лучше происходит преобразование сил.

Для движителя мы назвали эти коэффициенты отрицательными. Они тоже будут количественно выражаться как отношение сил тангенциального потока к силам нормального потока, то есть:

$$\text{КЭИС}_{\text{л}} = \frac{R_{\text{н}}}{R_{tg}}.$$

Для движителя – наоборот: как для отрицательного коэффициента стремление КЭИС к единице будет характеризовать более полное использование сил, а его стремление у нулю – худшее использование сил.

Далее мы должны будем действовать следующим образом. Получив величину тяги движителя $R_{\text{н.л}}$ и зная величину сил тангенциального потока $R_{tg\ \text{п}}$, которая равна заданной силе $R_{\Sigma}$, мы должны будем через их отношение получить коэффициент эффективного использования сил как:

$$КЭИС_д = \frac{R_{н.л}}{R_{tg\ п}}.$$

Затем разделить на этот коэффициент либо тангенциальную силу, либо заданную силу. Тогда мы получим действительную потребную силу $R_{дейс}$, которая по величине будет больше заданной силы $R_\Sigma$:

$$R_{дейс} = \frac{R_\Sigma}{КЭИС_д}.$$

После того, как мы получили действительную величину силы $R_{дейс}$, мы должны будем проделать расчёт движителя с самого начала, то есть подобрать силовую установку для действительной величины силы $R_{дейс}$, а затем уже относительно неё получить необходимые геометрические размеры движителя и его тягу. После такого просчёта величина тяги двигателя должна будет равняться величине заданной силы. В этом будет заключаться окончательный выбор силовой установки и расчёт гребного винта типа движителя “ТОЛИК”.

Отметим, что осевые лопастные движители и машины обладают сравнительно низким коэффициентом эффективного использования сил. По этой причине их нужно применять лишь в необходимых случаях.

## *Глава II*

## МЕХАНИЗМ МАШУЩЕГО ПОЛЁТА ПТИЦ

Здесь будет поставлена несколько иная задача по сравнению с предыдущей. В данной задаче нам придётся задаться определённым весом или силой веса $G$ твёрдого тела и затем найти для тела с таким весом потребную площадь крыльев и величину усилия, которое необходимо прилагать к ним в машущем полёте, чтобы сохранить движение тела в среде атмосферы. То есть, полагаем, что полёт должен происходить в атмосфере Земли.

Далее начинаем действовать в соответствии с законами и положениями механики безынертной массы. Начнём с того, что определим потребную площадь крыльев из условия минимальной скорости полёта. Как это сделать, нами дано при исследовании движения очень тонкой пластинки. В этом разделе есть неравенство (100), которое утверждает, что минимальная скорость горизонтального полёта должна хотя бы немного превышать скорость свободного падения пластинки. Ибо только в этом случае возникает достаточная несущая сила, которая удерживает пластинку в горизонтальном полёте. Величина сил давления при свободном падении пластинки определяется правой частью неравенства (96), то есть:

$$R_{в.п} = 2F_{пл} \cdot \rho W_в^2. \quad (1)$$

В этом равенстве цифра 2 обозначает то положение, что мы учитываем силы взаимодействия со средой лобовой и тыльной поверхностей. Площадь пластины обозначим как $F_{пл}$. Величина силы нам задана из условия задачи. По равенству (1) мы должны, задаваясь скоростью свободного падения пластинки, найти её площадь. Естественно, задаваться можно любой величиной скорости свободного падения, но мы должны исходить из реальных условий. Ведь по условиям нашей задачи вес пластинки нам известен. И, например, мы желаем выяснить, может ли человек с помощью своей мускульной силы летать, как птица. Тогда мы задаемся весом человека вместе с весом его оснащения для полёта, то есть принимаем суммарный вес $G$ равным 100 кг. Плотность воздуха $\rho$ принимаем равной 0,1 кг/м$^3$. Затем задаемся скоростью вертикального падения $W_в$ порядка 10 м/сек. Далее для вычисления площади пластинки, или площади крыльев, мы пользуемся равенством (1):

$$F_{пл} = \frac{G}{2\rho W_в^2} = \frac{100}{2 \cdot 0{,}1 \cdot 10^2} = 5 м^2.$$

В результате мы получили, что для человека весом 100 кг площадь пластинки $F_{пл}$ = 5м$^2$. Её площадь мы должны разделить пополам, чтобы получить площадь одного крыла $F_к$ = 2,5м$^2$. Если мы представим каждое крыло шириной в один метр и длиной 2,5 м, то они будут соизмеримы с габаритами человека, вес которого составляет 80-90 кг.

Скорость свободного падения пластины $W_в$, которую мы приняли равной 10 м/сек, означает, что начальная, или стартовая, скорость должна превышать эту скорость. Ибо она будет минимальной скоростью горизонтального полёта человека с крыльями площадью 5м$^2$, которая обеспечивает необходимые несущие силы горизонтального полёта.

Рассмотрим стартовую скорость человека, исходя из его физических возможностей. Мы приняли её в десять метров в секунду. Современные спринтеры пробегают сто метров больше, чем за десять секунд. Если мы примем стартовую скорость на один метр в секунду меньше, то площадь крыльев возрастет до шести с лишним квадратных метров, а если мы снизим её на два метра в секунду, то площадь крыльев возрастет до восьми квадратных метров. Таким образом, вы убедились, что снижать стартовую скорость нельзя. Поэтому мы оставляем площадь крыльев в пять квадратных метров и ещё раз рассмотрим стартовые возможности человека.

С полной оснасткой на ровном месте человек сможет развить скорость 6-8 м/сек. Ведь ему надо только развить такую скорость, а не бежать с подобной скоростью какое-то расстояние. Далее полагаем, что разогнавшись до скорости 6-8 м/сек человек может подпрыгнуть и тем самым достичь необходимой минимальной скорости горизонтального полёта. При разбеге вес оснастки будет уменьшаться по мере увеличения стартовой скорости за счёт несущей силы, что также повышает физические возможности человека. Кроме того, если человек использует попутный ветер скоростью 3-4 м/сек, то его скорости 6-8 м/сек будет достаточно для старта. Наконец, он может делать разбег по наклонной плоскости, например, с горы. В этом случае он имеет вполне реальную возможность набрать необходимую стартовую скорость. Человеку не обязательно весить 80-90 кг, он может иметь вес 50-60 кг. Для такого человека будет достаточна площадь крыльев 3-4 квадратных метра. Что тоже повышает шансы для набора необходимой стартовой скорости. В общем, будем считать, что стартовая скорость 10 м/сек доступна для человека и что человек сможет взлететь на собственных крыльях.

Подняться в воздух – это всего-навсего половина дела. Далее необходимо обеспечить горизонтальный полёт за счёт мускульной силы. Посмотрим с этой точки зрения на возможности человека, насколько крепки его руки.

Получив соответствующую площадь крыльев, мы можем мысленно скомпоновать их совместно с туловищем в один общий твёрдый предмет. Затем задаёмся скоростью полёта в 20-25 м/сек. После чего выделяем лобовую и тыльную поверхности в этом твёрдом предмете (крылья-человек) и придаём им наиболее обтекаемую форму. Как это сделать, мы показали выше. Исходя из заданной скорости горизонтального полёта ($W_г$ = 20-25 м/сек) и из обтекаемости лобовой и тыльной поверхности, мы получим величину суммарной силы $R_\Sigma$, которая необходима для обеспечения горизонтального полёта с заданной скоростью. После соответствующих просчётов мы получим величину этой силы порядка $R_\Sigma$ = 40-50 кг. Эту величину человек должен обеспечить за счёт машущего движения крыльев.

Мы получили исходные данные для расчёта движителя. В данном случае в качестве движителя будут служить два крыла, площадь которых нам известна. Машущие движения крыльев представляют собой тоже вращательное движение, которое совершается не на полный оборот, а в пределах 20°-30°. То есть крыло проходит по дуге 20°-30° из начального положения, а затем снова возвращается в начальное положение, совершив тот же путь в обратном направлении.

Так как крылья совершают вращательное движение и создают при этом определённую величину силы тяги, то принцип их работы как движителей тоже будет основан на принципе работы движителя "ТОЛИК". В основе этого принципа лежит плоский установившийся вид движения жидкостей и газов. С принципом работы движителя "ТОЛИК" мы познакомились при расчёте корабельного винта. Единственное, что мы не учли при его расчёте, – это сопротивление, которое возникает на винте как твёрдом теле при его движении относительно невозмущенной жидкости или газа среды. Об этом надо помнить.

Винты современных самолётов и кораблей делаются без знания законов движения жидкостей и газов. По этой причине их коэффициент эффективного использования сил очень низок. В то же время сопротивление подобных винтов тоже велико. Корабельные винты имеют сравнительно большой угол поворота лопастей. Лопасти авиационных винтов в своем большинстве имеют регулируемый угол поворота лопастей. Поэтому ими пользуются и для создания тяги, и для торможения самолёта.

Мы остановились на том, что в основе маховых движений крыльев лежит принцип работы движителя "ТОЛИК". Крылья у нас имеют определённую площадь, которую изменять нельзя, а также нам известна величина тяглового усилия, которое должны создавать крылья, работая как движитель. Вот эти особенности лежат в основе расчёта крыльев как движителей. Они вносят определённые трудности в расчёт, которые заключаются в том, что известная площадь крыла как площадь лопасти движителя располагается по поверхности логарифмической спирали с одной стороны и имеет определённый угол наклона $\alpha$. Это значит, что площадь лопасти будет больше площади сечения тангенциального потока. Насколько больше – конкретную величину мы можем установить лишь при знании геометрических размеров плоского установившегося потока крыла как движителя.

Как вы знаете, расчёт движителя "ТОЛИК" очень прост, но конкретная величина площади лопасти в данном случае требует метода приближенного решения. Это значит, что мы в конечном итоге должны будем получить действительные, или истинные, размеры плоского установившегося потока крыльев, создающего заданную величину тяги, с помощью нескольких приближений. Нам остается выбрать метод последовательных приближений и далее действовать в соответствии с этим методом.

Мы знаем, что начальные этапы расчёта движителя и центробежного насоса одинаковы. Поэтому мы примем, что площадь сечения тангенциального потока $F_{tg}$ = 1м$^2$. С этой величины тангенциальной площади начнём свои приближенные решения. Зная величину тангенциальной площади сечения потока и величину сил тяги $R_\Sigma$ = 20-25 кг (мы получили выше эту величину, которая в данном случае записана для одного крыла), мы по уравнению сил, которое имеет вид:

$$R_{tg} = F_{tg} \cdot \rho W_{tg}^2 ,$$

найдем величину тангенциальной скорости при плотности воздуха $\rho$ = 0,1 кг/м$^3$, как:

$$W_{tg} = \sqrt{\frac{25}{0{,}1 \cdot 1}} \approx 15 \text{ м/сек.}$$

Затем из равенства тангенциальной площади $F_{tg}$ цилиндрической поверхности внутренней границы потока с радиусом окружности равных скоростей, которое имеет вид:

$$F_{tg} = 2\pi r_{\text{р.с}} \cdot a,$$

мы найдем величину радиуса окружности равных скоростей, так как высота $a$ цилиндра нам известна как ширина крыла, или ширина тангенциальной площади, то есть $a$ = 1м. Тогда:

$$r_{\text{р.с}} = \frac{1}{2 \cdot 3{,}14 \cdot 1} = 0{,}16 \text{ м.}$$

Зная величину радиуса окружности равных скоростей и тангенциальную скорость ($W_{tg\,\text{д}} = 2\pi r_{\text{р.с}} \cdot n$), мы можем определить число оборотов лопасти движителя в секунду как:

$$15 = 2 \cdot 3{,}14 \frac{1}{2 \cdot 3{,}14} \cdot n.$$

Отсюда: $n$ = 15 об/сек. По полученным величинам мы уже можем построить расположение крыла как лопасти движителя. Сделаем это на рис. 28.

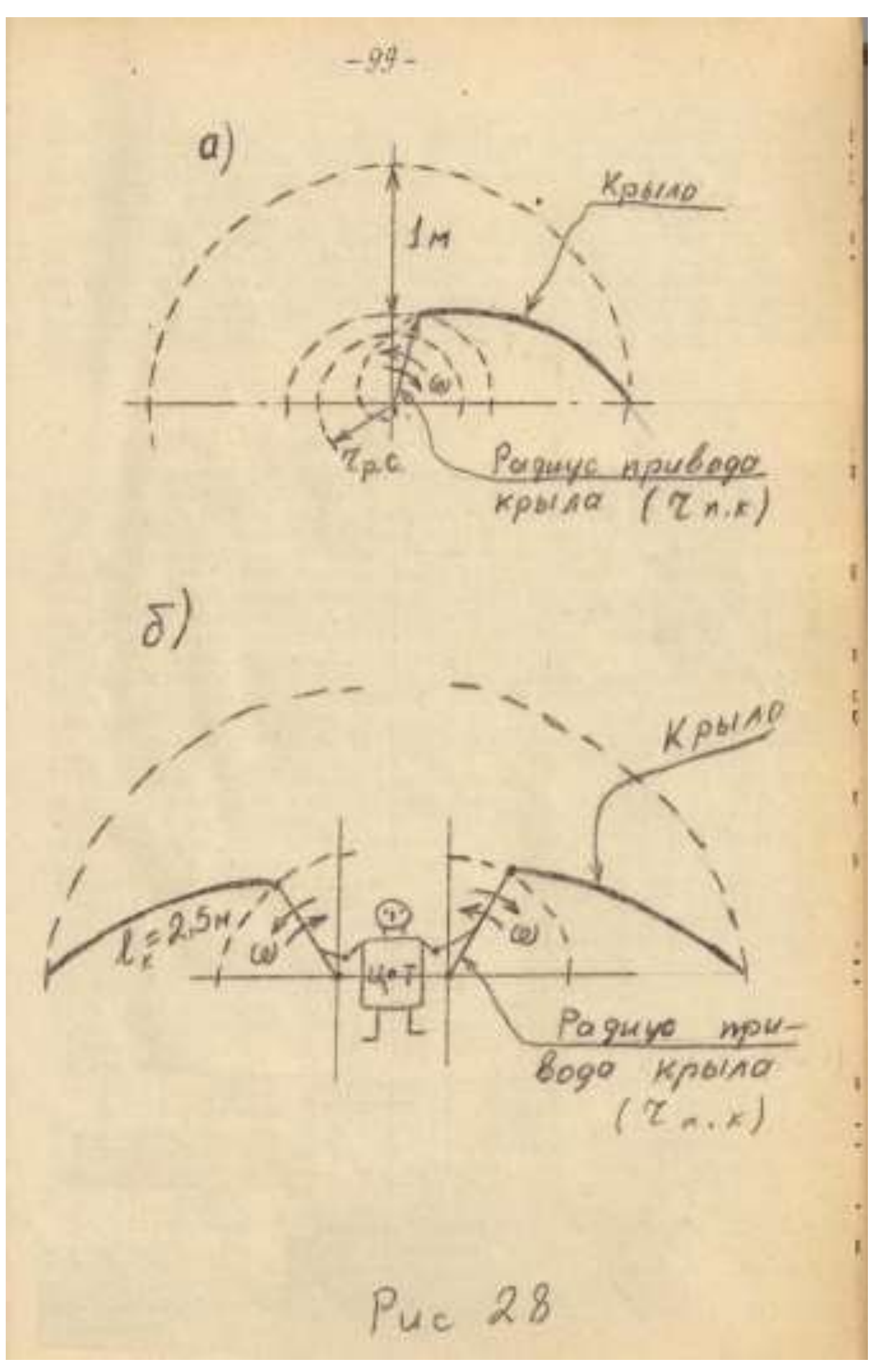

Рис 28

Окружность равных скоростей является исходным данным для построения логарифмической спирали. Мы получили радиус $r_{\text{р.с}}$ = 0,16 м. В соответствии с этим радиусом строим логарифмическую спираль. Затем, выше окружности равных скоростей располагаем тангенциальный поток таким образом, чтобы его границы выделили на логарифмической спирали участок длиной $l_{\text{к}}$ = 2,5 м. Подобное построение лучше выполнять графическим способом в соответствующем масштабе. Участок логарифмической спирали длиной $l_{\text{к}}$ = 2,5 м будет искомой поверхностью крыла. Машущие движения такого крыла будут создаваться с помощью движения радиуса привода крыла $r_{п.к.}$. Вот таким способом мы получим поверхность крыла как лопасти движителя.

В то же время, если этому крылу придать профилировку в нормальном направлении, то есть отклонить его на соответствующий угол $\alpha$, как мы это делали для лопасти корабельного движителя в предыдущей задаче, то данное крыло можно будет использовать в качестве движителя для человека весом в 100 кг. Это крыло будет отвечать условиям задачи. Приближенность решения данного крыла будет выражаться в том, что мы, задавшись площадью сечения тангенциального потока, в зависимости от него получили число оборотов $n$ и некоторую длину радиуса привода $r_{п.к.}$. В действующей конструкции радиус привода крыла должен выполняться из стержней. Величины числа оборотов и длины радиуса привода крыла могут нас не устроить. В этом заключается приближенность решения.

В данном случае эти величины нас не устраивают. Мы исходили из того условия, что крылья будут приводиться в движение мускульной силой человека. И хотя их надо будет отклонять всего на 20°-30°, а не на полный оборот, человек сможет сделать не более двух отклонений в секунду, в то время как требуется делать 15 отклонений в секунду. Это значит, что полученное нами число отклонений превосходит физические возможности человека во много раз. Да и длина радиуса привода крыльев, хотя она нам и неизвестна, так как она получается графическим способом, тоже должна быть слишком большой, то есть не устраивать нас своей величиной. Чтобы изменить эти величины, мы должны будем повторить вышеизложенный расчёт во втором приближении, задавшись другой величиной площади сечения тангенциального потока. В общем, мы должны будем проделать несколько таких приближений, пока не получим нужных для нас результатов. Все эти приближения делаются на этом этапе так же, как для лопасти центробежного насоса.

После того, как будут получены нужные результаты, крыло начинают профилировать, как лопасть движителя. Для этого полученную логарифмическую спираль используют зеркально отображенной. Затем поверхность крыла профилируют в нормальном направлении с соответствующим углом наклона $\alpha$. эти расчёты затем проверяют по коэффициенту эффективного использования сил (КЭИС). В зависимости от полученных результатов и условий либо оставляют полученные результаты, либо повторяют расчёты заново, как это было показано в предыдущей задаче. Мы же с вами не будем проделывать всех этих расчётов. По той причине, что в следующем приближении нам придется увеличить площадь сечения тангенциального потока в 2-3 раза. Увеличение площади сечения тангенциального потока повлечёт за собой увеличение длины крыла с 2,5 м до 4-5 м. Что, как и повышение числа оборотов, будет неприемлемо физически для человека. Отсюда мы просто сделаем заключение, что мускульная сила человека не может быть использована в качестве силового привода крыльев при желании человека летать, как птица. <...>

## ЗАКЛЮЧЕНИЕ

Мы закончили ещё одну работу по прикладному применению законов механики безынертной массы. Первой работой, написанной в этом плане, была работа под названием “Строение Солнца и планет солнечной системы с точки зрения механики безынертной массы”, где было установлено внутреннее строение Солнца и планет солнечной системы, дан расчёт центробежных насосов и компрессоров. В данной работе мы определили понятие обтекаемости твёрдых тел при их движении в жидкостной и газовой среде. Установили геометрические размеры потоков взаимодействия на лобовой и тыльной поверхности твёрдого тела. Выяснили очень важный вопрос, что газы в потоках взаимодействия при любых скоростях движения твёрдого тела ведут себя, как несжимаемые жидкости, в соответствии с изохорным процессом термодинамики.

Далее мы выяснили, что жидкости тоже подчиняются изохорному процессу, но ведут себя с определёнными особенностями, присущими только жидкостям. В термодинамике процессы для жидкостей не даны. В общем, до настоящего времени никто не знал, куда можно было бы приспособить изохорный процесс. Мы этот вопрос выяснили и установили, что он присутствует и при акустическом виде движения, и в потоках взаимодействия, и в сопле Лаваля, и в движителях, и в различных лопастных машинах. В связи с этим мы должны сделать уточнение. В работе "Строение Солнца и планет солнечной системы с точки зрения механики безынертной массы", следуя общепринятому мнению, мы написали, что газы в колесах центробежных компрессоров ведут себя в соответствии с адиабатическим процессом. После исследований, проведенных в данной работе это положение надо считать неверным, так как мы установили, что в потоках взаимодействия газы подчиняются изохорному процессу. По этой причине газы в колесах центробежных компрессоров тоже будут подчиняться изохорному процессу. Эта поправка очень существенна, и с ней нельзя не считаться.

Также мы выявили необходимость новой технической характеристики – коэффициента эффективного использования сил (КЭИС). В настоящее время для подобных целей существует только коэффициент полезного действия (КПД). Он характеризует эффективность использования энергии. Энергию мы не можем ощутить непосредственно, как материальное тело. Мы ощущаем её через действие сил, ибо она проявляет себя через действие сил. Поэтому, прибегая к коэффициенту полезного действия, мы можем лишь в общем плане установить проявление энергии через силы. Конкретные этапы силового взаимодействия остаются для нас неясными. В других случаях КПД просто не приемлем в качестве технической характеристики. Например, при движении корабля возникает энергия в потоках взаимодействия и энергия топлива, используемого в силовых установках типа двигателя внутреннего сгорания. С помощью только КПД мы не сможем установить связь между двумя этими энергиями. Следовательно, мы здесь не сможем применить закон сохранения энергии. Коэффициент эффективного использования сил (КЭИС) дает нам возможность связать между собой эти виды энергии. КЭИС приемлем также для характеристики любой другой техники и любого другого силового взаимодействия. В таких случаях он становится очень важным показателем. Например, современная техника характеризуется многоплановым силовым взаимодействием.

При помощи КПД мы можем только в общем плане охарактеризовать всё это силовое взаимодействие. Применив КЭИС для каждого участка силового взаимодействия, мы можем охарактеризовать в плане эффективного использования сил каждый силовой узел или участок любой машины и сказать, какой из этих узлов требует дальнейшего совершенствования, а какой не требует. Ведь КЭИС дает нам предельные показатели любого силового взаимодействия. Если мы для какого-то участка или узла не сможем составить коэффициент эффективного использования сил, то это говорит о том, что мы не знаем, какое силовое взаимодействие происходит на данном участке. Тогда данное силовое взаимодействие становится научной проблемой. В этом случае учёные будут знать конкретно, куда им направлять свои усилия. Это значит, что КЭИС дает возможность избежать совершенствования уже совершенного и одновременно он дает возможность выявлять несовершенное и сосредоточить на нём внимание людей. Вот таким важным показателем для техники и для её создателей является КЭИС.

Мы также дали количественное и качественное определение для несущих и подъёмных сил летательных аппаратов. Несущие силы определяются минимальной горизонтальной скоростью, а подъёмные силы – формой профиля крыла.

Также мы получили расчёт движителя типа "ТОЛИК", который может использоваться в современных условиях как предельно совершенная с точки зрения механики безынертной массы конструкция в качестве корабельных и авиационных винтов, разнотипных вентиляторов, осевых насосов и компрессоров. Применение более совершенных конструкций требует уже сегодняшний день нашей жизни, но когда данная работа увидит свет – сказать трудно. <...>

***

## ЛИТЕРАТУРА

1. *Шкурченко И.З.* “Механика жидкости и газа, или механика безынертной массы” // рукопись. Октябрь 1971г.

2. *Шкурченко И.З.* “Строение Солнца и планет солнечной системы с точки зрения механики безынертной массы” // рукопись. Март 1974 г.

3. *Шкурченко И.З.* Заявки на предполагаемое открытие законов механики безынертной массы. Декабрь 1969 г. – август 1970 г.

***

### ДОПОЛНИТЕЛЬНЫЕ МАТЕРИАЛЫ ОТ РЕДАКТОРА

Автор собирался сделать гребной винт для моторной лодки с мотором “Вихрь”. На фото уменьшенная модель движителя «ТОЛИК», изготовлённая автором из алюминиевого диска при помощи сверла и ножовки в домашних условиях. Настоящий движитель изготовить не удалось.

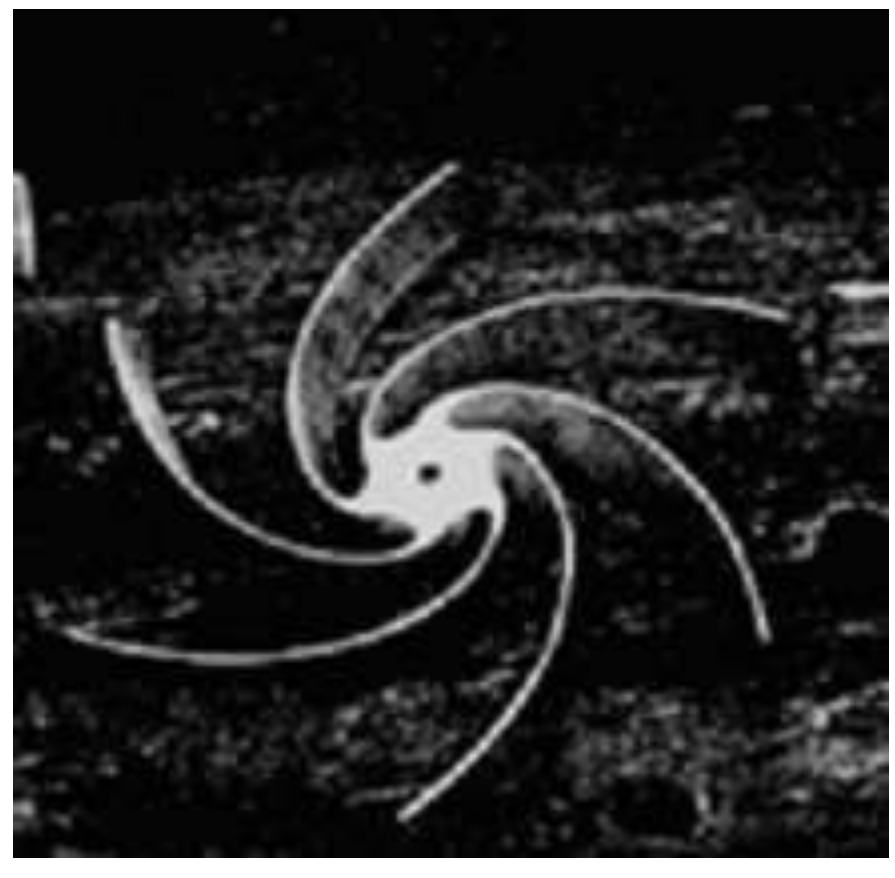

*модель движителя «ТОЛИК»*

***

*После того, как в 2004 году редактор оцифровал рукописи работ по механике безынертной массы и в какой-то степени стал их понимать, ему попались две статьи в журнале «Наука и жизнь», которые показывают актуальность механики безынертной массы. Сейчас, в сентябре 2022 года, редактор посчитал полезным привести два своих старых текста начала 2000-х годов по темам этих статей, т.к. на сайте журнала «Наука и жизнь» (https://www.nkj.ru/archive/#1998 ) в архиве нет нужных номеров. Возможно, кому-то удастся найти эти статьи.*

***

В журнале «Наука и жизнь» (№ 10, 1998 г.) в рубрике «Гипотезы, предположения, догадки» помещена статья д.т.н., академика Российской академии ракетно-артеллерийских наук В. Яворского, которая называется «Энергия “из ниоткуда”». Автор пишет, что в 70-е годы, работая в Научно-исследовательском машиностроительном институте (НИМИ) над средствами поражения брони, он обратил внимание на энергетический эффект в виде чрезвычайно большого выделения теплоты, происходящее при «внедрении длинного металлического, не снаряженного взрывчаткой стержня – бронебойного снаряда – в стальную бронеплиту большой толщины». Расчёты, сделанные автором статьи, показали, что количество выделявшейся тепловой энергии всегда существенно превосходило кинетическую энергию снаряда, которой тот обладал в момент удара (ниже будет приведен конкретный расчёт, данный в этой статье).

Эффект был замечен автором статьи в 70-е годы. «Всего» через 20 лет ему удалось привлечь внимание ученых мужей, к этому, как он называет, «энергетическому парадоксу». Собрался научно-технический совет НИМИ (июнь 1993 г.). В результате обсуждения «парадокса» «было принято решение получить достоверные данные при помощи специальной экспериментальной работы». Из-за недостатка средств, выделённых на эксперименты, были уменьшены некоторые параметры, например, масса снарядов и др., но всё равно результаты испытаний подтвердили существование «парадокса». При массе ударника 61,5 г превышение составило 20%, а при массе ударника 88,5 г – 48%. «Стабильность полученных результатов – пишет автор - дает основание говорить об их достаточной достоверности». «Научно-технический совет дал положительную оценку этой работе и назвал разность между затраченной и выделившейся энергией энергетическим дисбалансом». Объяснения же дано не было. Далее автор пишет, что «по мнению исследователей из Физического института им. П. Н. Лебедева (ФИАН), обнаруженный дисбаланс указывает на большую сложность процессов, сопровождающих внедрение снаряда в броню. Корректный их учёт представляет собой сложную задачу, весьма важную как в теоретическом, так и в практическом отношении. И хотя говорить о нарушении закона сохранения энергии нет никаких оснований, необходимо выяснить, что же всё-таки происходит в момент удара и откуда берётся «лишняя» энергия».

Статья помещена в журнале за 1998 год. Значит, вопрос - откуда берутся «излишки» и что происходит, оставался невыясненным ещё в течение 5 лет после признания факта существования «парадокса» и вывода, что его необходимо объяснить. Было ли дано объяснение «парадоксу» позже 1998 года – мне неизвестно. Сейчас, наверное, или самого автора нет в живых, или он уже давно находится не у дел по возрасту.

В статье помещен снимок 1972 года, где изображен ударник, пробитый образец стали и дано описание полноценного испытания того времени:

«Броня толщиной 400 мм пробита снарядом 125-мм пушки Д-2 при скорости встречи $V_{уд}$ = 1390 м/с и массе снаряда $M_{сн}$ = 4 кг. Объём зоны разогрева, исходя из диаметра круга на лицевой стороне плиты 300-350 мм и на тыльной стороне 90 мм, составляет 13,4 дм$^3$, а масса разогретого металла $M_{мет}$ = 105 кг.

Минимальная температура на границе зоны $t_{min}$ = 350°С. Минимальное количество теплоты в зоне разогрева

$$Q = M_{мет} \cdot t_{min} \cdot \xi,$$

где $\xi$ = 0,103 ккал/кг·град – коэффициент теплоемкости стали.

$$Q = 105 \cdot 350°C \cdot 0{,}103 = 3785 \text{ ккал}.$$

Переведем тепловую энергию в джоули, получим:

$$E_т = Q \cdot 4{,}2 \cdot 10^3 = 3785 \cdot 4{,}2 \cdot 10^3 = 15{,}9 \cdot 10^6 \, Дж.$$

Кинетическая энергия снаряда

$$E_{сн} = \frac{m_{сн} \cdot V_{уд}^2}{2} = 3{,}86 \cdot 10^6 \, Дж .$$

Отношение

$$\frac{E_т}{E_{сн}} = \frac{15{,}9 \cdot 10^6}{3{,}86 \cdot 10^6} = 4{,}12. .$$

(Кпд более 400%.)» (С)

В экспериментальных работах 1993 года калибр пушки был взят меньше в 5 с лишним раз, а масса снарядов была меньше почти в 60 раз. Скорость снарядов также была меньше, но не намного (1240 м/с). Автор связывает изменение величины «лишней энергии» с изменением параметров в опыте.

Если вспомнить Часть 2 («Движение твёрдых тел в среде, при котором вносимая их движением удельная энергия возмущения превышает по величине удельную энергию

среды»), то станет ясно, что так называемая «лишняя энергия» есть энергия термодинамических сил сопротивления, которая и разогрела металл.

Прирост термодинамических сил сопротивления зависит от параметров движения, в т.ч. от площади взаимодействия лобовой поверхности снаряда, от его массы, т.к. от массы зависит время торможения, а значит и время прироста, от скорости, т.к. прирост происходит не на любой скорости и т.д. Возможно, что прирост термодинамических сил сопротивления происходит при скорости снаряда меньше звуковой, по типу действия термодинамических сил сопротивления в водной среде. Ещё одна особенность в том, что отсутствуют силы тылового сопротивления.

Если применить положения механики безынертной массы в теории упругости и пластичности материалов, выявив особенности действия сил в твёрдой среде, то этот конкретный парадокс, который не давал уважаемому В. Яворскому покоя всю жизнь, перестанет быть парадоксом: никакого дисбаланса энергий нет. Тем более, в данном случае нет нарушения закона сохранения энергии, а только лишь его очередное подтверждение. Поэтому, когда теория упругости и пластичности материалов будет пересмотрена, как говорил автор, тогда можно будет вести «корректный учёт» «процессов большой сложности», но не ранее. Другого объяснения «парадокса» как не было, так и не будет, потому что другого объяснения нет.

Кстати, автор статьи в опытах 1993 года отмечает особенности процесса: он описывает вторую фотографию, приведенную в статье, так:

«Модели бронебойных снарядов – ударники массой 61,5 г и 88,5 г. и образцы броневой стали. При внедрении ударника в образец внутри материала развивались давления до $3{,}6 \cdot 10^4$ кгс/см$^2$ и возникали сквозные трещины. Чтобы избежать потерь энергии, образец был запрессован в массивное стальное кольцо». Если судить по фотографии, где показаны эти ударники различных масс, они различаются только по длине, а площадь их сечения и форма лобовой поверхности одинаковы.

***

Теперь возьмём другой номер журнала «Наука и жизнь» (№ 11 за 1994 год). Рубрика называется «Переписка с читателем». Другой неравнодушный человек – д.т.н. С. Смирнов - пытается докричаться, словно новый Левша, до «кого следует», используя в качестве рупора журнал (только рупор бессилен против глухих). Его статья называется «Доктрина треснула, но всё же уцелела». Т.е. это не первая статья на тему, которой он занимается. Первая была помещена в № 2 за тот же год, но этого номера у меня нет. Судя по названию второй статьи, нужный эффект первой статьей достигнут не был – всё осталось в том же положении, а «трещину» в доктрине, её, ведь, можно потихоньку и замазать со временем, что, видно, и произошло за 10 лет, прошедших со времени написания этой статьи, ибо «научные» представления так и не изменились.

Теперь о теме, ибо она действительно очень важная. Буквально жизненно важная.

Автор утверждает, после изложения ряда аргументов, что «необходимо понять, что до сих пор мы не располагаем объективной информацией о движении грунта при землетрясении». Он утверждает, что, кроме так называемых колебательных движений, существуют ударные волны. Эти волны не учитываются приборами, потому что приборы к такому учёту не приспособлены, но именно эти волны, которые считают несуществующими, потому что они не регистрируются, вызывают большинство разрушений.

Согласно положениям механики безынертной массы сейсмический вид движения в принципе не отличается от акустического вида движения. Следовательно, действие сил в перемещающейся от эпицентра зоне возмущения, направлено в двух взаимно перпендикулярных направлениях. Движение грунта в поступательном направлении образует трещины, которые вблизи эпицентра могут достигать ширины в несколько метров, т.е. по их ширине, плотности грунта и т.п. параметрам можно судить о скорости движения вещества в поступательном потоке, если, конечно, пользоваться теорией механики безынертной массы. А также это поступательное движение, или даже сдвиг, естественно, вызывает колебания зданий. Оно же и регистрируется приборами.

Но, согласно автору статьи, и то, что регистрируется, регистрируется в искаженном виде. В частности, он пишет, что «если верить сейсмограммам», а не своим глазам, то «амплитудные смещения не превышают 10-20 сантиметров и окончательная величина их всегда равна нулю <…>. Значит, зафиксированные колебания *совсем* не отражают реальной картины».

Автор статьи, конечно, тоже не мог знать реальной картины движения грунта, но он приблизился к пониманию действительности, идя по следу действия сил, каковым является характер разрушений.

Движение грунта в нормальном направлении, т.е. нормальные динамические силы давления, действуют на здание наподобие заложенной под фундаментом взрывчатки. Не располагая знанием о существовании этих сил, точнее - прямым доказательством их существования, а только косвенным - в виде характера разрушений, ибо современная акустическая теория и прямые доказательства – несовместимые вещи, автор статьи предлагает, например, «на первом этаже, сразу над тонкими стойками, разместить массивный элемент, отражающий и рассеивающий волны», а также «необходимо исключить нарушение связей фундамента и здания, используя с этой целью материал с высокой прочностью на растяжение (например, металл или древесину), но, ни в коем случае, не следует применять в таких случаях кирпич или даже бетон».

Предложения очень дельные, но всё равно повышение сейсмостойкости зданий, если и идет сейчас, то только за счёт экспериментальных работ (что, собственно, и есть замазывание трещины в доктрине, дабы она устояла). Итак, согласно автору статьи, теория, которая существует, помочь в деле конструирования сейсмостойких зданий, и даже нормальных регистрационных приборов, не способна.
Здесь может помочь только теория механики безынертной массы, в том числе в своем будущем приложении к теории упругости и пластичности материалов

Я также думаю, что теория механики безынертной массы способна на большее, чем конструирование сейсмостойких зданий, а именно – она поможет найти способ нейтрализации энергии сейсмической волны, может быть, используя принцип глушителя, или ещё какой-либо. Принцип работы глушителя заключается в том, что от избытка энергии освобождаются при помощи сброса части массы, т.е. данный принцип «глушения» не универсальный в практическом отношении (на практике применяются разные методы «глушения», но они не могут быть доведены до совершенства без помощи новой теории). Если окружить город, стоящий в сейсмоопасной зоне, защитным сооружением (не считают же роскошью строить плотины против наводнений, так как это дешевле, нежели периодическое восстановление руин), то ему будут угрожать только те землетрясения, эпицентр которых находится непосредственно под городом. Или слишком близко от города, так как защита может не выдержать, в смысле - нейтрализовать не всю энергию сейсмической волны. Если бы такая возможность была полностью исключена, то можно было бы строить обычные, не сейсмостойкие, сооружения и дома внутри защищенной зоны.

Если же по-прежнему утверждать, что волна – это сжатие и разрежение вещества среды, то никогда человечество не будет находиться в безопасности, как оно не находится сегодня и не находилось «третьего дня». Стало быть, ничего не изменится ни завтра, ни послезавтра, никогда. Поэтому человечество, по-прежнему, будет приносить жертвы своим заблуждениям.

***

По всем вопросам можно обращаться к редактору по адресу: arodionova1@rambler.ru